\documentclass[aps,reprint,twocolumn,superscriptaddress,nofootinbib,longbibliography,notitlepage,floatfix]{revtex4-2}
\usepackage{graphicx}
\usepackage{tabularx}
\usepackage{amsmath}
\usepackage{amstext}
\usepackage{amssymb}
\usepackage{xfrac}
\usepackage[colorlinks,citecolor=blue]{hyperref}

\usepackage{graphicx}
\usepackage{amsmath}
\usepackage{amstext}
\usepackage{amssymb}
\usepackage{amsfonts}
\usepackage{longtable,booktabs}
\usepackage{hyperref}
\usepackage{url}
\usepackage{subfigure}
\usepackage{dsfont}
\usepackage{booktabs}
\usepackage{amsbsy}
\usepackage{dcolumn}
\usepackage{amsthm}
\usepackage{bm}
\usepackage{esint}
\usepackage{multirow}
\usepackage{hyperref}
\usepackage{cleveref}
\usepackage{amsmath}
\usepackage{amsfonts}
\usepackage{amssymb}
\usepackage{mathrsfs}
\usepackage{amsfonts}
\usepackage{amsbsy}
\usepackage{dcolumn}
\usepackage{bm}
\usepackage{multirow}
\usepackage{color}
\usepackage{extarrows}
\usepackage{datetime}
\usepackage{comment}
\usepackage[super]{nth}
\usepackage{tikz-cd}

\usepackage{makecell}
\usepackage{booktabs}
\usepackage{threeparttablex}

\usepackage{scalerel}
\usepackage{tikz}
\usetikzlibrary{svg.path}

\usepackage{xr}

\definecolor{orcidlogocol}{HTML}{A6CE39}
\tikzset{
orcidlogo/.pic={
	\fill[orcidlogocol] svg{M256,128c0,70.7-57.3,128-128,128C57.3,256,0,198.7,0,128C0,57.3,57.3,0,128,0C198.7,0,256,57.3,256,128z};
	\fill[white] svg{M86.3,186.2H70.9V79.1h15.4v48.4V186.2z}
	svg{M108.9,79.1h41.6c39.6,0,57,28.3,57,53.6c0,27.5-21.5,53.6-56.8,53.6h-41.8V79.1z M124.3,172.4h24.5c34.9,0,42.9-26.5,42.9-39.7c0-21.5-13.7-39.7-43.7-39.7h-23.7V172.4z}
	svg{M88.7,56.8c0,5.5-4.5,10.1-10.1,10.1c-5.6,0-10.1-4.6-10.1-10.1c0-5.6,4.5-10.1,10.1-10.1C84.2,46.7,88.7,51.3,88.7,56.8z};
}
}

\newcommand\orcidicon[1]{\href{https://orcid.org/#1}{\mbox{\scalerel*{
			\begin{tikzpicture}[yscale=-1,transform shape]
				\pic{orcidlogo};
			\end{tikzpicture}
		}{|}}}}

\hypersetup{
colorlinks=magenta,
linkcolor=blue,
filecolor=magenta,
urlcolor=magenta,
}
\def\Z{\mathbb{Z}}

\newtheorem{theorem}{Theorem}

\newtheorem*{theorem*}{Theorem}

\def\S{\mathsf{S}}

\def\OS{\mathsf{OS}}
\def\D{\mathsf{D}}
\def\OD{\mathsf{OD}}

\newcommand{\cat}{|\psi\rangle_{\text{cat}}}
\newcommand{\cluster}{|\psi\rangle_{\text{cluster}}}
 
\newcommand{\zq}{|\psi\rangle_{\mathbb{Z}_q}}   
\newcommand{\threedtc}{|\psi\rangle_{\text{3D TC}}}
\newcommand{\xcube}{|\psi\rangle_{\text{X-cube}}}

\def\Z{\mathbb{Z}}

\usepackage{extarrows}
\usepackage{datetime}

\usepackage{comment}

\usepackage{lineno}

\usepackage{booktabs}

\begin{document}

%\title{Subdimensional Entanglement Entropy: From Geometric-Topological Response to Mixed-State Holography}
\title{Subdimensional Entanglement Entropy: from Geometric-Topological Response to Entanglement-Induced Mixed-State Holography}

\author{Meng-Yuan Li\orcidicon{0000-0001-8418-6372}}
\affiliation{School of Physics, State Key Laboratory of Optoelectronic Materials and Technologies, and Guangdong Provincial Key Laboratory of Magnetoelectric Physics and Devices, Sun Yat-sen University, Guangzhou, 510275, China}
\affiliation{Institute for Advanced Study, Tsinghua University, Beijing, 100084, China}
\author{Peng Ye\orcidicon{0000-0002-6251-677X}}
\email{yepeng5@mail.sysu.edu.cn}
\affiliation{School of Physics, State Key Laboratory of Optoelectronic Materials and Technologies, and Guangdong Provincial Key Laboratory of Magnetoelectric Physics and Devices, Sun Yat-sen University, Guangzhou, 510275, China}

%\thanks{}

\date{{\color{cyan}\textbf{\today}}}
\begin{abstract}
	We introduce the subdimensional entanglement entropy (SEE), defined on subdimensional entanglement subsystems (SESs) embedded in the bulk, as an entanglement-based probe of how geometry and topology jointly shape universal properties of quantum matter. By varying the dimension, geometry, and topology of the SES, we show that the subleading term of SEE exhibits sharply distinct responses in a series of stabilizer states representing fixed points of different phases, including cluster states, $\mathbb{Z}_q$ topological orders, and fracton orders. Treating the reduced density matrix of an SES as a many-body mixed state supported on the SES manifold, we further establish a general correspondence between bulk stabilizers and mixed-state symmetries on SESs, separating them into strong and weak classes, and use it to identify strong-to-weak spontaneous symmetry breaking within SESs. Finally, for SESs with nontrivial SEE, we show that a set of weak symmetries acts as transparent patch operators of the corresponding strong symmetries. This motivates the notion of transparent composite symmetry, which remains robust under bipartition-factorized finite-depth quantum circuits that preserve SEE, and implies that a $D$-dimensional SES may holographically encode a $(D+1)$-dimensional topological order. These results establish SEE not only as a sharp probe of geometric-topological response, but also as a route from bulk pure-state entanglement to mixed-state symmetry and holography on subdimensional manifolds.
\end{abstract}

%\keywords{}

\maketitle
\tableofcontents

\section{Introduction}

Recently, a broad class of gapped phases, including subsystem symmetry-protected topological (SSPT) phases and fracton orders, has been shown to exhibit unconventional entanglement entropy (EE) scaling beyond standard constant topological entanglement entropy (TEE) or size-dependent topological ground-state degeneracy~\cite{Levin2006,Kitaev2006,Haah2011,Vijay2015,Vijay2016,Zou2016,Shirley2019c,Williamson2019a,Ma2018a,Shi2018,Zhang2024,You2018,Stephen2019,Kato2020}. 
These phenomena indicate that, in such systems, infrared orders are shaped not only by topology but also in an essential way by geometry. 
This motivates the search for an entanglement-based framework that can cleanly distinguish geometric and topological contributions.

In this paper, we develop such a framework by introducing the \textit{subdimensional entanglement entropy} (SEE), defined on \textit{subdimensional entanglement subsystems} (SESs). 
An SES is typically a manifold embedded in the bulk, with \textit{well-defined} intrinsic dimension and codimension in the continuum limit. 
This distinguishes SESs from both standard bulk-like subsystems and more general graph-like skeletal subsets of zero volume~\cite{Berthiere2022,Lyu2025}. 
Specifically, since an SES admits a well-defined continuum manifold interpretation, its geometric and topological attributes can be varied in a controlled way, allowing SEE to probe geometric-topological response directly. 

\begin{figure}
	%\centering
	\includegraphics[width=1.0\linewidth]{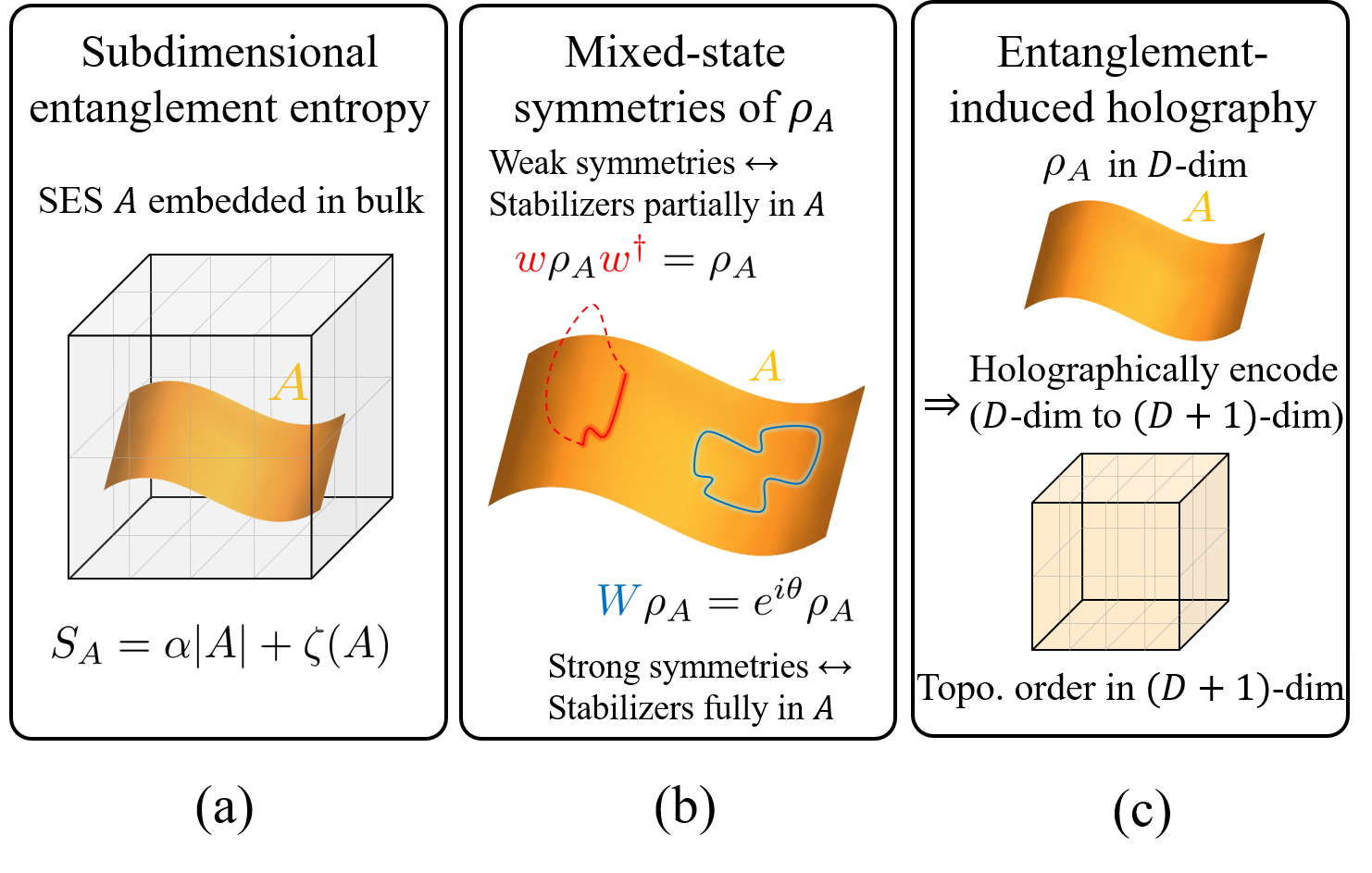}
	\caption{An overview of the main results of this work. (a) Schematic of a subdimensional entanglement subsystem (SES) $A$, whose subdimensional entanglement entropy (SEE) is composed of a volume law term $\alpha|A|$ and a subleading term $\zeta(A)$. The topological and geometric responses of $\zeta(A)$ characterize various kinds of orders. (b) Mixed-state symmetries of the reduced density matrix $\rho_A$ of SES $A$. The strong symmetries $W$ and weak symmetries $w$ of $\rho_A$ are microscopically originated from stabilizers fully and partially embedded in $A$, respectively. Interestingly, strong-to-weak spontaneous symmetry breaking (SW-SSB) is observed for SESs with nontrivial SEE. (c) Mixed-state holography induced by SEE: for an arbitrary $D$-dimensional SES with nontrivial SEE considered in this work, its strong and weak symmetries compose transparent composite symmetries (TCS), whose algebra holographically encodes a topological order in $(D+1)$ dimensions.}
	\label{fig:pic}
\end{figure}

SEE thus provides a response framework for quantum many-body states that is information-theoretic in nature. 
This is qualitatively different from conventional geometric-topological response theories, where one probes quantum phases by deforming the entire spacetime manifold or the bulk lattice~\cite{Qi2008a,Ryu2012b,Stone2012a,Ye2013c,Cheng2014a,Lapa2017a,Nissinen2019,Han2019}. 
While conventional EE extracts limited information from bulk bipartitions, SEE gives rise to a \textit{structured} family of entanglement data associated with SESs of \textit{different dimensions, geometries, and topologies}. 
As we show below, this framework does more than refine entanglement diagnostics: once an SES is specified, its reduced density matrix naturally defines a mixed many-body state supported on that submanifold,  opening a direct route from pure-state entanglement to mixed-state physics including \textit{mixed-state symmetry structures} and \textit{topological holography} on SESs~\cite{Buca2012,deGroot2022,Ma2023a,Ma2023,Lieu2020,Albert2014,Ellison2025,Wang2025b,Zhang2025,Lessa2025,Lee2022,Sala2024,Lu2025,Luo2025,Chatterjee2023,Ji2019,Kong2020a,Inamura2023}. For clarity, an overview of the main results of this work is provided in Fig.~\ref{fig:pic}. 

Quantitatively, for an SES $A$, we define the SEE through $S_A=-\mathrm{Tr}\,\rho_A\log_2\rho_A$ which typically takes the form (Fig.~\ref{fig:pic}a)
\begin{align}
	S_A=\alpha |A|+\zeta(A),
\end{align}
where $|A|$ denotes the volume of $A$, $\rho_A$ is the reduced density matrix on $A$, $\alpha$ is a nonuniversal coefficient, and $\zeta(A)$ is a subleading term that can contain universal information. Note that since $A$ is subdimensional, its volume is equivalently its surface measure. 

In this paper, we restrict our computation of SEE to stabilizer states, which allows us to consider various kinds of phases, including SSPT orders, $\mathbb{Z}_q$ topological orders, and fracton orders, on the same footing. Such stabilizer states realize the ground states of a series of representative models for these phases, and lead to a wealth of important insights into them~\cite{OHern1999,Mishra2004,Batista2005,Nussinov2005,Seiberg2021a,Distler2022,Devakul2018b,Zhang2024,Dennis2002,Kitaev2003,Bullock2007,Watanabe2023,Hamma2005,Vijay2016}. By analytically computing SEE, we demonstrate that SEE resolves geometric and topological structures invisible to conventional EE. In particular, SSPT phases~\cite{You2018} and topological orders can exhibit similar EE scaling~\cite{Zou2016,Williamson2019a,Stephen2019,Kato2020}, yet SEE distinguishes them sharply: in SSPT phases, $\zeta(A)$ depends on the \textit{geometry and orientation} of $A$, whereas in topological orders it depends only on the \textit{topology} of $A$. 
This motivates two complementary notions, namely \textit{geometric SEE} (gSEE) and \textit{topological SEE} (tSEE), which can coexist in fracton orders.

A second central aspect of this work is to view $\rho_A$ not merely as an auxiliary reduced density matrix, but as the density matrix of a \textit{bona fide} mixed many-body state living on the SES (Fig.~\ref{fig:pic}b). 
Surprisingly, this perspective allows us to study mixed-state symmetries on SESs while varying the dimension, geometry, and topology of the underlying submanifold, which leads to a bridge connecting pure-state entanglement in bulk and mixed-state physics in lower dimensions. 
In open quantum systems, symmetries are commonly divided into \textit{strong} and \textit{weak} classes~\cite{Buca2012,deGroot2022,Ma2023a,Ma2023}: for a mixed state $\rho$, a strong symmetry $W$ satisfies $W\rho=e^{i\theta}\rho$ for some phase $\theta$, whereas a weak symmetry $w$ satisfies $w\rho w^\dagger=\rho$. 
For stabilizer states, we  establish a rigorous correspondence between stabilizers and mixed-state symmetries on an SES $A$: stabilizers fully supported on $A$ generate strong symmetries, while stabilizers partially supported on $A$ generate weak symmetries, as stated in \textbf{Theorem}~\ref{Theorem:strong_weak_stabilizers}. 
In this way, both strong and weak symmetries of the mixed state on the SES acquire a concrete microscopic origin in the stabilizer structure of the bulk pure state. 

Furthermore, we investigate strong-to-weak spontaneous symmetry breaking (SW-SSB)~\cite{Lieu2020,Albert2014,Wang2025b,Ellison2025,Zhang2025,Lee2022,Lessa2025,Sala2024}, finding that SW-SSB occurs in a series of SESs with nontrivial SEE, for both global and $1$-form strong symmetries. 
These results connect SES physics to broader developments in mixed-state phases, including thermal states, intrinsically mixed topological orders and decohered topological codes~\cite{Wang2025b,Ma2023,Ellison2025,Zhang2025,Sohal2025,Lessa2025}.

Remarkably, a topological holography between mixed-state symmetry structures on SESs and higher dimensional topological orders induced by SEE is also identified (Fig.~\ref{fig:pic}c). The algebra generated by mixed-state symmetries of SESs aligns naturally with the transparent patch (t-patch) operator framework~\cite{Chatterjee2023} developed in the study of topological holography by Wen \textit{et al.}~\cite{Ji2019,Kong2020a,Chatterjee2023,Inamura2023}. 
In that framework, t-patch operators in a $D$-dimensional system organize local and extended symmetric operators whose algebra holographically encodes a $(D{+}1)$-dimensional topological order. Related developments extend topological holography to fracton stabilizer codes~\cite{HuZhangYe2026}. 
For SESs with nontrivial SEE, namely $\zeta(A)\neq 0$, we find that a set of weak symmetries acts as t-patch operators associated with the corresponding strong symmetries. 
This correspondence remains invariant under bipartition-factorized FDQCs preserving SEE, which motivates us to define a composite of strong and weak symmetries, dubbed \textit{transparent composite symmetry} (TCS) [see Table~\ref{tab:comp_sym} and Fig.~\ref{fig:com_sym}]. 
As a consequence, the mixed state supported on a $D$-dimensional SES may holographically encode a $(D+1)$-dimensional topological order.

This paper is organized as follows. 
In Sec.~\ref{sec:pre}, we briefly review the entanglement entropy of Pauli stabilizer states and transparent-patch operators, both of which are crucial for the subsequent discussion. 
Useful notations are introduced in Sec.~\ref{subsec:notations}. 
In Sec.~\ref{sec:response}, we compute SEE for SESs of various topologies and geometries in stabilizer states with different kinds of order, including spontaneous subsystem symmetry breaking, topological order, SSPT order, and fracton order. 
In Sec.~\ref{sec:ses_sym}, we establish a general relation between stabilizers and symmetries of SES mixed states by proving \textbf{Theorem}~\ref{Theorem:strong_weak_stabilizers}, and demonstrate the SW-SSB properties of SESs with either global or $1$-form strong symmetries. 
In Sec.~\ref{sec:tcs}, we explicitly compute the strong and weak symmetries for a series of SESs with nontrivial SEE, and show that they form transparent composite symmetries, which holographically encode various topological orders (TO) in one higher dimension, with the results summarized in Table~\ref{tab:comp_sym}. Finally, we summarize our findings and discuss possible future directions in Sec.~\ref{sec:outlook}.

\section{Preliminaries}
\label{sec:pre}

In this section, we briefly introduce several topics that are essential to this work, including the computation of the entanglement entropy of stabilizer states and the framework of transparent patch operators. Finally, in Sec.~\ref{subsec:notations}, we introduce some useful notations and conventions.

\subsection{Entanglement entropy of stabilizer states}

As we focus on the stabilizer formalism, where SEE is computed for specific stabilizer states, we first review the derivation of a general formula for the entanglement entropy of stabilizer states (Eq.~\eqref{eq:EE}) following Refs.~\cite{Fattal2004,Hamma2005a,Ma2018a}. In this paper, we analyze the SEE of stabilizer states composed of $q$-dimensional qudits, focusing on prime $q$ for simplicity. A qudit of dimension $q$ generalizes the qubit, which is recovered in the case $q=2$. We denote the generalized Pauli operators for qudits by $\tilde{X}$ and $\tilde{Z}$ (reducing to standard Pauli operators $X,Z$ for qubits). A stabilizer state $|\psi\rangle$ is defined as the unique eigenstate with eigenvalue $+1$ of a complete set of commuting products of generalized Pauli operators, referred to as stabilizer generators.

To derive the entanglement entropy for these states, we begin with a set of independent and mutually commuting stabilizer generators $\mathcal{S}=\{O_{i}\}$. There are $N$ qudits in total, where each qudit can be viewed as a spin-$(q-1)/2$ degree of freedom. We require the existence of a unique stabilizer state $|\psi\rangle$ stabilized by all $O_{i}$, i.e., $O_{i}|\psi\rangle=|\psi\rangle$ for all $i$. Equivalently, $|\psi\rangle$ is the unique ground state of the parent Hamiltonian $H=-\sum_{i}O_{i}$ (or $H=-\sum_{i}(O_{i}+O_{i}^{\dagger})$ for non-Hermitian $O_i$). We note that the uniqueness of $|\psi\rangle$ may require some generators $O_{i}$ to be nonlocal. Since a simultaneous eigenstate of all $O_i$ is completely determined by their eigenvalues, we denote such an eigenstate by $|\vec{k}\rangle$, where $\vec{k}=(k_{1},k_{2},\cdots,k_{N})$ with $k_{l}\in \{0,1,\cdots, q-1\}$ specifying the eigenvalue of $O_{l}$, i.e., $O_{l}|\vec{k}\rangle=\omega^{k_l}|\vec{k}\rangle$, with $\omega = e^{i\frac{2\pi }{q}}$.

Consider the Abelian group $G$ multiplicatively generated by the generator set $\mathcal{S}$. Since all stabilizers in $\mathcal{S}$ are independent and satisfy $O_i^q = I$ (where $I$ is the identity operator), $G$ is isomorphic to $(\mathbb{Z}_{q})^{|\mathcal{S}|}$, where $|\mathcal{S}|$ denotes the cardinality of $\mathcal{S}$. Furthermore, an element $g\in G$ is uniquely determined by a sequence $\vec{n}=(n_{1},n_{2},\cdots,n_{|\mathcal{S}|})$, where $n_{i}\in \{0,1,\cdots, q-1\}$ denotes the exponent of $O_{i}$ in the product $g$ (e.g., $n_{i}=2$ implies a factor $O_{i}^2$). Its action on a simultaneous eigenstate $|\vec{k}\rangle$ is given by $g(\vec{n})|\vec{k}\rangle=\omega^{\vec{k}\cdot\vec{n}}|\vec{k}\rangle$.

Because $|\psi\rangle$ can be uniquely determined by eigenvalues of $\{O_i\}$, we have $|\mathcal{S}| = N$. Consequently, it is straightforward to find that the projector onto the state $|\psi\rangle$ takes the form $|\psi\rangle\langle\psi|=\frac{1}{|G|}\sum_{\vec{n}}g(\vec{n})$. Similar to the ground state projector, the density matrix for a general eigenstate is $\rho=|\vec{k}\rangle\langle\vec{k}|=\frac{1}{|G|}\sum_{\vec{n}}\omega^{-\vec{k}\cdot\vec{n}}g(\vec{n})$. Therefore, given a bipartition of the $N$ qudits into subsystems $A$ and $B$, the reduced density matrix $\rho_{A} = \text{Tr}_B \rho$ is given by $\rho_{A}=\frac{1}{|G|}\sum_{\vec{n}}\omega^{-\vec{k}\cdot\vec{n}}\text{Tr}_{B}g(\vec{n})$. If $g(\vec{n})$ acts nontrivially on any qudit in subsystem $B$, $\text{Tr}_{B}g(\vec{n})$ vanishes; otherwise, $\text{Tr}_{B}g(\vec{n})=q^{N_{B}}$. Therefore, we have
\begin{equation}
	\begin{aligned}
		\rho_{A}
		=\frac{|G_A|}{q^{N_A}}\left(\frac{1}{|G_A|}\sum_{\vec{n}_{A}}\omega^{-\vec{k}_{A}\cdot\vec{n}_{A}}g(\vec{n}_{A})\right),
		\label{eq:RDM}
	\end{aligned}
\end{equation}
where $\frac{q^{N_{B}}}{|G|} = \frac{|G_A|}{q^{N_A}} \times \frac{1}{|G_A|}$ is utilized, $N_{A}$ ($N_{B}$) is the number of qudits in subsystem $A$ ($B$), and $\vec{n}_{A}$ denotes a sequence such that $g(\vec{n}_A)$ is supported entirely within $A$. Here, $G_{A}$ is the subgroup of $G$ generated by stabilizers fully supported in $A$, and $\vec{k}$ is formally replaced by $\vec{k}_{A}$ to emphasize that distinct global states $|\vec{k}\rangle$ may become indistinguishable in $A$. Physically, $\vec{k}_{A}$ labels a subspace of the Hilbert space of $A$ in which every $g(\vec{n}_A)\in G_A$ has eigenvalue $\omega^{\vec{k}_{A}\cdot\vec{n}_{A}}$. Therefore, the term in parentheses in Eq.~\eqref{eq:RDM} acts as a projector onto the subspace specified by $\vec{k}_{A}$.

As $\rho_A$ has trace $1$ and the subspace specified by $\vec{k}_{A}$ has dimension $q^{N_A}/|G_A|$, the reduced density matrix $\rho_A$ is diagonal with $q^{N_A}/|G_A|$ non-zero eigenvalues, all equal to $|G_A|/q^{N_A}$. The von Neumann entropy of $A$ given by $S_{A}=-\text{Tr}\rho_{A}\log_{2}\rho_{A}$ is thus:
\begin{equation}
	\begin{aligned}
		S_{A}=N_{A} \log_2 q-\log_{2}|G_{A}|.
		\label{eq:EE}
	\end{aligned}
\end{equation}
Here, $|G_A|$ is the order of the subgroup of stabilizers fully supported within $A$ (including the identity).

Crucially, to characterize the SEE in a specific ground state of a given model (i.e., a stabilizer state), one must specify not only the local stabilizer generators defining the parent Hamiltonian but also a set of nonlocal stabilizers that complete the full stabilizer group. This complete set uniquely selects a representative ground state, whose SEE can be computed to probe the associated geometric and topological responses. As implied by the derivation, starting from a local Hamiltonian $H$ compatible with different choices of nonlocal stabilizers may yield eigenstates with distinct entanglement entropies for the same subsystem, such that degenerate ground states of the same Hamiltonian may also be distinguished by SEE.

\subsection{Topological holography and transparent patch operators}
\label{sec:rev_of_tpatch}

In this subsection, we briefly review the concept of transparent patch operators (t-patch operators), following Ref.~\cite{Chatterjee2023}. Conventionally, symmetries are described by groups of symmetry transformations. More recently, however, generalized symmetries---such as higher-form symmetries and subsystem symmetries---have challenged this paradigm. Because these symmetries are supported on subsystems of varying topology and geometry, their action cannot always be fully captured by groups. Alternatively, a symmetry can be physically understood as a set of constraints on local operators (e.g., Hamiltonian terms). The relevant physical properties are then encoded in the algebra formed by the operators satisfying these constraints, termed \emph{local symmetric operators} (LSOs). This algebraic perspective naturally accommodates generalized symmetries beyond the group description. Furthermore, in cases where distinct symmetry transformations impose identical constraints on operators, the algebra of LSOs may offer a more physically relevant description than the transformation group itself.

Although LSOs are local by definition, they can algebraically generate extended operators, such as string-like and membrane-like operators. These spatially extended operators must be included in the closure of the LSO algebra. Among these local and extended symmetric operators, a subclass consisting of \emph{patch operators} satisfying a transparency property is of special interest. Given a set of symmetry transformations, transparent patch (t-patch) operators are generated by corresponding LSOs and thus naturally encode the symmetry. Moreover, the algebra of t-patch operators encodes a braided fusion (higher) category via the fusion and braiding of their boundaries; consequently, the isomorphism classes of such algebras are referred to as \emph{categorical symmetries} in Ref.~\cite{Chatterjee2023}.

Concretely, a patch operator is a tensor network operator of the form:
\begin{equation}
	O_{P}=\sum_{{a_{i}}}\Phi(\{a_{i}\})\prod_{i\in P}O_{i}^{a_{i}},
\end{equation}
where $i$ denotes a site, $P$ is a patch (a region with the topology of an $n$-dimensional disk, such as a point, an open string, or an open membrane), $O^{a_{i}}_i$ is a local operator near site $i$ labeled by $a_i$ (including the identity $\text{id}_{i}$), and $\Phi(\{a_{i}\})$ is a complex weight determined by the configuration $\{a_{i}\}$.

Given a set of symmetry transformations $\{W_i\}$, patch operators $\{O_{P}\}$ are said to be \emph{transparent} if they are invariant under $\{W_i\}$ and satisfy the following condition:
\begin{equation}
	O_{P_{1}}O_{P_{2}}=O_{P_{2}}O_{P_{1}},\label{eq:transparent}
\end{equation}
provided the boundaries of the patches $P_{1}$ and $P_{2}$ (denoted $\partial P_{1}$ and $\partial P_{2}$) are not linked and are spatially well-separated.

Furthermore, t-patch operators can be classified into two categories: \emph{patch charge operators} and \emph{patch symmetry operators}. A t-patch operator $O_{P}$ is classified as a patch charge operator if it has an empty bulk (i.e., $O_{i}^{a_{i}}=\text{id}_{i}$ for all $i$ far from $\partial P$); otherwise, it is classified as a patch symmetry operator.

In this paper, to focus on the physical applications, we adopt the t-patch operators derived for specific symmetry transformations in Ref.~\cite{Chatterjee2023}. These t-patch operators have been proven to generate algebras corresponding to braided fusion (higher) categories and topological orders; see Sec.~\ref{sec:tcs}.

\subsection{Notations and conventions}
\label{subsec:notations}

For convenience, we introduce the following notations for subdimensional manifolds. An $n$-dimensional object is denoted by $\mathsf{S}^n$ (e.g., $\mathsf{S}^1$ for a string), and its dual-lattice counterpart by $\mathsf{D}^n$ (see Fig.~\ref{fig:2dtc} (b)). In particular, open objects are denoted by $\mathsf{OS}^n$ and $\mathsf{OD}^n$. Furthermore, we use $\hat{k}$ to represent the unit vector along the $k$-direction, and $v,l,p,c$ to denote a vertex, a link, a plaquette, and a cube, respectively.

Unless otherwise specified, our analysis is restricted to non-self-intersecting objects that contain no closed dual objects (e.g., a dual string that contains no loop in 2D), ensuring that such nonuniversal contributions can be safely ignored. For convenience, we adopt the convention $|A|\equiv N_A$ and assume periodic boundary conditions (PBC) by default.

\section{Geometric-topological response theory}
\label{sec:response}

In this section, we compute the SEE of various specific many-body states in stabilizer formalism to demonstrate its diagnostic power. For each system, we provide general results and detailed calculations.

\subsection{2D subsystem cat state}

\begin{figure}
	\centering
	\includegraphics[width=1.0\linewidth]{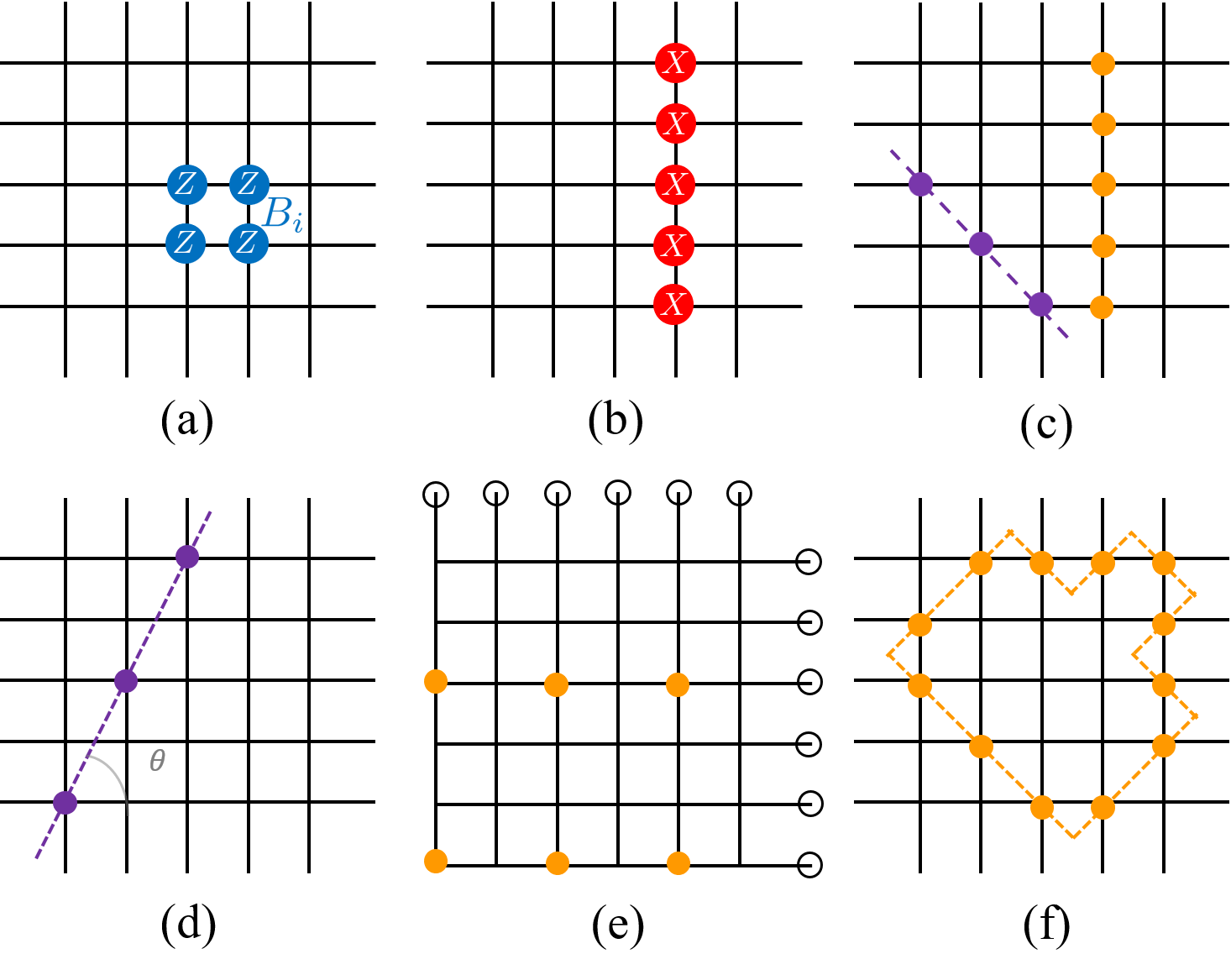}
	\caption{Representative Hamiltonian terms, subsystem symmetry and SESs of the 2D plaquette Ising model. Panel (a) shows a representative Hamiltonian term $B_i$ in Eq.~\eqref{eq:ham_pim}, where $i$ denotes a site, and the remaining terms follow by lattice translations. Panel (b) shows a typical subsystem symmetry generator $\prod_i X_i$ supported on an axial line. Panel (c)-(f) illustrate SESs with nonzero (orange) and zero (purple) subleading terms $\zeta(A)$. Panel (c) and (d) show segments of a straight line at angle $\theta$ to $x$-axis. Formally, a line at some special angles can reduce to a rectangular array of sites with nonzero $\zeta(A)$ as shown in panel (e), where circles highlight the identification of sites under PBC. Finally, panel (f) shows a loop-like SES.}
	\label{fig:pim}
\end{figure}

We first examine the subsystem cat state $|\text{cat}\rangle$ as an example of 2D systems with subsystem symmetries. Such symmetries interpolate between global and local symmetries, as they are supported on rigid subsystems of intermediate dimensionality. This state is realized as a ground state of the 2D plaquette Ising model~\cite{Xu2004}, which is a representative lattice model with spontaneous subsystem symmetry breaking order~\cite{OHern1999,Mishra2004,Batista2005,Nussinov2005,Seiberg2021a,Distler2022}. The Hilbert space consists of spin-$1/2$ degrees of freedom (qubits) on the sites of an $L\times L$ square lattice with PBC. The parent Hamiltonian is [Fig.~\ref{fig:pim}(a)]:
\begin{equation}
\label{eq:ham_pim}
	H=-\sum_{i}B_{i},
\end{equation}
where $B_{i}=Z_{i}Z_{i+\hat{x}}Z_{i+\hat{y}}Z_{i+\hat{x}+\hat{y}}$. This model possesses linear subsystem symmetries generated by $W(\mathsf{S}^1) = \prod_{j\in {\mathsf{S}^1}}X_j$, where $\mathsf{S}^1$ denotes a straight line along the $x$ or $y$ direction [Fig.~\ref{fig:pim}(b)]. There are $2L-1$ such independent generators in total.

We define the subsystem cat state by adopting the linear subsystem symmetry generators $W(\mathsf{S}^1)$ as nonlocal stabilizers. Explicitly, $\cat=2^{\frac{-2L+1}{2}}(|000\cdots0\rangle+W(\mathsf{S}^1)|000\cdots0\rangle+\cdots)$, where the summation runs over all $2^{2L-1}$ elements in the subsystem symmetry group. Thus, the complete stabilizer set includes the local terms $B_i$ and the nonlocal operators $W(\mathsf{S}^1)=\prod_{i\in \mathsf{S}^1}X_i$ for straight lines $\mathsf{S}^1$ aligned with the $x$ or $y$ axis.

For the subsystem cat state, we find that the subleading term $\zeta(A)$ in the SEE is sensitive to the geometric orientation of the SES, thus it should be regarded as a geometric SEE (gSEE) term. For a general straight line SES $A$ closed under PBC oriented at an angle $\theta$ to the $x$-axis, the subleading term is given by [Fig.~\ref{fig:pim}(c)]:
\begin{align}
	\zeta(A)=-\delta(\theta,0)-\delta\!\left(\theta,\tfrac{\pi}{2}\right),
\end{align}
where $\delta(x,y)$ is the Kronecker delta. This formula indicates that a non-zero correction appears only when the line-like SES aligns with the $x$ or $y$ axes.

This formula is obtained by  explicitly calculating SEE for SESs of various kinds:
\begin{itemize}
	\item For a straight line of $L$ spins along the $x$ direction, we have $N_{A}=L$, $|G_{A}|=2$, so $S_{A}=N_{A}-\log_{2}|G_{A}|=L-1 = |A| - 1$.
	\item For a straight line of $L$ spins along diagonal direction, we have $N_{A}=L$, $|G_{A}|=1$, so $S_{A}=N_{A}-\log_{2}|G_{A}|=L = |A|$.
	\item For a straight line closed under PBC at an angle $\theta$ to the $x$ axis ($0<\theta<\frac{\pi}{2}$), we have $N_{A}=l$, $|G_{A}|=1$, so $S_{A}=N_{A}-\log_{2}|G_{A}|=l = |A|$ [Fig.~\ref{fig:pim}(d)]. Seemingly, $\zeta(A)\neq 0$ at special commensurate angles satisfying $\tan(\theta)=\frac{q_y}{q_x}$, where $q_x$ and $q_y$ are coprime integers and simultaneously satisfy $\gcd(q_x,L)>1,\gcd(q_y,L)>1$. Nevertheless, for such special angles $A$ actually becomes  a rectangular array of sites due to PBC and is no longer line-like [Fig.~\ref{fig:pim}(e)], thus such special cases can be ignored when we focus on line-like SESs.
	\item For a closed loop of $l$ spins as in Fig.~\ref{fig:pim} (f), we have $N_{A}=l$, $\log_{2}|G_{A}|=3$, so $S_{A}=N_{A}-\log_{2}|G_{A}|=l-3 = |A| - 3$.
\end{itemize}

Notably, the constant term in the SEE of the loop SES is \textit{not} a consequence of the loop topology. Rather, it arises because the specific geometry of the loop fully contains three independent ``discrete rectangle'' stabilizers. Each discrete rectangle involves four vertices of a rectangle and contributes a $-1$ to the SEE, as it supports a stabilizer; for a subsystem consisting solely of such a discrete rectangle ($N_A=4$), one finds $|G_A|=2$ and thus $S_A = 4-1=3$.

\subsection{2D cluster state}

\begin{figure}
	\centering
	\includegraphics[width=1.0\linewidth]{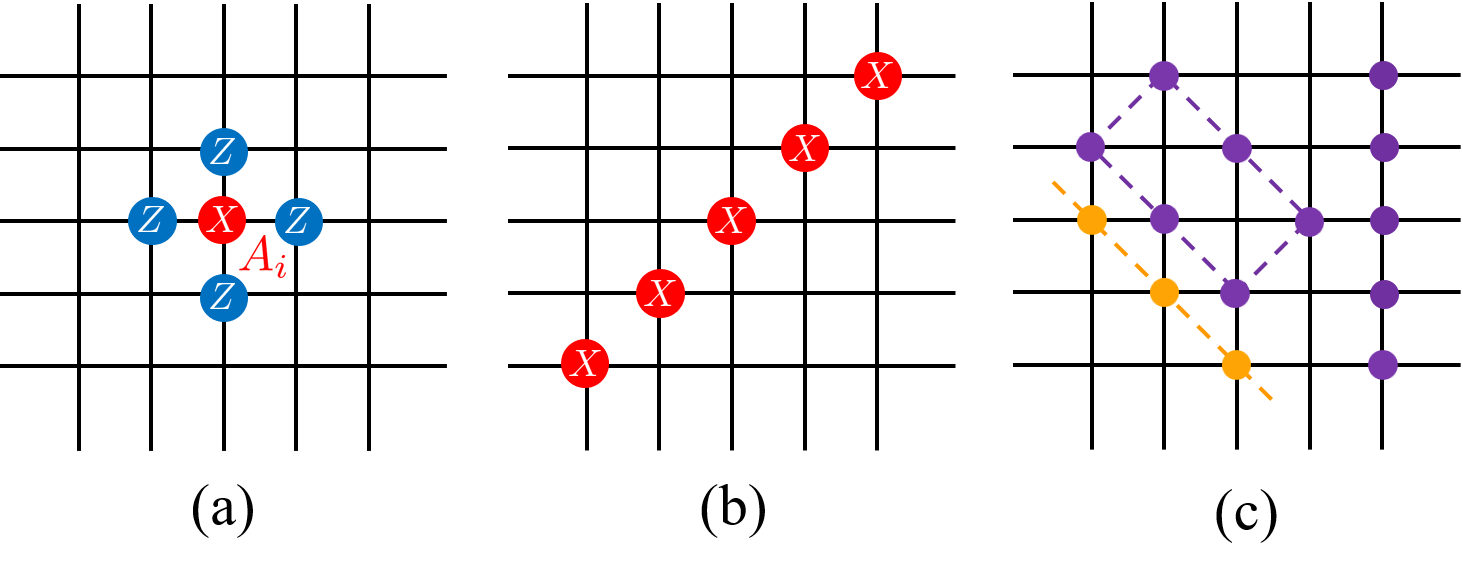}
	\caption{Representative Hamiltonian terms, subsystem symmetry and SESs of the 2D cluster model. Panel (a) shows a representative Hamiltonian term $A_i$ in Eq.~\eqref{eq:ham_cluster}, where $i$ denotes a site. Panel (b) shows a typical subsystem symmetry generator $\prod_i X_i$ supported on a diagonal line. Panel (c) illustrates SESs with nonzero (orange) and zero (purple) subleading terms $\zeta(A)$.}
	\label{fig:cluster}
\end{figure}

Next, we consider a 2D cluster state $\cluster$ as a representative example of subsystem symmetry protected topological (SSPT) orders~\cite{Raussendorf2001,You2018,Devakul2018a,Burnell2022,Zhou2022a}. The cluster state model considered here has linear subsystem symmetries supported on line-like subsystems, while it is straightforward to apply the SEE diagnostic to SSPT with subsystem symmetries supported on fractal, chaotic and more general patterns~\cite{Devakul2018b,Zhang2024}. The Hilbert space consists of qubits on the sites of an $L\times L$ square lattice. The parent Hamiltonian is [Fig.~\ref{fig:cluster}(a)]:
\begin{equation}
\label{eq:ham_cluster}
	H=-\sum_i A_i,
\end{equation} 
where $A_i = Z_{i-\hat{x}}Z_{i-\hat{y}}X_{i}Z_{i+\hat{x}}Z_{i+\hat{y}}$.
This model hosts SSPT order protected by linear subsystem symmetries. These symmetries are generated by $W(\mathsf{S}^1)=\prod_{i\in \mathsf{S}^1}X_i$, where $\mathsf{S}^1$ is a straight diagonal line [Fig.~\ref{fig:cluster}(b)]. Under PBC, SSPT phases possess a unique ground state, analogous to global SPT phases. Thus, we denote the unique ground state simply by $\cluster$ without the need to introduce additional nonlocal stabilizers.

For the 2D cluster state, the gSEE term appears only when the SES lies strictly along a linear subsystem that supports a symmetry generator. This orientation dependence is summarized as [Fig.~\ref{fig:cluster}(c)]:
\begin{align}
	\zeta(A)=-\delta\!\left(\theta,\tfrac{\pi}{4}\right)-\delta\!\left(\theta,\tfrac{3\pi}{4}\right),
\end{align}
where $A$ is a straight line closed under PBC. This formula is obtained by  explicitly calculating SEE for SESs of various kinds:

\begin{itemize}
	\item For a straight line of $L$ spins along
	$x$-axis, we have $N_{A}=L$, $|G_{A}|=1$, so $S_{A}=N_{A}-\log_{2}|G_{A}|=L = |A|$.
	\item For a straight line of $L$ spins along diagonal direction, we have $N_{A}=L$, $|G_{A}|=2$, so $S_{A}=N_{A}-\log_{2}|G_{A}|=L-1$.
    \item For a straight line closed under PBC at an angle $\theta$ to the $x$ axis ($-\frac{\pi}{4}<\theta<\frac{\pi}{4}$), we have $N_{A}=l$, $|G_{A}|=1$, so $S_{A}=N_{A}-\log_{2}|G_{A}|=l = |A|$.
	\item For a contractible closed loop of $l$ spins (extends only diagonally), we have $N_{A}=l$, $|G_{A}|=1$, so $S_{A}=N_{A}-\log_{2}|G_{A}|=l$.
	\item For a straight open string of $l$ spins (extends only diagonally), we have $N_{A}=l$, $|G_{A}|=1$, so $S_{A}=N_{A}-\log_{2}|G_{A}|=l$.
\end{itemize}

The distinct orientation dependence of $\zeta(A)$ in 2D subsystem cat and cluster states exemplifies the diagnostic utility of gSEE.

\subsection{2D $\Z_q$ toric code models}

In contrast to subsystem-symmetric systems, the SEE of the 2D $\mathbb{Z}_q$ topological order depends solely on topology. To demonstrate how SEE identifies topological order, we consider the 2D $\Z_q$ toric-code model as a representative example of an Abelian topological order~\cite{Dennis2002,Kitaev2003,Bullock2007,Watanabe2023}. Here qubits in the original 2D toric code are replaced by $q$-level qudits, and we assume $q$ is prime for simplicity. The Hilbert space is composed of qudits on links of a square lattice of the size $L\times L$. Here, a $q$-level qudit is defined by a $q$-dimensional local Hilbert space with an orthonormal basis $|0\rangle, |1\rangle, \cdots , |q-1\rangle$. For a qudit, we have generalized Pauli operators defined as follows:
\begin{align*}
	\tilde{X}|i\rangle &= |i + 1 \mod\  q\rangle,\\
	\tilde{Z}|i\rangle &= \omega^i |i\rangle,
\end{align*}
where $i=0,1,\cdots,q-1$, $\omega = e^{i\frac{2\pi}{q}}$. These operators satisfy $\tilde{Z}\tilde{X}=\omega \tilde{X}\tilde{Z}$.

\begin{figure}
	\centering
	\includegraphics[width=1.0\linewidth]{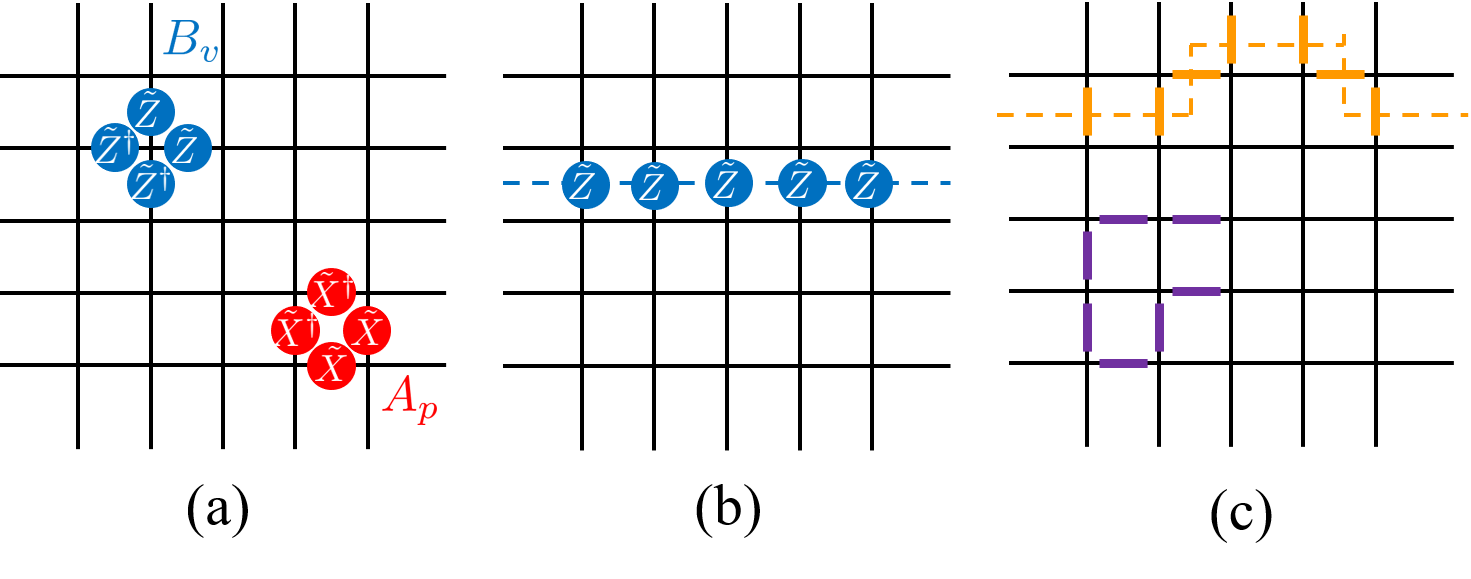}
	\caption{Representative Hamiltonian terms, logical operator and SESs of a 2D $\Z_q$ toric code model. Panel (a) shows representative Hamiltonian terms in Eq.~\eqref{eq:ham_2dtc}. Panel (b) shows a typical logical operator supported on a noncontractible dual loop. Panel (c) illustrates SESs with nonzero (orange) and zero (purple) subleading terms $\zeta(A)$.}
	\label{fig:2dtc}
\end{figure}

The Hamiltonian is given by [Fig.~\ref{fig:2dtc}(a)]:
\begin{equation}
\label{eq:ham_2dtc}
	H=-\sum_p (A_p + A_p^{\dagger}) -\sum_v (B_v + B^{\dagger}_v),
\end{equation}
where $A_p = \tilde{X}_{p+\frac{\hat{x}}{2}} \tilde{X}_{p-\frac{\hat{y}}{2}} \tilde{X}^{\dagger}_{p-\frac{\hat{x}}{2}} \tilde{X}^{\dagger}_{p+\frac{\hat{y}}{2}}$,  $B_v = \tilde{Z}_{v+\frac{\hat{x}}{2}} \tilde{Z}_{v+\frac{\hat{y}}{2}} \tilde{Z}^{\dagger}_{v-\frac{\hat{x}}{2}} \tilde{Z}^{\dagger}_{v-\frac{\hat{y}}{2}}$, $\hat{x}$ ($\hat{y}$) is the unit vector along the $x$ ($y$) direction .

With PBC, the ground states of a 2D $\Z_q$ toric-code model are $q^2$-fold degenerate. To specify a concrete ground state as a stabilizer state, we choose the nonlocal stabilizers to be the logical operators $W(\mathsf{D}^1_x)=\prod_{l\in\mathsf{D}^1_x}\tilde{Z}_l$ and $W(\mathsf{D}^1_y)=\prod_{l\in\mathsf{D}^1_y}\tilde{Z}_l$, where $\mathsf{D}^1_{x(y)}$ denotes a noncontractible dual loop along the $x$ ($y$) direction and $l$ labels a link [Fig.~\ref{fig:2dtc}(b)]. We denote the corresponding stabilizer state by $\zq$. Here, we choose straight representatives of the nonlocal stabilizers only for simplicity; their supports can be freely deformed by multiplying them by $B_v$ operators.

For this state, we find that the subleading term $\zeta(A)$ depends solely on the topology of the SES. Let $[A]$ denote the cobordism class of an SES $A$. In the convention used here, $[A]=[\emptyset]$ means that $A$ is closed and contractible. Let $\partial A$ denote the boundary of $A$. The results can be summarized as follows. For a string SES $A$, we have [Fig.~\ref{fig:2dtc}(c)]:
\begin{align}
	\zeta(A)=
	\begin{cases}
		-\log_2 q, & [A]=[\emptyset] ,\\
		0, & [A]\neq[\emptyset] \ \text{or}\ \partial A\neq \emptyset.
	\end{cases}
\end{align}
For a dual string $A$ (a string living on the dual lattice), we have:
\begin{align}
	\zeta(A)=
	\begin{cases}
		-\log_2 q, & \partial A=\emptyset,\\
		0, & \partial A\neq\emptyset .
	\end{cases}
\end{align}
These results are consistent with Refs.~\cite{Hamma2005a,Berthiere2022}, and indicate a purely topological SEE (tSEE).

These formulas are obtained by  explicitly calculating SEE for SESs of various kinds:
\begin{itemize}
	\item For a noncontractible loop of $L$ spins along the $x$ direction, we have $N_{A}=L$, $|G_{A}|=1$, so $S_{A}=N_{A} \log_2 q-\log_{2}|G_{A}|=L\log_2 q = |A|\log_2 q$.
	\item For a noncontractible dual loop of $L$ spins along the $x$ direction, we have $N_{A}=L$, $|G_{A}|=q$, so $S_{A}=N_{A}\log_2 q-\log_{2}|G_{A}|=L\log_2 q-\log_2 q= |A|\log_2 q-\log_2 q$.
	\item For a contractible closed loop of $l$ spins, we have $N_{A}=l$, $|G_{A}|=q$, so $S_{A}=N_{A}\log_2 q-\log_{2}|G_{A}|=l\log_2 q-\log_2 q= |A|\log_2 q-\log_2 q$.
	\item For an open string of $l$ spins, we have $N_{A}=l$, $|G_{A}|=1$, so $S_{A}=N_{A}\log_2 q-\log_{2}|G_{A}|=l\log_2 q= |A|\log_2 q$.
\end{itemize}

The topological invariance of tSEE terms for contractible loops can be proven as follows: firstly, for a contractible loop SES $A$ composed of $l$ qudits, as we assume SESs to be connected and contain no dual loop, no $B_v$ stabilizer or their products can be fully supported on $A$. Then, for a product of $A_p$ stabilizers $P=\prod_{p\in \OS^2} A_p$ (where $\OS^2$ is an open membrane) to be fully supported on $A$, $A$ must be the boundary of $\OS^2$, thus $\OS^2$ is uniquely determined up to complementation. Since $\prod_{p\in \bar{\OS^2}} A_p= (\prod_{p\in \OS^2} A_p)^{\dagger} = P^{q-1}$, where $\bar{\OS^2}$ is the complement of $\OS^2$, $P$ is the only nontrivial generator of the stabilizer group fully supported on $A$, $G_A$ is the cyclic group generated by $P$, and $|G_A| = q$. As a result, we have $S_A = |A|\log_2 q - \log_2 q$ for any contractible loop $A$, which means the $-\log_2 q$ subleading term is topologically invariant. For other SESs, the topological invariance of tSEE terms can be proved similarly. 

Here we can notice an obvious difference between subdimensional and conventional entanglement entropies: When $A$ is an open string, $S_A$ contains only the area (volume) law term without constant topological entanglement entropy (TEE) term, while for a rectangle entanglement subsystem with finite width we would have the TEE term. Intuitively, it may be traced back to a significant difference between 1D SES and conventional 2D bulk-like entanglement subsystems: the 1-form symmetry in 2D toric code model is supported on closed loops, thus only for 1D SES the 1-form symmetry can reduce to a global symmetry. For a finite-size disk SES, it would simply inherit the $1$-form symmetry of the total system. This observation leads to the further exploration of symmetries of SESs with nontrivial SEE in this work, as presented in Sec.~\ref{sec:ses_sym}
and Sec.~\ref{sec:tcs}.

Moreover, comparing the results of noncontractible loop and dual loop SESs, we can notice that the constant term in the SEE of dual loop SES comes from the nonlocal stabilizers we chose to specify a unique stabilizer state. Therefore, for a different ground state of 2D toric code model we may have different SEE for such noncontractible SESs. It shows that topologically degenerate ground states can also be distinguished by SEE.

\subsection{3D toric code model}

\begin{figure}
	\centering
	\includegraphics[width=1.0\linewidth]{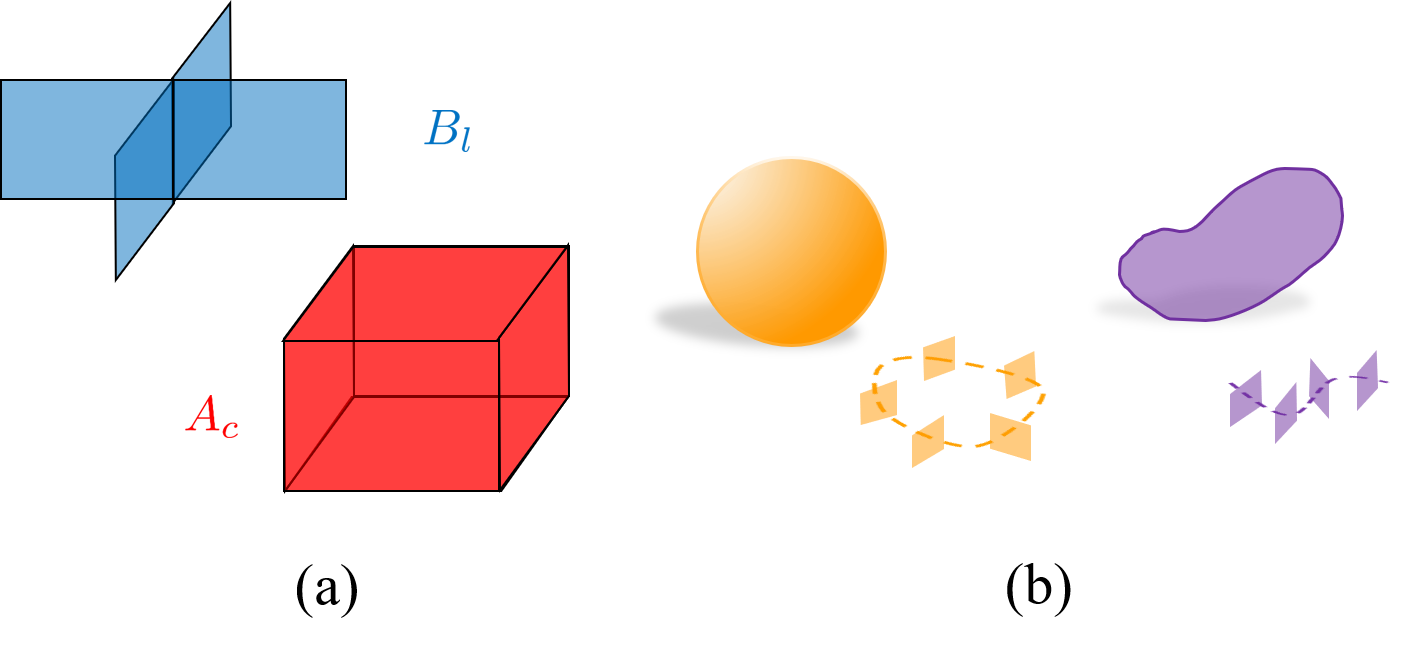}
	\caption{Representative Hamiltonian terms and SESs of the 3D toric code model. Panel (a) shows representative Hamiltonian terms in Eq.~\eqref{eq:ham_3dtc}, where plaquettes supporting spins involved in a $B_l$ and an $A_c$ terms are respectively colored blue and red. Panel (b) schematically illustrates SESs with nonzero (orange) and zero (purple) subleading terms $\zeta(A)$.}
	\label{fig:3dtc}
\end{figure}

We consider the 3D toric code model as a representative example of higher-dimensional topological orders, which allows for the exploration of both 1D and 2D SESs. These higher-dimensional topological orders exhibit a variety of interesting phenomena, including the existence of spatially extended excitations like strings and membranes, which have been extensively studied in the literature~\cite{Hamma2005,Bravyi2011,Walker2012,Lan2018,Wen2018a,Reiss2019,Lan2019,Zhang2021,chenyuan4d_2023,Huang2025jhep}. Here, we focus on the $\Z_2$ topological order for simplicity. The Hilbert space is composed of $\frac{1}{2}$-spins on plaquettes of a cubic lattice of the size $L\times L\times L$. The topologically ordered 3D toric code has the parent Hamiltonian~\cite{Hamma2005} [Fig.~\ref{fig:3dtc}(a)]:
\begin{equation}
\label{eq:ham_3dtc}
	H=-\sum_c A_c - \sum_l B_l,
\end{equation}
where $A_c = \prod_{p\in\partial c} X_p$ is the product of the Pauli X components of all spins nearest to cube $c$, and $B_l = \prod_{p\in\partial l} Z_p$ is the product of the Pauli $Z$ components of all spins nearest to link $l$, where $p$ refers to a plaquette.

With PBC, the ground states of the 3D toric-code model are $8$-fold degenerate. To specify a concrete ground state as a stabilizer state, we choose the nonlocal stabilizers to be the logical operators $W(\mathsf{D}^1_x)=\prod_{p\in\mathsf{D}^1_x}Z_p$, $W(\mathsf{D}^1_y)=\prod_{p\in\mathsf{D}^1_y}Z_p$, and $W(\mathsf{D}^1_z)=\prod_{p\in\mathsf{D}^1_z}Z_p$, where $\D^1_{x}$ ($\D^1_{y}$, $\D^1_{z}$) is a noncontractible dual loop along the $x$ ($y$, $z$) direction. Up to normalization, the corresponding stabilizer state is $\threedtc\propto\prod_{c}(1+A_{c})|000\cdots0\rangle$. As in the 2D toric-code model, degenerate ground states can be distinguished by the SEE of suitable noncontractible SESs.

In this model, the SEE demonstrates a purely topological response, independent of the local geometry of SESs. We summarize the results for the subleading term $\zeta(A)$ as follows. For a membrane SES $A$, we have:
\begin{align}
	\zeta(A)=
	\begin{cases}
		-1,&\ [A]=[\emptyset],\\
		0,&\ [A]\neq[\emptyset].
	\end{cases}
\end{align}
For a dual string SES $A$, we have:
\begin{align}
	\zeta(A)=
	\begin{cases}
		-1,&\ \partial A=\emptyset,\\
		0,&\ \partial A\neq\emptyset .
	\end{cases}
\end{align}
These results confirm the topological nature of the $\zeta(A)$ subleading SEE terms in the 3D toric code model, as illustrated in Fig.~\ref{fig:3dtc}(b).

These formulas are obtained by  explicitly calculating SEE for SESs of various kinds:
\begin{itemize}
	\item For a contractible dual loop composed of $l$ spins, we have $N_{A}=l$ , $|G_{A}|=2$, so $S_{A}=N_{A}-\log_{2}|G_{A}|=l-1=|A|-1$.
	\item For an open dual string composed of $l$ spins, we have $N_{A}=l$ , $|G_{A}|=1$, so $S_{A}=N_{A}-\log_{2}|G_{A}|=l=|A|$.
	\item For a cubic closed membrane of the size $6\times l\times l$ (i.e., the surface of an $l\times l\times l$ cube), we have $N_{A}=6l^{2}$ , $|G_{A}|=2$, so $S_{A}=N_{A}-\log_{2}|G_{A}|=6l^{2}-1=|A|-1$.
	\item For a flat open membrane composed of $6l^{2}$ spins, chosen to have the same number of spins as the cubic closed membrane above, we have $N_{A}=6l^{2}$, $|G_{A}|=1$, so $S_{A}=N_{A}-\log_{2}|G_{A}|=6l^{2}=|A|$.
	\item For a noncontractible closed membrane composed
	of $L^{2}$ spins perpendicular to direction $x$, we have $N_{A}=L^{2}$ , $|G_{A}|=1$, so $S_{A}=N_{A}-\log_{2}|G_{A}|=L^{2}=|A|$.
\end{itemize}

The topological invariance of tSEE terms for contractible closed membranes can be proved as follows: First, for a contractible closed membrane SES $A$ composed of $N_A$ spins, as we assume SESs to be connected and contain no dual loop, no $B_l$ stabilizer or their products can be fully supported on $A$. Then, for a product of $A_c$ stabilizers $C=\prod_{c\in \OS^3} A_c$ (where $\OS^3$ is a connected 3D region) to be fully supported on $A$, $A$ must be the boundary of $\OS^3$, thus $\OS^3$ is uniquely determined up to complementation; because $C=\prod_{c\in \OS^3} A_c = \prod_{c\in \bar{\OS}^3} A_c$, where $\bar{\OS}^3$ is the complement of $\OS^3$, $C$ is the only nontrivial element of the stabilizer group fully supported on $A$, so $|G_A| = 2$. As a result, we have $S_A = |A| - 1$ for any contractible closed membrane $A$, which means the $-1$ subleading term is topologically invariant. For other SESs, the topological invariance of tSEE terms can be proved similarly.

\subsection{3D X-cube model}

\begin{figure}
	\centering
	\includegraphics[width=1.0\linewidth]{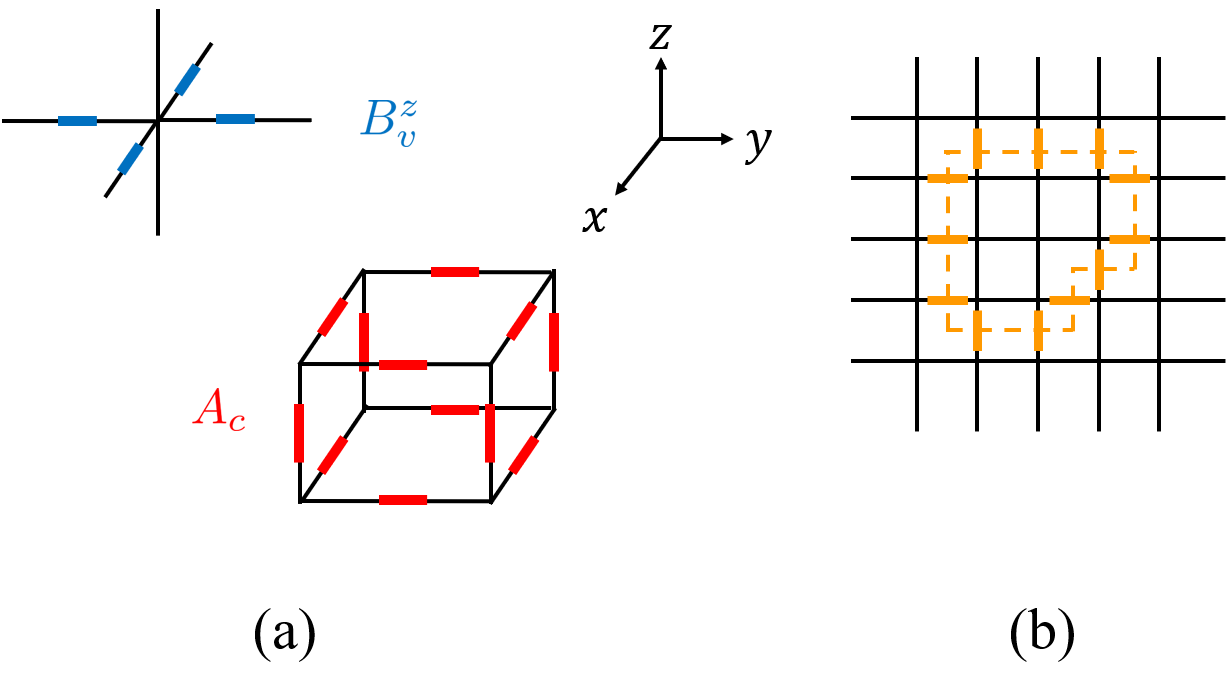}
	\caption{Representative Hamiltonian terms and SES of the X-cube model. Panel (a) shows representative Hamiltonian terms in Eq.~\eqref{eq:ham_xcube}, where links supporting spins involved in an $B^z_v$ and an $A_c$ terms are respectively colored blue and red. Panel (b) illustrates a typical planar contractible dual loop SES with nonzero subleading term $\zeta(A)$.}
	\label{fig:xcube}
\end{figure}

As another example in 3D, we consider the X-cube model as a representative model of fracton order~\cite{Vijay2016}, which can display a mixture of tSEE and gSEE for certain SESs. Fracton orders are a class of topological orders characterized by topological excitations with restricted mobility and have attracted attention from various communities, including quantum information and high-energy physics~\cite{Chamon2005,Haah2013,Vijay2015,Pretko2018,Ma2018a,Shirley2018,Slagle2018,Prem2019a,Seiberg2020c,Yuan2020,Seiberg2021a}. Though X-cube is a typical 3D Type-I fracton order, it is straightforward to apply the SEE diagnostic to Type-II and higher dimensional fracton orders~\cite{Haah2011,Vijay2015,Li2020,Li2021,Shen2022,Li2023,Hu2025,Li2026}. Again, here we can also consider both 1D and 2D SESs. The Hilbert space is obtained by assigning one $\frac{1}{2}$-spin to each link of a cubic lattice of the size $L\times L\times L$ with PBC. The parent Hamiltonian is given by [Fig.~\ref{fig:xcube}(a)]:
\begin{equation}
\label{eq:ham_xcube}
	H=-\sum_{c}A_{c}-\sum_v\sum_{k=x,y,z}B^k_v,
\end{equation} 
where $A_c = \prod_{l\in c} X_l$ is the product of the Pauli $X$ components of all spins nearest to cube $c$, and $B^k_v = \prod_{l\in{v^k}} Z_l$ is the product of the Pauli $Z$ components of all spins that are simultaneously a) nearest to vertex $v$ and b) embedded in a plane perpendicular to direction $k$ ($v^k$ simply denotes such a set of spins), $k=x,y,z$, $l$ refers to a link.

With PBC, the ground states of the X-cube model defined above are $2^{6L-3}$-fold degenerate. To specify a concrete ground state as a stabilizer state, we choose a complete independent set of nonlocal Pauli-$Z$ stabilizers supported on noncontractible dual loops. We denote these operators collectively by $W(\mathsf{D}^1_k)=\prod_{l\in \mathsf{D}^1_k} Z_l$, where $\D^1_k$ is a noncontractible dual loop along the $k$ direction, with $k=x,y,z$. For each $k$, there are two types of such dual loops, each composed of links oriented along one of the two directions perpendicular to $k$, and their transverse positions are chosen to provide a complete independent set. Up to normalization, the corresponding stabilizer state is $\xcube\propto\prod_{c}(1+A_{c})|000\cdots0\rangle$.

Unlike in previous cases, the subleading term $\zeta(A)$ in SEE results for the X-cube model can be sensitive to both the topology and the geometry of the SES, which is consistent with the coexistence of topological and geometric effects in fracton orders. As a concrete example, for a 1D dual string SES $A$, we find that $\zeta(A)$ is non-zero only when $A$ is both closed (topological condition) and planar (geometric condition) [Fig.~\ref{fig:xcube}(b)]:
\begin{align}
	\zeta(A)=
	\begin{cases}
		-1, & (\partial A=\emptyset) \ \text{and}\ (\exists \textsf{P}\ \text{s.t.}\ A\subseteq \textsf{P}),\\[3pt]
		0, & \text{otherwise},
	\end{cases}
\end{align}
where $\textsf{P}$ denotes a lattice plane (perpendicular to $x, y,$ or $z$). 

This formula is obtained by  explicitly calculating SEE for 1D SESs of various kinds:
\begin{itemize}
	\item For an arbitrary open string composed of $l$ spins, we have $N_{A}=l$ , $|G_{A}|=1$, so $S_{A}=N_{A}-\log_{2}|G_{A}|=l=|A|$.
	\item For a contractible loop composed of $l$ spins in a plane, we have $N_{A}=l$ , $|G_{A}|=1$, so $S_{A}=N_{A}-\log_{2}|G_{A}|=l=|A|$.
	\item For an arbitrary open dual string composed of $l$ spins, we have $N_{A}=l$ , $|G_{A}|=1$, so $S_{A}=N_{A}-\log_{2}|G_{A}|=l=|A|$.
	\item For a contractible dual loop composed of $l$ spins not fully embedded in any planes, we have $N_{A}=l$ , $|G_{A}|=1$, so $S_{A}=N_{A}-\log_{2}|G_{A}|=l=|A|$.
	\item For a contractible dual loop composed of $l$ spins in a plane, we have $N_{A}=l$, $|G_{A}|=2$, so $S_{A}=N_{A}-\log_{2}|G_{A}|=l-1=|A|-1$.
	\item For a noncontractible straight loop composed
	of $L$ spins, we have $N_{A}=L$ , $|G_{A}|=1$, so $S_{A}=N_{A}-\log_{2}|G_{A}|=L=|A|$.
	\item For a noncontractible straight dual loop composed
	of $L$ spins not fully embedded in any planes, we have $N_{A}=L$, $|G_{A}|=1$, so $S_{A}=N_{A}-\log_{2}|G_{A}|=L=|A|$.
	\item For a noncontractible straight dual loop composed
	of $L$ spins in a plane, we have $N_{A}=L$, $|G_{A}|=2$, so $S_{A}=N_{A}-\log_{2}|G_{A}|=L-1=|A|-1$.
\end{itemize}

We can prove the conditioned topological invariance (i.e., topologically invariant as long as certain geometric conditions are satisfied) of subleading SEE terms for planar contractible dual loop SESs as follows: first, for a planar contractible dual loop SES $A$ composed of $l$ spins and embedded in plane $\textsf{P}$ perpendicular to direction $z$, we take the dual lattice in $\textsf{P}$, where $A$ becomes a loop of the same length, and the support of a $B^z_v$ term in $\textsf{P}$ becomes a plaquette (i.e., a minimal loop). Obviously, no other kinds of local stabilizer terms can be fully supported in $A$, thus we can focus on the $B^z_v$ terms in $\textsf{P}$. Then, the conditioned invariance of the subleading SEE term of $A$ follows directly from the proof method used for 2D toric code model. 

Further structure arises for 2D SESs. 
For example, consider a membrane and a dual-membrane SES, both noncontractible and flat, each of size $L\times L$ (with $L$ the linear system size). 
The former yields $S_A=L^{2}-1=|A|/2-1$, while the latter gives $S_A=L^{2}-2L+1=|A|-2L+1$ (note that the two membranes respectively contain $2L^2$ and $L^2$ spins). The subextensive subleading term for the dual membrane resembles that in the conventional EE of a 3D subsystem in the X-cube model~\cite{Ma2018a}, but here it originates from noncontractible dual loops within the dual membrane and thus depends on the specific ground state chosen as $\xcube$. 

\subsection{A heuristic example distinguishing TEE and tSEE}
\label{subsec:tee_and_tsee}

The behavior of nontrivial tSEE terms is reminiscent of TEE. Nevertheless, tSEE terms depend on the topology of the SES and can be zero for SESs with trivial topology, which distinguishes them from TEE. Here, we use the standard 2D $\Z_2$ toric code as a heuristic model to further clarify the relation and difference between tSEE and TEE.

For a disk region $A_D$ in the 2D toric code with a smooth boundary, its reduced density matrix $\rho_{A_D}$ is~\cite{Levin2006}:
\begin{align}
	\label{eq:ses_disk}
	\rho_{A_D} = C\sum_{\tilde{p}} |\tilde{p}\rangle \langle \tilde{p}|,
\end{align}
where $C$ is a normalization constant, $\tilde{p}$ represents a spin configuration in the Pauli $Z$ basis along the boundary of $A_D$ with an even number of $|1\rangle$ spins, and $|\tilde{p}\rangle$ is the superposition of all closed-string configurations with the same boundary configuration $\tilde{p}$. Meanwhile, according to Eq.~\eqref{eq:RDM}, the reduced density matrix of the boundary of $A_D$, which is a contractible loop, is $\rho_A\propto (I_A+W_A)$, where $W_A = \prod_{i\in A} X_i$. In the Pauli $X$ basis, $\rho_A$ is diagonal:
\begin{align}
	\label{eq:ses_disk_boundary}
	\rho_{A} = C \sum_{\tilde{c}} |\tilde{c}\rangle \langle \tilde{c}|,
\end{align}
where $\tilde{c}$ represents a spin configuration of $A$ with an even number of spins in the $-1$ eigenstate of $X$. Equations~\eqref{eq:ses_disk} and \eqref{eq:ses_disk_boundary} show that the two reduced density matrices have the same form, although they act on different Hilbert spaces. Therefore, the SEE of a general contractible-loop SES $A$ in the 2D toric code equals the conventional entanglement entropy of the region $A_D$ that contains $A$ as its boundary. In this case, the nontrivial tSEE for a contractible-loop SES equals the TEE of disk regions, whereas the fact that open strings have trivial tSEE can be traced back to the topological fact that an open string cannot serve as the boundary of any region.

This heuristic example reveals the intrinsic difference between bulk-like entanglement subsystems and SESs: although both a 1D open string and a 2D disk are topologically trivial, and a 1D open string can be regarded as the shrinking limit of a disk, they are associated with different EE subleading terms. In contrast, a 1D contractible loop has the same EE subleading term as the disk surrounded by it. In this case, whether the SEE subleading term can be reduced to a conventional TEE depends on the topology of the SES, and this topology-dependent reducibility gives rise to the overall dependence of tSEE on the topology of the SES, which constitutes the characteristic feature of tSEE and underlies the diagnostic power of SEE for topological orders.

Intuitively, in gapped topological orders we always expect a bulk-like region to entangle with the rest of the system through its boundary. Therefore, for an SES $A$ that serves as the boundary of a bulk-like region, we expect a corresponding TEE contribution to give rise to nontrivial tSEE for $A$, which would make the topology-dependent behavior of tSEE robust in the topological phase. Though a comprehensive analysis of the robustness of SEE across various kinds of phases is beyond our scope, this observation inspires us to further investigate the reduced density matrices associated with SESs, as presented in Secs.~\ref{sec:ses_sym} and \ref{sec:tcs}.

\section{Mixed-state symmetries}
\label{sec:ses_sym}

To explore further physical consequences of SEE, we regard the reduced density matrix $\rho_A$ of an SES $A$ as the density matrix of a mixed state supported on $A$. Through this identification, we find that SEE provides a broad unifying framework connecting several diverse forefront directions, including stabilizer codes, topological holography, and mixed-state symmetries. In this section, we study mixed-state symmetries, laying the foundation for the topological holography framework discussed in the next section.

Symmetries of a mixed state $\rho$ are classified into strong and weak ones, where a strong symmetry $W$ satisfies $W \rho = e^{i\theta}\rho$ with a phase $\theta$, while a weak symmetry $w$ satisfies $w \rho w^\dagger = \rho$. These two classes of symmetries underlie many important phenomena, such as symmetry protected topological orders in open systems, nontrivial steady-state dynamics, and symmetry breaking in mixed states~\cite{Lieu2020,Albert2014,Wang2025b,Ellison2025,Zhang2025,Lee2022,Lessa2025,Sala2024,Buca2012,deGroot2022,Ma2023a,Ma2023,Castelnovo2007,Helmes2015}.

Furthermore, we analyze strong-to-weak spontaneous symmetry breaking (SW-SSB) of strong symmetries within SESs, which is another major focus in current research on mixed-state topological phases~\cite{Lieu2020,Albert2014,Wang2025b,Ellison2025,Zhang2025,Lee2022,Lessa2025,Sala2024}. We obtain a series of preliminary results showing that SW-SSB occurs in several SESs with nontrivial SEE.

\subsection{Deriving mixed-state symmetries of SESs from stabilizers}

First, in this subsection, we prove the following general relation for stabilizer states, which directly connects the strong and weak symmetries of entanglement subsystems to bulk stabilizers:
\begin{theorem}
	\label{Theorem:strong_weak_stabilizers}
	Given a Pauli stabilizer state $|\psi\rangle$ and an entanglement subsystem $A$ composed of a set of spins, any stabilizer fully supported on $A$ is a strong symmetry of $\rho_A$. Moreover, if the support of a stabilizer intersects both $A$ and its complement, its restriction to $A$ is a weak symmetry of $\rho_A$.
\end{theorem}

\begin{proof}
	
	First, we consider the Schmidt decomposition of $|\psi\rangle$:
\begin{equation*}
	|\psi\rangle=\sum_{i}a_{i}|\psi\rangle_{A}^{i}\otimes|\psi\rangle_{B}^{i},
\end{equation*}
where $A$ is the SES, $B$ is the remainder of the system, $\{|\psi\rangle_{A}^{i}\}$ and $\{|\psi\rangle_{B}^{i}\}$ are orthonormal sets of states for $A$ and $B$, respectively, and $a_{i}$ are normalized, real, and non-negative coefficients.
	
	Then, the reduced density matrix of
	$A$ can be formally expressed as:
	\begin{equation*}
		\rho_{A}=\sum_{i}a_{i}^{2}|\psi\rangle_{A}^{i}\langle\psi|_{A}^{i}.
	\end{equation*}
	
	For a Pauli stabilizer $W_{A}$ fully supported in $A$, we have $W_A |\psi\rangle = |\psi\rangle$. Substituting $|\psi\rangle$ by its Schmidt decomposition, we obtain:
	\begin{align*}
		\sum_{i}a_{i}(W_{A}|\psi\rangle_{A}^{i})\otimes|\psi\rangle_{B}^{i}= \sum_{i}a_{i}|\psi\rangle_{A}^{i}\otimes|\psi\rangle_{B}^{i},
	\end{align*}
	which requires $W_{A}|\psi\rangle_{A}^{i}=|\psi\rangle_{A}^{i}$
for every $i$ with $a_i\neq0$; thus, we have
	\begin{align}
		W_{A}\rho_{A}=\sum_{i}a_{i}^{2} (W_{A} |\psi\rangle_{A}^{i})\langle\psi|_{A}^{i}=\rho_{A},
	\end{align}
	which means $W_{A}$ is a strong symmetry of $\rho_{A}$.
	
	For a Pauli stabilizer $W_{AB}$ whose support intersects both $A$
and $B$, the on-site tensor-product structure of Pauli operators allows
us to decompose it as $W_{AB}=w_{A}\otimes w_{B}$, where $w_{A(B)}$
is a product of Pauli operators on different spatial sites, not necessarily
a stabilizer, fully supported in $A(B)$. Then, the condition $W_{AB}|\psi\rangle = |\psi\rangle$ can be written explicitly using the basis expansion:
	\begin{equation*}
		\sum_{i}a_{i}(w_{A}|\psi\rangle_{A}^{i})\otimes(w_{B}|\psi\rangle_{B}^{i}) = |\psi\rangle.
	\end{equation*}
	Consequently, proper substitution into the reduced density matrix definition $\rho_{A} = \mathrm{Tr}_{B}(W_{AB}\rho W^{\dagger}_{AB})$ yields
	\begin{align*}
		\rho_{A} &= \sum_{c_{B}} \langle c_{B}| \bigg[ \Big( \sum_{i}a_{i}w_{A}|\psi\rangle_{A}^{i}\otimes w_{B}|\psi\rangle_{B}^{i} \Big) \notag \\
		&\quad \times \Big( \sum_{i'}a_{i'}\langle\psi|_{A}^{i'}w^{\dagger}_{A}\otimes\langle\psi|_{B}^{i'}w^{\dagger}_{B} \Big) \bigg] |c_{B}\rangle \notag \\
		&= \sum_{i,i'} a_{i}a_{i'} \bigg( \sum_{c_{B}}\langle c_{B}|w_{B}|\psi\rangle_{B}^{i}\langle\psi|_{B}^{i'}w^{\dagger}_{B}|c_{B}\rangle \bigg) \notag \\
		&\quad \times (w_{A}|\psi\rangle_{A}^{i})(\langle\psi|_{A}^{i'}w^{\dagger}_{A}),
	\end{align*}
	where $c_B$ refers to a spin configuration in $B$, thus the summation denotes the trace over a complete basis of subsystem $B$.
	
	Note that $|\psi\rangle_{B}^{i}\langle\psi|_{B}^{i'}$ is traceless when $i'\neq i$. Based on the unitarity of $w_B$ and the cyclic invariance of the trace, the inner term for subsystem $B$ simplifies to $\sum_{c_{B}}\langle c_{B}|w_{B}|\psi\rangle_{B}^{i}\langle\psi|_{B}^{i'}w^{\dagger}_{B}|c_{B}\rangle = \mathrm{Tr}( |\psi\rangle_{B}^{i}\langle\psi|_{B}^{i'} ) = \delta_{ii'}$,
	where $\delta_{ii'}$ refers to the Kronecker delta.

	Therefore, we obtain
	\begin{equation}
		\rho_{A}=\sum_{i}a_{i}^{2}(w_{A}|\psi\rangle_{A}^{i})(\langle\psi|_{A}^{i}w^{\dagger}_{A})=w_{A}\rho_{A}w^{\dagger}_{A},
	\end{equation}
	which implies that $w_{A}$ is a weak symmetry of $\rho_{A}$. 
	
\end{proof}

As a concrete example, consider the flat noncontractible membrane SES $A$ in the X-cube model, as shown in Fig.~\ref{fig:com_sym}. Since the product of local stabilizers $B^k_v$ and non-local stabilizers $W(\D^1_{x(y,z)})$ fully embedded in $A$ can form an arbitrary Pauli $Z$ dual loop operator therein, $W_{\tilde{c}}=\prod_{l\in\tilde{c}} Z_l$ constitutes a strong symmetry of $A$ for any configuration $\tilde{c}$ composed of dual loops. Meanwhile, the restriction to $A$ of a product of general $W(\D^1_{x(y,z)})$ and $B^k_v$ stabilizers can yield an arbitrary Pauli $Z$ open dual string operator, and the restriction of $A_c$ stabilizers can yield an arbitrary Pauli $X$ contractible loop operator. Consequently, these two types of operators serve as weak symmetries of $A$. Finally, note that the above theorem does not guarantee that all symmetries of $\rho_A$ can be obtained from the stabilizers. For example, spatial symmetries such as translations obviously cannot be derived in this manner.

\subsection{SW-SSB signaled by nontrivial SEE}
\label{subsec:swssb_global}

Unlike conventional spontaneous symmetry breaking in pure states, SW-SSB is characterized not by the explicit absence of symmetry, but by long-range correlations of charged operators. To diagnose SW-SSB in an SES $A$, we employ the \textit{fidelity correlator} $F_X(i,j)$ and the \textit{Choi-state order parameter} $O_s$ of $W_{\mathsf{S}^1} = \prod_{k\in \mathsf{S}^1} X_k$ (see Sec.~\ref{subsec:swssb_global} and Sec.~\ref{subsec:swssb_1form} for detailed definitions) for global and 1-form strong symmetries, respectively, where $i,j$ are two sites in $A$ and $\mathsf{S}^1$ is a loop~\cite{Lee2022,Ellison2025,Lessa2025,Zhang2025}. For the stabilizer SESs considered here, only SESs with nontrivial SEE are relevant to this analysis, since $\rho_A$ is maximally mixed when $S_A=|A|$ and therefore has no nontrivial strong Pauli symmetry.

First we consider global strong symmetries, for which the fidelity correlator diagnostic is applied. For a mixed state $\rho$ and a charged local operator $O$, the fidelity correlator $F_O(x,y)$ is defined as: $F_O(x,y) = \langle O(x)O^{\dagger}(y)\rangle_F$, where $\langle \cdot \rangle_F$ denotes the fidelity average: $\langle O \rangle_F = \text{Tr}\sqrt{\sqrt{\rho} O\rho O^{\dagger}\sqrt{\rho}}$.

Consider the 1D (planar) contractible dual loop SES in $|\psi\rangle_{\Z_2}$ and $\xcube$. The reduced density matrix is $\rho_A=C(I_A+W_A)$ according to Eq.~\eqref{eq:RDM}, where $C$ is a normalization constant and $W_A = \prod_{i\in A} Z_i$ represents a strong symmetry of $\rho_A$. This provides a prototypical example of SW-SSB. We calculate the fidelity correlator of Pauli $X$ operators between two different sites $i,j\in A$:
\begin{align*}
	F_X(i,j) &=\text{Tr}\sqrt{\sqrt{\rho_A} X_i X_j\rho_A X_j X_i\sqrt{\rho_A}}\\
	&=\text{Tr}\sqrt{\sqrt{\rho_A} X_i X_j \cdot C(I_A + W_A) \cdot X_j X_i\sqrt{\rho_A}}\\
	&=\text{Tr}\sqrt{\sqrt{\rho_A} \cdot \rho_A \cdot \sqrt{\rho_A}}=1,
\end{align*}
where we utilized the commutation relation $[W_A, X_i X_j]=0$, which implies $X_i X_j W_A X_j X_i = W_A$. In contrast, the standard two-point correlation function vanishes: $\text{Tr}(\rho_A X_i X_j) = 0$ for $i \neq j$. The non-vanishing fidelity correlator combined with vanishing standard correlation indicates that the strong global $\Z_2$ symmetry of $\rho_A$ undergoes spontaneous strong-to-weak breaking. For the 2D contractible closed membrane SES in $\threedtc$, the reduced density matrix has the same form; thus, the strong global $\Z_2$ symmetry in 2D similarly exhibits SW-SSB.

\subsection{SW-SSB of $1$-form strong symmetries}
\label{subsec:swssb_1form}

Turning to the 2D noncontractible flat closed membrane SES in $\xcube$, we employ the Choi-state order parameter diagnostic to analyze the SW-SSB of 1-form strong symmetries. According to Eq.~\eqref{eq:RDM}, the reduced density matrix of region $A$ takes the form $\rho_{A}=C'\sum_{\tilde{c}}W_{\tilde{c}}$, where the sum runs over all dual loop patterns $\tilde{c}\subset A$ (including the empty set), $W_{\tilde{c}}=\prod_{i\in\tilde{c}}Z_{i}$ applies Pauli $Z$ on each spin in $\tilde{c}$, and $C'$ is a normalization constant. Each $W_{\tilde{c}}$ represents a strong symmetry transformation of $\rho_A$.

To construct the Choi state, we first diagonalize $\rho_A$. Since all operators $W_{\tilde{c}}$ are diagonal in the computational (Pauli $Z$) basis, we work with computational basis states $\{|s\rangle\}$, where $|s\rangle = |s_1 s_2 \dots s_{N_A}\rangle$ with $s_i \in \{0,1\}$. Each basis state is an eigenvector of all $W_{\tilde{c}}$: $W_{\tilde{c}} |s\rangle = \left(\prod_{i\in\tilde{c}}Z_{i}\right) |s\rangle = (-1)^{\sum_{i\in\tilde{c}}s_i} |s\rangle$ for any $\tilde{c}$. The operators $\{W_{\tilde{c}}\}$ form an Abelian group under multiplication. For each configuration $|s\rangle$, the mapping $\chi_s: W_{\tilde{c}} \mapsto (-1)^{\sum_{i\in\tilde{c}}s_i}$ defines a character of this group. By the orthogonality theorem for group characters, the sum $\sum_{\tilde{c}} (-1)^{\sum_{i\in\tilde{c}}s_i}$ is non-zero only when $(-1)^{\sum_{i\in\tilde{c}}s_i} = 1$ for all $\tilde{c}$, which requires the number of spins with $s_i=1$ on any dual loop to be even. Denoting such configurations by $\tilde{s}$, we obtain:
$\rho_A = C''\sum_{\tilde{s}} |\tilde{s}\rangle\langle \tilde{s}|$,
where $C''$ is a normalization constant. The corresponding Choi state is:
$|\rho_A\rangle\rangle = \sqrt{C''} \sum_{\tilde{s}}  |\tilde{s}_u\rangle \otimes |\tilde{s}_l\rangle$, 
where subscripts $u$ and $l$ denote the ``upper'' and ``lower'' Hilbert spaces, respectively.

Then, we consider the order parameter for the strong $1$-form symmetry in this Choi state:
$O_s = \lim_{r\rightarrow\infty} \langle \langle \rho_A | W_{\S^1,u} W_{\S^1,l}|\rho_A \rangle\rangle$, 
where $\S^1$ is a contractible loop of radius $r$, and $W_{\S^1,u(l)}$ denotes $\prod_{i\in\S^1} X_i$ acting on the upper (lower) Hilbert space. This operator qualifies as an order parameter as its corresponding anyon braids nontrivially with the anyon corresponding to the strong symmetry $W_{\tilde{c}}$.
The calculation yields:
\begin{align*}
	O_s &= \lim_{r\rightarrow\infty} C''( \sum_{\tilde{s}'}  \langle\tilde{s}'_l| \otimes \langle \tilde{s}'_u|) W_{\S^1,u} W_{\S^1,l} ( \sum_{\tilde{s}}  |\tilde{s}_u\rangle \otimes |\tilde{s}_l\rangle)\\
	& = \lim_{r\rightarrow\infty} C'' \sum_{\tilde{s}'}  \sum_{\tilde{s}}  \delta(\tilde{s}',\tilde{s} + \S^1) \langle(\tilde{s}'+\S^1)_u |\tilde{s}_u\rangle\\
	& = \lim_{r\rightarrow\infty} C'' \sum_{\tilde{s}}  \langle \tilde{s}_u|\tilde{s}_u\rangle = 1,
\end{align*} 
where $\tilde{s}'+\S^1$ refers to the $\Z_2$ addition of these two patterns (i.e., the symmetric difference between the two sets of spins). Similarly, we find that the order parameter for the weak $1$-form symmetry $O_w = \lim_{r\rightarrow \infty} \langle \langle \rho_A | W_{\S^1,u} |\rho_A \rangle\rangle$ vanishes, as $\lim_{r\rightarrow\infty} C'' \sum_{\tilde{s}'}  \sum_{\tilde{s}}  \langle(\tilde{s}'+\S^1)_u| \delta(\tilde{s}',\tilde{s}) |\tilde{s}_u\rangle = 0$. Consequently, we conclude that the strong $1$-form symmetry of the 2D noncontractible flat closed membrane SES in $\xcube$ is spontaneously broken to a weak symmetry.

\subsection{Discussion}

In summary, for a series of SESs with nontrivial SEE, our analytical calculations reveal that the SESs exhibit SW-SSB of the corresponding strong symmetry (global or 1-form). Specifically, the diagnostics $F_X(i,j)$ and $O_s(W_{\mathsf{S}^1})$ for the respective symmetries approach nonzero constants for arbitrarily distant $i,j$ and arbitrarily large loops $\mathsf{S}^1$. Notably, the SW-SSB of 1-form symmetries signals intrinsically mixed topological orders that are inequivalent to pure-state topological orders~\cite{Wang2025b,Ma2023,Ellison2025,Zhang2025,Sohal2025}; this is exactly the case for the 2D noncontractible flat closed membrane SES in the X-cube model. Therefore, such SW-SSB behavior shows that entanglement in the total system can induce nontrivial mixed-state phases in SESs.

It has been proved that SW-SSB is a general property of thermal states within a fixed charge sector of a global symmetry~\cite{Lessa2025,Liu2025}. Since the 1D and 2D SESs with global strong symmetries considered in this section can be regarded as such thermal states in the $T\rightarrow \infty$ limit, our results are consistent with this theorem. Furthermore, we expect this relation between SW-SSB and nontrivial SEE subleading terms to hold for general stabilizer states: according to Eq.~\eqref{eq:RDM}, for an arbitrary SES $A$ in a stabilizer state with a volume-law leading term $|A|$ and a constant subleading term in the SEE, assuming translation symmetry, this subleading term must result from stabilizers that are fully supported on $A$ as global operators; thus, $\rho_A$ can be regarded as a thermal state within a fixed charge sector and would exhibit SW-SSB. While for an SES $A$ with trivial SEE $S_A = |A|$, $\rho_A$ would reduce to the identity, and no nontrivial strong symmetry could be present. Given that the volume-law scaling of the SEE is also reminiscent of thermal states, such SW-SSB suggests an intriguing connection between entanglement-induced phenomena and thermal physics beyond the stabilizer formalism. 

In particular, SW-SSB of $1$-form symmetries can also describe the decohered 2D toric code~\cite{Zhang2025}, which suggests an SEE-induced relation between noiseless complicated codes and noisy simple codes. Thus, the SW-SSB exhibited by SESs illuminates the rich structure of SEE and associated SESs from a distinct perspective.

\section{Transparent composite symmetries and mixed-state topological holography}
\label{sec:tcs}

\begin{table*}[t]
	\caption{Transparent composite symmetries (TCS) of subdimensional entanglement subsystems (SESs). 
		In each case, both the strong and weak symmetries are entirely supported within the SES $A$. 
		Here, $i$ labels a site (spin) in the underlying model, and $P_{\mathsf{OS}^n}^{X}\!\equiv\!\prod_{i\in\mathsf{OS}^n}X_i$ applies Pauli-$X$ operators on all spins in region $\mathsf{OS}^n$; 
		other weak symmetry operators are defined analogously. 
		In 1D SESs, $i{+}1$ denotes a nearest neighbor of $i$; 
		in the 2D SES of the X-cube model, $\tilde{c}$ denotes a configuration composed of dual loops $\D^1$ in $A$, so $W_{\tilde{c}}$ is a 1-form symmetry transformation. 
		For instance, $P_{\partial\mathsf{OS}^1}^{Z}=Z_jZ_k$, where $j,k$ are the two endpoints of the open string $\mathsf{OS}^1$ (i.e., $\partial\mathsf{OS}^1$). 
		In each SES with nontrivial SEE, the weak symmetries serve as transparent patch (t-patch) operators of the strong symmetry, and their algebra holographically encodes a topological order (TO) of one higher dimension. Such a composite of strong and weak symmetries forms a transparent composite symmetry. In the 2D $\mathbb{Z}_2$ toric code (in purple), an open dual string SES with trivial SEE is included for comparison.}
	\label{tab:comp_sym}
	\centering
	\setlength{\tabcolsep}{5pt}
	\renewcommand{\arraystretch}{1.15}
	\begin{tabular*}{\textwidth}{@{\extracolsep{\fill}} l l l l l}
		\toprule
		Model & SES $A$ & Strong symmetry & Weak symmetry & Encoded TO \\ 
		\midrule
		{\color{purple}2D $\mathbb{Z}_2$ toric code} & {\color{purple}Open dual string}& {\color{purple}Identity}& 
		{\color{purple}\makecell{0D: $Z_i$, $X_i$} }& 
		{\color{purple}Trivial }\\[0.2em]
		
		2D $\mathbb{Z}_2$ toric code & Contractible dual loop & Global: $W_A=\prod_{i\in A} Z_i$ & 
		\makecell{0D: $Z_i$, $X_i X_{i+1}$\\ 1D: $P^Z_{\mathsf{OS}^1}$, $P^X_{\partial \mathsf{OS}^1}$} & 
		2D $\mathbb{Z}_2$ TO \\[1.0em]
		
		3D $\mathbb{Z}_2$ toric code & Contractible membrane & Global: $W_{A}=\prod_{i\in A}X_{i}$ & 
		\makecell{0D: $X_{i}$, $Z_{i}Z_{i+\hat{x}(\hat{y},\hat{z})}$\\ 1D: $P_{\partial \mathsf{OS}^1}^{Z}$\\ 2D: $P_{\mathsf{OS}^2}^{X}$} & 
		3D $\mathbb{Z}_2$ TO \\[1.5em]
		
		X-cube & Planar contractible dual loop & Global: $W_A=\prod_{i\in A} Z_i$ & 
		\makecell{0D: $Z_i$, $X_i X_{i+1}$\\ 1D: $P^Z_{\mathsf{OS}^1}$, $P^X_{\partial \mathsf{OS}^1}$} & 
		2D $\mathbb{Z}_2$ TO \\[1.0em]
		
		X-cube & Flat noncontractible membrane & $1$-form: $W_{\tilde{c}}=\prod_{i\in\tilde{c}}Z_{i},\ \forall \tilde{c}$ & 
		\makecell{0D: $Z_i$\\ 1D: $P_{\mathsf{OD}^1}^{Z}$\\ 2D: $P_{\partial\mathsf{OS}^2}^{X}$} & 
		3D $\mathbb{Z}_2$ TO \\
		\bottomrule
	\end{tabular*}
\end{table*}

Remarkably, for the SESs with nontrivial SEE ($\zeta(A)\!\neq\!0$) considered below, we identify a set of strong and weak symmetries that combine to form what we call \textit{transparent composite symmetries} (TCSs). In each such composite, the listed weak symmetries form t-patch operators~\cite{Chatterjee2023} associated with the strong symmetry transformations (see Sec.~\ref{sec:rev_of_tpatch} for a review), hence the terminology. Because the algebra of t-patch operators can encode data of topological orders, TCS induces a form of mixed-state topological holography that relates mixed-state symmetries of lower-dimensional subsystems to topological orders in higher dimensions.

In this section, we explicitly show the existence of TCS in a series of concrete SESs, utilizing results of t-patch operators associated with various symmetry transformations derived in Ref.~\cite{Chatterjee2023}. We use Eq.~\eqref{eq:RDM}, the definition of reduced density matrices of stabilizer states, to explicitly find the strong and weak symmetries of SESs, and show that our results are consistent with \textbf{Theorem}~\ref{Theorem:strong_weak_stabilizers}. We notice that SESs in various models may be the same, and the difference between orders are reflected in the partition generating the SES, such as requiring the SESs to satisfy certain geometric or topological constraints. In general, based on the explicit examples, we conjecture that the correspondence between nontrivial SEE and TCS extends to systems beyond stabilizer code constructions, which is further supported by the robustness of TCS demonstrated in Sec.~\ref{subsec:tpatch_robust}.

\begin{figure}[t]
	\centering
	\includegraphics[width=0.9\linewidth]{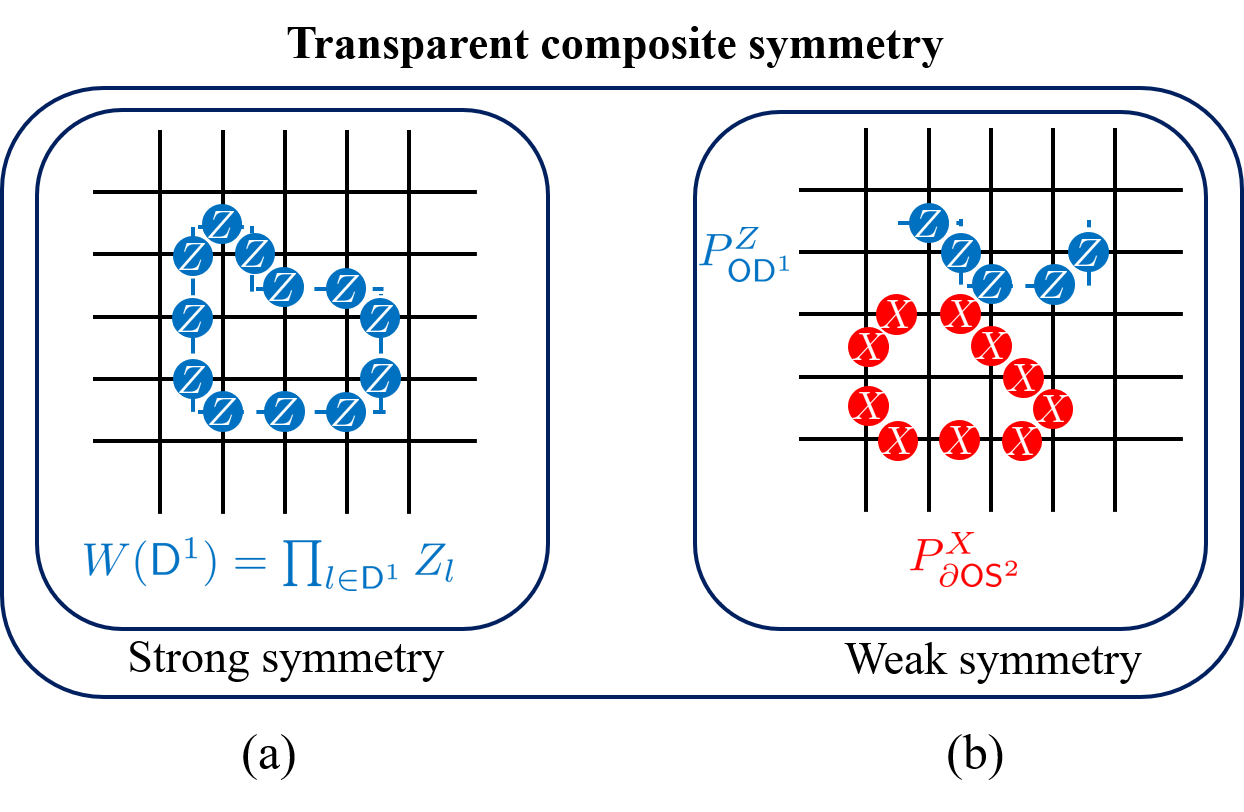}
	\caption{TCS of a flat, noncontractible 2D closed membrane SES $A$ in the X-cube model. 
		Panels~(a) and~(b) show a portion of $A$. 
		The operators generating the strong and weak symmetries are depicted in (a) and (b), respectively, together constituting the TCS of this SES.}
	\label{fig:com_sym}
\end{figure}

The main results of this section are summarized in Table~\ref{tab:comp_sym}. In each nontrivial case involving an SES of \emph{a given dimension}, the listed weak symmetries serve as the t-patch operators of the corresponding strong symmetries, forming an algebraic structure that holographically encodes a topological order in \emph{one higher dimension}~\cite{Chatterjee2023,Ji2019,Kong2020a,Inamura2023}. As a concrete example, Fig.~\ref{fig:com_sym} shows the TCS of a membrane SES in the X-cube model from Table.~\ref{tab:comp_sym} (detailed in Sec.~\ref{subsec:2D_1form_TCS}), where the nontrivial commutation relation between the t-patch operators $P^Z_{\mathsf{OD}^1}$ and $P^X_{\partial\mathsf{OS}^2}$ reproduces the braiding phase of the two kinds of fundamental topological excitations in 3D $\mathbb{Z}_2$ topological order. For SESs with trivial SEE ($S_A=|A|$), the corresponding TCSs are trivial: no strong symmetry exists, and the weak symmetries are generated by all local Pauli operators, matching the trivial categorical symmetry~\cite{Chatterjee2023}. Interestingly, the order encoded in an SES can evidently differ from that of the total system, as shown in Table~\ref{tab:comp_sym}, indicating that TCS encodes complementary information compared with the t-patch operators~\cite{Chatterjee2023}.

\subsection{TCS of 1D (planar) contractible dual loop SES in $\xcube$ and $|\psi\rangle_{\Z_2}$}
\label{subsec:1D_global_TCS}

First, we consider the TCSs of a 1D planar contractible dual-loop SES in $\xcube$ and a 1D contractible dual-loop SES in $|\psi\rangle_{\Z_2}$ simultaneously, because Eq.~\eqref{eq:RDM} gives the same reduced density matrix for both SESs:
\begin{equation}
	\rho_{A}=C'(I_{A}+W_{A}),
\end{equation}
where $I_{A}$ is the identity matrix supported on $A$, $W_{A}=\prod_{i\in A}Z_{i}$ is uniformly acting Pauli $Z$ on each spin in $A$, $C'$ is a normalization constant. For this $\rho_{A}$, we have $W_{A}\rho_{A}=C'(W_{A}+I_{A})=\rho_{A}W_{A}=\rho_{A}$, that means $W_{A}$ is a strong symmetry of $\rho_{A}$.

In the Pauli $X$ basis, we can expand $\rho_{A}$ as follows:
\begin{align*}
	\rho_{A} & =C(\sum_{c}|c\rangle\langle c|+\sum_{c}W_{A}|c\rangle\langle c|),\\
	& =C\sum_{c}(|c\rangle+W_{A}|c\rangle)\langle c|,
\end{align*}
where $c$ denotes a spin configuration of $A$ in Pauli $X$ basis, $|c\rangle$ is the corresponding configuration state, $C$ is a normalization constant. Here, we would like to mention that up to a basis transformation, the above $\rho_{A}$ also describes some other SESs, such as straight line SES along the direction of linear subsystem symmetry in $\cat$ and $\cluster$, and contractible dual loop SES in $\threedtc$.

Given the strong symmetry transformation $W_{A}=\prod_{i\in A}Z_{i}$, its associated t-patch operators of various dimensions are as follows:
\begin{itemize}
	\item $0$-dimensional t-patch operators: $Z_{i}$ and $X_{i}X_{i+1}$.
	\item $1$-dimensional t-patch operators: patch symmetry operator $P_{ij}^{Z}=Z_{i+1}Z_{i+2}\cdots Z_{j}$
	and patch charge operator $P_{ij}^{X}=X_{i}X_{j}$, where $i<j$ is assumed, and we use
	a convention to start the $P_{ij}^{Z}$ string from site $i+1$.
\end{itemize}

Note that $P_{ij}^{X}$ is regarded as an open string operator connecting site $i$ and $j$ with empty bulk, thus it can also be denoted as $P^X_{\partial \OS^1}$. Similarly, $P_{ij}^{Z}$ can be denoted as $P^Z_{\OS^1}$.

Then, we prove that the t-patch operators of $W_{A}$ are the weak symmetries of $\rho_{A}$:
\begin{itemize}
	\item $P^Z_{\OS^1}$: for an arbitrary $1$-dimensional
	patch symmetry operator $P^Z_{\OS^1}=Z_{i+1}Z_{i+2}\cdots Z_{j}$, where $\OS^1$ is an open string connecting site $i$ and $j$,
	we have
	\begin{align*}
		P^Z_{\OS^1}\rho_{A}P^Z_{\OS^1} & =C\sum_{c}(P^Z_{\OS^1}|c\rangle+W_{A}P^Z_{\OS^1}|c\rangle)\langle c|P_{\OS^1}^{Z},\\
		& =C\sum_{c}(|c\rangle+W_{A}|c\rangle)\langle c|,\\
		& =\rho_{A},
	\end{align*}
	where we used $[W_{A},P^Z_{\OS^1}]=0$ and the fact we can replace
	$|c\rangle$ by $P^Z_{\OS^1}|c\rangle$ inside the summation. Therefore,
	$P^Z_{\OS^1}$ is a weak symmetry of $\rho_{A}$.
	\item $P^X_{\partial \OS^1}$: for an arbitrary $1$-dimensional patch
	charge operator $P^X_{\partial \OS^1}=X_{i}X_{j}$, where $\OS^1$ is an open string connecting site $i$ and $j$, $\partial \OS^1$ is the boundary of $\OS^1$ (i.e., site $i$ and $j$), we have
	\begin{align*}
		P^X_{\partial \OS^1}\rho_{A}P^X_{\partial \OS^1}
		=C\sum_{c}(P^X_{\partial \OS^1}|c\rangle+W_{A}P^X_{\partial \OS^1}|c\rangle)\langle c|P^X_{\partial \OS^1},
	\end{align*}
	where we used $[W_{A},P^X_{\partial \OS^1}]=0$. Furthermore, noting that the action
	of $P^X_{\partial \OS^1}$ on $|c\rangle$ can only give a $1$ or $-1$ phase
	depending on $|c\rangle$, and the phase from $P^X_{\partial \OS^1}|c\rangle$
	and $\langle c|P^X_{\partial \OS^1}$ would cancel each other, we obtain $P^X_{\partial \OS^1}\rho_{A}P^X_{\partial \OS^1}=\rho_{A}$.
	Therefore, $P^X_{\partial \OS^1}$ is a weak symmetry of $\rho_{A}$.
	\item The $0$-dimensional t-patch operators $Z_i$ and $X_i X_{i+1}$ have the same form as the above
	spatially extended operators, thus they are also the weak symmetries
	of $\rho_{A}$.
\end{itemize}

In summary, we see that for either a 1D planar contractible dual loop SES in $\xcube$ or a 1D contractible dual loop SES in $|\psi\rangle_{\Z_2}$, its weak symmetries form the t-patch operators of its strong symmetry, thus the strong and weak symmetries together form a TCS. 

Finally, we demonstrate how such a TCS encodes a topological order. Because the two $0$-dimensional patch operators can be respectively expressed as $1$-dimensional patch operators $P_{i(i+1)}^{Z}$ and $P_{i(i+1)}^{X}$, and t-patch operators composed of the same type of Pauli operators must commute with each other, we can see the only important nontrivial commutation relation of the t-patch operators is as follows:
\begin{equation}
	P_{ij}^{X}P_{kl}^{Z}=-P_{kl}^{Z}P_{ij}^{X},\ i\leq k < j < l,
\end{equation}
which reproduces the braiding phase between the two kinds of fundamental topological excitations in 2D $\Z_2$ topological order. Without loss of generality, here we assume the $P_{kl}^{Z}$ operator is at ``the right side'' of the $P_{ij}^{X}$ operator, i.e., $k\geq i,l> j,l>k,j>i$. The complete algebra of these t-patch operators corresponds to a braided fusion category that describes the $\Z_{2}$ topological order in 2D toric code model~\cite{Chatterjee2023}.

\subsection{TCS of 2D contractible closed membrane SES in $\threedtc$}
\label{subsec:2D_global_TCS}

According to Eq.~\eqref{eq:RDM}, the reduced density matrix of this SES is:
\begin{equation}
	\rho_{A}=C'(I_{A}+W_{A}),
\end{equation}
where $I_{A}$ is the identity matrix supported on $A$, $W_{A}=\prod_{i\in A}X_{i}$ is uniformly acting Pauli $X$ on each spin in $A$, $C'$ is a normalization constant. In the Pauli $Z$ basis, we can expand $\rho_{A}$ as follows:
\begin{align*}
	\rho_{A} & =C(\sum_{c}|c\rangle\langle c|+\sum_{c}W_{A}|c\rangle\langle c|),\\
	& =C\sum_{c}(|c\rangle+W_{A}|c\rangle)\langle c|,
\end{align*}
where $c$ denotes a configuration in $A$, $|c\rangle$ is the corresponding configuration state, and $C$ is a normalization constant. Clearly, $W_{A}$ is a strong symmetry of $\rho_{A}$.

Given the strong symmetry transformation $W_{A}=\prod_{i\in A}X_{i}$, the associated t-patch operators of various dimensions are as follows:
\begin{itemize}
	\item $0$-dimensional t-patch operators: $X_{i}$ and $Z_{i}Z_{i+\hat{k}}$,
	where $\hat{k}=\hat{x},\hat{y}$ is the unit vector along spatial
	direction $x$ or $y$.
	\item $1$-dimensional t-patch operators: patch charge operator $P_{\partial\OS^1}^{Z} = P_{ij}^{Z}=Z_{i}Z_{j}$.
	\item $2$-dimensional t-patch operators: patch symmetry operator $P_{\OS^2}^{X}=\prod_{i\in \OS^2}X_{i}$, where
	$\OS^2$ is a 2D open disk.
\end{itemize}

Then, we prove that the t-patch operators of $W_{A}$ are the weak symmetries of $\rho_{A}$:
\begin{itemize}
	\item $P_{\partial \OS^1}^{Z}$: for an arbitrary $1$-dimensional patch
	charge operator $P_{\partial \OS^1}^{Z}=Z_{i}Z_{j}$, where $\OS^1$ is an open string connecting site $i$ and $j$, $\partial \OS^1$ is the boundary of $\OS^1$ (i.e., site $i$ and $j$), we have
	\begin{equation*}
		P_{\partial \OS^1}^{Z}\rho_{A}P_{\partial \OS^1}^{Z}=C\sum_{c}(P_{\partial \OS^1}^{Z}|c\rangle+W_{A}P_{\partial \OS^1}^{Z}|c\rangle)\langle c|P_{\partial \OS^1}^{Z},
	\end{equation*}
	where we used $[W_{A},P_{\partial \OS^1}^{Z}]=0$; then, notice that the action
	of $P_{\partial \OS^1}^{Z}$ on $|c\rangle$ can only give a $1$ or $-1$ phase
	depending on $|c\rangle$, and the phase from $P_{\partial \OS^1}^{Z}|c\rangle$
	and $\langle c|P_{\partial \OS^1}^{Z}$ would cancel each other, we obtain $P_{\partial \OS^1}^{Z}\rho_{A}P_{\partial \OS^1}^{Z}=\rho_{A}$.
	Therefore, $P_{\partial \OS^1}^{Z}$ is a weak symmetry of $\rho_{A}$.
	\item $P_{\OS^2}^{X}=\prod_{i\in \OS^2}X_{i}$: for an arbitrary $2$-dimensional
	patch symmetry operator $P_{\OS^2}^{X}=\prod_{i\in \OS^2}X_{i}$, where $\OS^2$ refers to a 2D open disk, we have
	\begin{align*}
		P_{\OS^2}^{X}\rho_{A}P_{\OS^2}^{X} & =C\sum_{c}(P_{\OS^2}^{X}|c\rangle+W_{A}P_{\OS^2}^{X}|c\rangle)\langle c|P_{\OS^2}^{X},\\
		& =C\sum_{c}(|c\rangle+W_{A}|c\rangle)\langle c|,\\
		& =\rho_{A},
	\end{align*}
	where we used $[W_{A},P_{\OS^2}^{X}]=0$ and the fact we can replace
	$|c\rangle$ by $P_{\OS^2}^{X}|c\rangle$ inside the summation. Therefore,
	$P_{\OS^2}^{X}$ is a weak symmetry of $\rho_{A}$.
	\item The $0$-dimensional t-patch operators $X_i$ and $Z_i Z_{i+\hat{k}}$ ($\hat{k}$ is a unit vector extending inside the membrane SES) have the same form as the above
	spatially extended operators, thus they are also the weak symmetries
	of $\rho_{A}$.
\end{itemize}

In summary, we see that for a 2D contractible closed membrane SES $\rho_{A}$ in $\threedtc$, a series of its weak symmetries form the t-patch operators of its strong symmetry, thus these strong and weak symmetries form a TCS.

Here, the nontrivial commutation relation between spatially extended t-patch operators is:
\begin{equation}
	P_{ij}^{Z}P_{\OS^2}^{X}=-P_{\OS^2}^{X}P_{ij}^{Z},
\end{equation}
where exactly one of site $i,j$ belongs to the disk $\OS^2$. The algebra of these t-patch operators corresponds to a braided fusion $2$-category that describes the topological order of the 3D $\Z_{2}$ toric code~\cite{Chatterjee2023}.

\subsection{TCS of 2D flat noncontractible closed membrane SES in $\xcube$}
\label{subsec:2D_1form_TCS}

According to Eq.~\eqref{eq:RDM}, the reduced density matrix of this SES is:
\begin{equation}
	\rho_{A}=C'\sum_{\tilde{c}}W_{\tilde{c}},
\end{equation}
where $I_{A}$ is the identity matrix supported on $A$, $W_{\tilde{c}}=\prod_{i\in\tilde{c}}Z_{i}$ is uniformly acting Pauli $Z$ on each spin in a pattern $\tilde{c}\subset A$ composed of dual loops and $C'$ is a normalization constant. Note that here $\tilde{c}$ is actually a set of spins, and when $\tilde{c}$ is empty (i.e., it contains no spin) $W_{\tilde{c}}$ reduces to $I_{A}$. Again, we can notice that for an arbitrary dual loop pattern $\tilde{c}'$:
\begin{equation*}
	W_{\tilde{c}'}\rho_{A}=C'\sum_{\tilde{c}}W_{\tilde{c}'}W_{\tilde{c}}=C'\sum_{\tilde{c}''}W_{\tilde{c}''}=\rho_{A},
\end{equation*}
since $W_{\tilde{c}'}W_{\tilde{c}}=W_{\tilde{c}''}$ has the same form as $W_{\tilde{c}}$, and we can freely exchange $\tilde{c}$ and $\tilde{c}''$ in the summation over all dual loop patterns. Therefore, for an arbitrary dual loop configuration $\tilde{c}'$, $W_{\tilde{c}'}$ is a strong symmetry of $\rho_{A}$.

In the Pauli $X$ basis, we can still expand $\rho_{A}$ as follows:
\begin{align*}
	\rho_{A} & =C\sum_{\tilde{c}}\sum_{c}W_{\tilde{c}}|c\rangle\langle c|,
\end{align*}
where $c$ denotes a configuration of $A$ in Pauli $X$ basis, $|c\rangle$ is the corresponding configuration state, $C$ is a normalization constant.

Given the $1$-form strong symmetry transformations $W_{\tilde{c}}=\prod_{i\in\tilde{c}}Z_{i}$, t-patch operators of various dimensions are as follows:
\begin{itemize}
	\item $0$-dimensional t-patch operators: $Z_{l}$, where $l$ denotes a link.
	\item $1$-dimensional t-patch operators: patch symmetry operator $P_{\OD^{1}}^{Z}=\prod_{l\in \OD^{1}}Z_{l}$,
	where $\OD^{1}$ is an open dual string;
	\item $2$-dimensional t-patch operators: patch charge operator $P_{\partial \OS^2}^{X}=\prod_{l\in\partial \OS^2}X_{l}$,
	where $\OS^2$ is a 2D open disk, $\partial \OS^2$ as a loop is the boundary of $\OS^2$.
\end{itemize}

Then, we prove that the t-patch operators of $W_{\tilde{c}}$ are the weak symmetries of $\rho_{A}$: 
\begin{itemize}
	\item $P_{\OD^{1}}^{Z}=\prod_{i\in \OD^{1}}Z_{i}$: for an arbitrary $1$-dimensional
	patch symmetry operator $P_{\OD^{1}}^{Z}=\prod_{i\in \OD^{1}}Z_{i}$,
	we have
	\begin{equation*}
		P_{\OD^{1}}^{Z}\rho_{A}P_{\OD^{1}}^{Z}=C\sum_{\tilde{c}}\sum_{c}W_{\tilde{c}}P_{\OD^{1}}^{Z}|c\rangle\langle c|P_{\OD^{1}}^{Z},
	\end{equation*}
	where we used $[W_{\tilde{c}},P_{\OD^{1}}^{Z}]=0$ and the fact we can
	replace $|c\rangle$ by $P_{\OD^{1}}^{Z}|c\rangle$ inside the summation
	over $c$. Therefore, $P_{\OD^{1}}^{Z}$ is a weak symmetry of $\rho_{A}$.
	\item $P_{\partial\OS^2}^{X}=\prod_{i\in\partial \OS^2}X_{i}$: for an arbitrary $2$-dimensional
	patch charge operator $P_{\partial\OS^2}^{X}=\prod_{i\in\partial \OS^2}X_{i}$, where $\OS^2$ refers to a 2D open disk, $\partial \OS^2$ is a loop, we
	have
	\begin{align*}
		P_{\partial\OS^2}^{X}\rho_{A}P_{\partial\OS^2}^{X} & =C\sum_{\tilde{c}}\sum_{c}W_{\tilde{c}}P_{\partial\OS^2}^{X}|c\rangle\langle c|P_{\partial\OS^2}^{X}=\rho_{A},
	\end{align*}
	where we used $[W_{\tilde{c}},P_{\partial\OS^2}^{X}]=0,\forall\tilde{c},\OS^2$. Notice
	that the action of $P_{\partial\OS^2}^{X}$ on $|c\rangle$ can only give a $1$
	or $-1$ phase depending on $|c\rangle$, and the phase from $P_{\partial\OS^2}^{X}|c\rangle$
	and $\langle c|P_{\partial\OS^2}^{X}$ would cancel each other, we obtain $P_{\partial\OS^2}^{X}\rho_{A}P_{\partial\OS^2}^{X}=\rho_{A}$.
	Therefore, $P_{\partial\OS^2}^{X}$ is a weak symmetry of $\rho_{A}$.
	\item The $0$-dimensional t-patch operators $Z_i$ have the same form as the above
	spatially extended operators, thus they are also the weak symmetries
	of $\rho_{A}$.
\end{itemize}

In summary, we see that for a noncontractible closed membrane SES $A$ in $\xcube$, a series of its weak symmetries form the t-patch operators of its strong symmetry, thus these strong and weak symmetries form a TCS.

Here, the nontrivial commutation relation between spatially extended t-patch operators is:
\begin{equation}
	P_{\OD^{1}}^{Z}P_{\partial \OS^2}^{X}=-P_{\partial \OS^2}^{X}P_{\OD^{1}}^{Z},
\end{equation}
when exactly one endpoint of $\OD^{1}$ belongs to the disk $\OS^2$. The algebra of these t-patch operators also corresponds to a braided fusion $2$-category that describes the topological order of the 3D $\Z_{2}$ toric code, as in the case of a 2D SES with global $\Z_{2}$ symmetry. Therefore, the $\Z_{2}$ $1$-form symmetry here is physically equivalent to the $\Z_{2}$ global symmetry in the sense that they impose the same constraints on local operators, i.e., the algebras of local operators symmetric under the two symmetries are equivalent, despite the apparent difference between the symmetry transformations~\cite{Chatterjee2023}.

\subsection{Robustness of TCS under FDQC}
\label{subsec:tpatch_robust}

In this subsection, we prove that TCS is robust under bipartition-factorized finite-depth quantum circuits (FDQCs) which preserve the entanglement entropy of the chosen entanglement subsystem.

Consider a pure state $|\psi\rangle$ on a total system $\mathcal{M}$. For a subsystem $A$, let $\rho_A = \text{Tr}_{\bar{A}}(|\psi\rangle\langle\psi|)$ be the corresponding reduced density matrix. Suppose $\rho_A$ is equipped with a TCS composed of strong symmetries $\{W_i\}$ and weak symmetries $\{w_j\}$, where $i$ and $j$ are indices labeling the symmetry transformations. We consider an FDQC $U$ acting on $|\psi\rangle$ that has a tensor product structure with respect to the partition into $A$ and its complement $\bar{A}$, i.e., $U = U_A \otimes U_{\bar{A}}$. Here, $U_A$ and $U_{\bar{A}}$ are FDQCs supported entirely on $A$ and $\bar{A}$, respectively. A circuit of this bipartition-factorized form preserves the entanglement entropy $S_A$. By contrast, a circuit that does not factorize in this way does not, in general, preserve SEE. To see this, consider a universal gate set consisting of single-qubit gates and CNOT gates. A circuit that violates the factorized form must contain at least one CNOT gate crossing the bipartition, which could make a stabilizer originally fully supported on $A$ acquire support in $\bar A$ under a cross-boundary CNOT. This can change the SEE. Thus, we focus on bipartition-factorized FDQCs. Under such a transformation, the reduced density matrix on $A$ becomes $\rho_A^a = U_A \rho_A U_A^\dagger$, where superscript $a$ is for ``altered''.

First, we show that $U$ preserves the strong and weak symmetries of $\rho_A$ up to an isomorphism. We begin with the weak symmetries. For an arbitrary weak symmetry $w_j$ supported entirely on $A$, it automatically commutes with $U_{\bar{A}}$. We define its transformed counterpart as $w_j^a = U_A w_j U_A^\dagger$. We can verify that $w_j^a$ is a symmetry of the new state $\rho_A^a$:
\begin{align*}
	w_j^a \rho_A^a (w_j^a)^\dagger 
	&= (U_A w_j U_A^\dagger) (U_A \rho_A U_A^\dagger) (U_A w_j U_A^\dagger)^\dagger \\
	&= U_A (w_j \rho_A w_j^\dagger) U_A^\dagger \\
	&= U_A \rho_A U_A^\dagger = \rho_A^a.
\end{align*}
Thus, $\{w_j^a\}$ form a set of weak symmetries of $\rho_A^a$.
Similarly, for a strong symmetry $W_i$ satisfying $W_i \rho_A = e^{i\theta} \rho_A$, we define its transformed counterpart as $W_i^a = U_A W_i U_A^\dagger$. It acts on $\rho_A^a$ as:
\begin{align*}
	W_i^a \rho_A^a &= (U_A W_i U_A^\dagger) (U_A \rho_A U_A^\dagger) \\
	&= U_A (W_i \rho_A) U_A^\dagger \\
	&= U_A (e^{i\theta} \rho_A) U_A^\dagger = e^{i\theta} \rho_A^a.
\end{align*}
Thus, $\{W_i^a\}$ form a set of strong symmetries of $\rho_A^a$. In summary, $U$ induces an isomorphism between the symmetries of $\rho_A$ and $\rho_A^a$, explicitly given by:
\begin{align}
	\omega \rightarrow U_A \omega U_A^{\dagger},
\end{align}
where $\omega \in (\{W_i\}\cup\{w_j\})$ is a symmetry transformation of $\rho_A$.

Next, we prove that the TCS structure itself is also preserved. This requires showing that the transformed weak symmetries are still transparent with respect to the transformed strong symmetries, and that their algebra is invariant. Since $U_A$ is an FDQC and the original symmetries are locally generated, the transformed symmetries $\{W_i^a\}$ and $\{w_j^a\}$ are also locally generated patch operators. The commutativity between transformed strong and weak symmetries, $[W_i^a, w_j^a] = 0$, is verified as follows:
\begin{align*}
	[W_i^a, w_j^a] &= [U_A W_i U_A^\dagger, U_A w_j U_A^\dagger] \\
	&= U_A [W_i, w_j] U_A^\dagger \\
	&= 0,
\end{align*}
where the last equality holds because the original symmetries commute, $[W_i, w_j] = 0$.

Finally, we demonstrate that the algebra of the weak symmetries is also preserved. The commutator of any pair of transformed weak symmetries is given by:
\begin{align}
	[w_j^a, w_k^a] = [U_A w_j U_A^\dagger, U_A w_k U_A^\dagger] = U_A [w_j, w_k] U_A^\dagger .
\end{align}
Furthermore, as an FDQC, $U_A$ is locality-preserving and can enlarge the support of an operator only by a finite distance determined by the circuit depth and gate range. Therefore, for patch operators whose boundaries are separated by distances larger than this finite dressing scale, topological relations between their supports, such as linking or nonlinking, are preserved. This implies that the transformed weak symmetries still form t-patch operators of the transformed strong symmetries and span an equivalent algebra up to the isomorphism given by conjugation by $U_A$. In conclusion, an FDQC of the form $U = U_A \otimes U_{\bar{A}}$ preserves the transparent composite symmetries of $\rho_A$.

\section{Summary and outlook}
\label{sec:outlook}
In this paper, we establish subdimensional entanglement entropy (SEE) as a sharp probe of quantum many-body orders through its geometric and topological responses, and as a route from bulk pure-state entanglement to mixed-state symmetry and holography on subdimensional manifolds. By systematically analyzing subdimensional manifolds in a range of quantum phases, we show that SEE resolves structures that are invisible to conventional entanglement diagnostics. We further prove a general relation between subdimensional entanglement subsystems (SESs) and mixed-state symmetries in stabilizer states, and use it to uncover strong-to-weak spontaneous symmetry breaking (SW-SSB) in SESs with nontrivial SEE. Most notably, we identify an entanglement-induced mixed-state holography: for each SES considered in this work, weak and strong symmetries combine into a transparent composite symmetry (TCS), whose algebra holographically encodes a topological order of one higher dimension. We further show that TCS remains robust under the bipartition-factorized finite-depth quantum circuits which preserve the SEE of the chosen SES. This robustness suggests that the structure may extend beyond exactly solvable points.

These results open several directions for future work. First, throughout this paper, we focus on SEE of stabilizer states, which serve as representative models for various phases but do not allow us to investigate SEE in general states. Thus, a natural direction for future work is to explore SEE in broader settings beyond stabilizer states, including fracton phases and cluster states in the presence of external fields, using numerical methods such as tensor-network wavefunctions~\cite{Zhu2022} and quantum Monte Carlo~\cite{Zhou2022,Zhou2022a,Ding2025,Mao2025}. Analytically exploring the SEE in each phase requires a concrete discussion within the specific theoretical framework of that phase, which would require extensive further work~\cite{Calabrese2004,Castelnovo2007a,Castelnovo2008,Eisert2010,Kalis2012,Ma2018a,Stephen2019,Li2019,Williamson2019a,Reiss2019,Muehlhauser2020}. Free-fermion studies on fractal and hyperbolic lattices provide further examples of geometry-dependent entanglement scaling~\cite{ZhouYe2024,HuangZhouYe2025}, motivating extensions of SEE beyond the stabilizer setting. For SESs, another important direction is to investigate their singular-value spectrum and Schmidt vectors, in order to test possible extensions of the correspondence between entanglement spectrum and physical edge spectrum~\cite{Li2008}. 

Furthermore, it is interesting to extend the idea of TCS to general composite symmetries composed of strong and weak symmetries satisfying certain relations. We expect such general composite symmetries and their breaking to provide another perspective for understanding mixed-state phases. Finally, volume-law scaling of SEE and SW-SSB at the exactly solvable points both show the similarity between thermal states within a certain symmetry sector and SESs with nontrivial SEE~\cite{Popescu2006,Lessa2025,Liu2025}, which inspires us to further explore the connection between thermalized systems and complicated many-body entangled states.

\acknowledgments
We thank Z.-X. Luo, M. Li, G. Yue, and R. Luo for helpful discussions. This project was initiated and primarily carried out at Sun Yat-sen University with support from the Research Center for Magnetoelectric Physics of Guangdong Province (Grant No.~2024B0303390001), the Guangdong Provincial Key Laboratory of Magnetoelectric Physics and Devices (Grant No.~2022B1212010008), and the National Natural Science Foundation of China (Grant Nos.~12474149, 12074438, and 11847608). Additional support was provided by the Shanghai Committee of Science and Technology (Grant No.~25LZ2600800). M.-Y. Li was also supported by the Shuimu Tsinghua Scholar Program.

%\bibliography{REF}

\begin{thebibliography}{104}%
\makeatletter
\providecommand \@ifxundefined [1]{%
 \@ifx{#1\undefined}
}%
\providecommand \@ifnum [1]{%
 \ifnum #1\expandafter \@firstoftwo
 \else \expandafter \@secondoftwo
 \fi
}%
\providecommand \@ifx [1]{%
 \ifx #1\expandafter \@firstoftwo
 \else \expandafter \@secondoftwo
 \fi
}%
\providecommand \natexlab [1]{#1}%
\providecommand \enquote  [1]{``#1''}%
\providecommand \bibnamefont  [1]{#1}%
\providecommand \bibfnamefont [1]{#1}%
\providecommand \citenamefont [1]{#1}%
\providecommand \href@noop [0]{\@secondoftwo}%
\providecommand \href [0]{\begingroup \@sanitize@url \@href}%
\providecommand \@href[1]{\@@startlink{#1}\@@href}%
\providecommand \@@href[1]{\endgroup#1\@@endlink}%
\providecommand \@sanitize@url [0]{\catcode `\\12\catcode `\$12\catcode `\&12\catcode `\#12\catcode `\^12\catcode `\_12\catcode `\%12\relax}%
\providecommand \@@startlink[1]{}%
\providecommand \@@endlink[0]{}%
\providecommand \url  [0]{\begingroup\@sanitize@url \@url }%
\providecommand \@url [1]{\endgroup\@href {#1}{\urlprefix }}%
\providecommand \urlprefix  [0]{URL }%
\providecommand \Eprint [0]{\href }%
\providecommand \doibase [0]{https://doi.org/}%
\providecommand \selectlanguage [0]{\@gobble}%
\providecommand \bibinfo  [0]{\@secondoftwo}%
\providecommand \bibfield  [0]{\@secondoftwo}%
\providecommand \translation [1]{[#1]}%
\providecommand \BibitemOpen [0]{}%
\providecommand \bibitemStop [0]{}%
\providecommand \bibitemNoStop [0]{.\EOS\space}%
\providecommand \EOS [0]{\spacefactor3000\relax}%
\providecommand \BibitemShut  [1]{\csname bibitem#1\endcsname}%
\let\auto@bib@innerbib\@empty
%</preamble>
\bibitem [{\citenamefont {{Levin}}\ and\ \citenamefont {{Wen}}(2006)}]{Levin2006}%
  \BibitemOpen
  \bibfield  {author} {\bibinfo {author} {\bibfnamefont {M.}~\bibnamefont {{Levin}}}\ and\ \bibinfo {author} {\bibfnamefont {X.-G.}\ \bibnamefont {{Wen}}},\ }\bibfield  {title} {\bibinfo {title} {{Detecting Topological Order in a Ground State Wave Function}},\ }\href {https://doi.org/10.1103/PhysRevLett.96.110405} {\bibfield  {journal} {\bibinfo  {journal} {Phys. Rev. Lett.}\ }\textbf {\bibinfo {volume} {96}},\ \bibinfo {pages} {110405} (\bibinfo {year} {2006})}\BibitemShut {NoStop}%
\bibitem [{\citenamefont {{Kitaev}}\ and\ \citenamefont {{Preskill}}(2006)}]{Kitaev2006}%
  \BibitemOpen
  \bibfield  {author} {\bibinfo {author} {\bibfnamefont {A.}~\bibnamefont {{Kitaev}}}\ and\ \bibinfo {author} {\bibfnamefont {J.}~\bibnamefont {{Preskill}}},\ }\bibfield  {title} {\bibinfo {title} {{Topological Entanglement Entropy}},\ }\href {https://doi.org/10.1103/PhysRevLett.96.110404} {\bibfield  {journal} {\bibinfo  {journal} {\prl}\ }\textbf {\bibinfo {volume} {96}},\ \bibinfo {eid} {110404} (\bibinfo {year} {2006})},\ \Eprint {https://arxiv.org/abs/hep-th/0510092} {arXiv:hep-th/0510092 [hep-th]} \BibitemShut {NoStop}%
\bibitem [{\citenamefont {Haah}(2011)}]{Haah2011}%
  \BibitemOpen
  \bibfield  {author} {\bibinfo {author} {\bibfnamefont {J.}~\bibnamefont {Haah}},\ }\bibfield  {title} {\bibinfo {title} {{Local stabilizer codes in three dimensions without string logical operators}},\ }\href {https://doi.org/10.1103/PhysRevA.83.042330} {\bibfield  {journal} {\bibinfo  {journal} {Phys. Rev. A}\ }\textbf {\bibinfo {volume} {83}},\ \bibinfo {pages} {042330} (\bibinfo {year} {2011})}\BibitemShut {NoStop}%
\bibitem [{\citenamefont {Vijay}\ \emph {et~al.}(2015)\citenamefont {Vijay}, \citenamefont {Haah},\ and\ \citenamefont {Fu}}]{Vijay2015}%
  \BibitemOpen
  \bibfield  {author} {\bibinfo {author} {\bibfnamefont {S.}~\bibnamefont {Vijay}}, \bibinfo {author} {\bibfnamefont {J.}~\bibnamefont {Haah}},\ and\ \bibinfo {author} {\bibfnamefont {L.}~\bibnamefont {Fu}},\ }\bibfield  {title} {\bibinfo {title} {{A new kind of topological quantum order: A dimensional hierarchy of quasiparticles built from stationary excitations}},\ }\href {https://doi.org/10.1103/PhysRevB.92.235136} {\bibfield  {journal} {\bibinfo  {journal} {Phys. Rev. B}\ }\textbf {\bibinfo {volume} {92}},\ \bibinfo {pages} {235136} (\bibinfo {year} {2015})}\BibitemShut {NoStop}%
\bibitem [{\citenamefont {Vijay}\ \emph {et~al.}(2016)\citenamefont {Vijay}, \citenamefont {Haah},\ and\ \citenamefont {Fu}}]{Vijay2016}%
  \BibitemOpen
  \bibfield  {author} {\bibinfo {author} {\bibfnamefont {S.}~\bibnamefont {Vijay}}, \bibinfo {author} {\bibfnamefont {J.}~\bibnamefont {Haah}},\ and\ \bibinfo {author} {\bibfnamefont {L.}~\bibnamefont {Fu}},\ }\bibfield  {title} {\bibinfo {title} {{Fracton topological order, generalized lattice gauge theory, and duality}},\ }\href {https://doi.org/10.1103/PhysRevB.94.235157} {\bibfield  {journal} {\bibinfo  {journal} {Phys. Rev. B}\ }\textbf {\bibinfo {volume} {94}},\ \bibinfo {pages} {235157} (\bibinfo {year} {2016})}\BibitemShut {NoStop}%
\bibitem [{\citenamefont {{Zou}}\ and\ \citenamefont {{Haah}}(2016)}]{Zou2016}%
  \BibitemOpen
  \bibfield  {author} {\bibinfo {author} {\bibfnamefont {L.}~\bibnamefont {{Zou}}}\ and\ \bibinfo {author} {\bibfnamefont {J.}~\bibnamefont {{Haah}}},\ }\bibfield  {title} {\bibinfo {title} {{Spurious long-range entanglement and replica correlation length}},\ }\href {https://doi.org/10.1103/PhysRevB.94.075151} {\bibfield  {journal} {\bibinfo  {journal} {Phys. Rev. B}\ }\textbf {\bibinfo {volume} {94}},\ \bibinfo {pages} {075151} (\bibinfo {year} {2016})}\BibitemShut {NoStop}%
\bibitem [{\citenamefont {{Shirley}}\ \emph {et~al.}(2019)\citenamefont {{Shirley}}, \citenamefont {{Slagle}},\ and\ \citenamefont {{Chen}}}]{Shirley2019c}%
  \BibitemOpen
  \bibfield  {author} {\bibinfo {author} {\bibfnamefont {W.}~\bibnamefont {{Shirley}}}, \bibinfo {author} {\bibfnamefont {K.}~\bibnamefont {{Slagle}}},\ and\ \bibinfo {author} {\bibfnamefont {X.}~\bibnamefont {{Chen}}},\ }\bibfield  {title} {\bibinfo {title} {{Universal entanglement signatures of foliated fracton phases}},\ }\href {https://doi.org/10.21468/SciPostPhys.6.1.015} {\bibfield  {journal} {\bibinfo  {journal} {SciPost Phys.}\ }\textbf {\bibinfo {volume} {6}},\ \bibinfo {pages} {015} (\bibinfo {year} {2019})}\BibitemShut {NoStop}%
\bibitem [{\citenamefont {{Williamson}}\ \emph {et~al.}(2019)\citenamefont {{Williamson}}, \citenamefont {{Dua}},\ and\ \citenamefont {{Cheng}}}]{Williamson2019a}%
  \BibitemOpen
  \bibfield  {author} {\bibinfo {author} {\bibfnamefont {D.~J.}\ \bibnamefont {{Williamson}}}, \bibinfo {author} {\bibfnamefont {A.}~\bibnamefont {{Dua}}},\ and\ \bibinfo {author} {\bibfnamefont {M.}~\bibnamefont {{Cheng}}},\ }\bibfield  {title} {\bibinfo {title} {{Spurious Topological Entanglement Entropy from Subsystem Symmetries}},\ }\href {https://doi.org/10.1103/PhysRevLett.122.140506} {\bibfield  {journal} {\bibinfo  {journal} {Phys. Rev. Lett.}\ }\textbf {\bibinfo {volume} {122}},\ \bibinfo {pages} {140506} (\bibinfo {year} {2019})}\BibitemShut {NoStop}%
\bibitem [{\citenamefont {Ma}\ \emph {et~al.}(2018)\citenamefont {Ma}, \citenamefont {Schmitz}, \citenamefont {Parameswaran}, \citenamefont {Hermele},\ and\ \citenamefont {Nandkishore}}]{Ma2018a}%
  \BibitemOpen
  \bibfield  {author} {\bibinfo {author} {\bibfnamefont {H.}~\bibnamefont {Ma}}, \bibinfo {author} {\bibfnamefont {A.~T.}\ \bibnamefont {Schmitz}}, \bibinfo {author} {\bibfnamefont {S.~A.}\ \bibnamefont {Parameswaran}}, \bibinfo {author} {\bibfnamefont {M.}~\bibnamefont {Hermele}},\ and\ \bibinfo {author} {\bibfnamefont {R.~M.}\ \bibnamefont {Nandkishore}},\ }\bibfield  {title} {\bibinfo {title} {{Topological entanglement entropy of fracton stabilizer codes}},\ }\href {https://doi.org/10.1103/PhysRevB.97.125101} {\bibfield  {journal} {\bibinfo  {journal} {Phys. Rev. B}\ }\textbf {\bibinfo {volume} {97}},\ \bibinfo {pages} {125101} (\bibinfo {year} {2018})}\BibitemShut {NoStop}%
\bibitem [{\citenamefont {Shi}\ and\ \citenamefont {Lu}(2018)}]{Shi2018}%
  \BibitemOpen
  \bibfield  {author} {\bibinfo {author} {\bibfnamefont {B.}~\bibnamefont {Shi}}\ and\ \bibinfo {author} {\bibfnamefont {Y.-M.}\ \bibnamefont {Lu}},\ }\bibfield  {title} {\bibinfo {title} {{Deciphering the nonlocal entanglement entropy of fracton topological orders}},\ }\href {https://doi.org/10.1103/PhysRevB.97.144106} {\bibfield  {journal} {\bibinfo  {journal} {Phys. Rev. B}\ }\textbf {\bibinfo {volume} {97}},\ \bibinfo {pages} {144106} (\bibinfo {year} {2018})}\BibitemShut {NoStop}%
\bibitem [{\citenamefont {{Zhang}}\ \emph {et~al.}(2024)\citenamefont {{Zhang}}, \citenamefont {{Li}},\ and\ \citenamefont {{Ye}}}]{Zhang2024}%
  \BibitemOpen
  \bibfield  {author} {\bibinfo {author} {\bibfnamefont {J.-Y.}\ \bibnamefont {{Zhang}}}, \bibinfo {author} {\bibfnamefont {M.-Y.}\ \bibnamefont {{Li}}},\ and\ \bibinfo {author} {\bibfnamefont {P.}~\bibnamefont {{Ye}}},\ }\bibfield  {title} {\bibinfo {title} {{Higher-Order Cellular Automata Generated Symmetry-Protected Topological Phases and Detection Through Multi Point Strange Correlators}},\ }\href {https://doi.org/10.1103/PRXQuantum.5.030342} {\bibfield  {journal} {\bibinfo  {journal} {PRX Quantum}\ }\textbf {\bibinfo {volume} {5}},\ \bibinfo {pages} {030342} (\bibinfo {year} {2024})}\BibitemShut {NoStop}%
\bibitem [{\citenamefont {You}\ \emph {et~al.}(2018)\citenamefont {You}, \citenamefont {Devakul}, \citenamefont {Burnell},\ and\ \citenamefont {Sondhi}}]{You2018}%
  \BibitemOpen
  \bibfield  {author} {\bibinfo {author} {\bibfnamefont {Y.}~\bibnamefont {You}}, \bibinfo {author} {\bibfnamefont {T.}~\bibnamefont {Devakul}}, \bibinfo {author} {\bibfnamefont {F.~J.}\ \bibnamefont {Burnell}},\ and\ \bibinfo {author} {\bibfnamefont {S.~L.}\ \bibnamefont {Sondhi}},\ }\bibfield  {title} {\bibinfo {title} {{Subsystem symmetry protected topological order}},\ }\href {https://doi.org/10.1103/PhysRevB.98.035112} {\bibfield  {journal} {\bibinfo  {journal} {Phys. Rev. B}\ }\textbf {\bibinfo {volume} {98}},\ \bibinfo {pages} {035112} (\bibinfo {year} {2018})}\BibitemShut {NoStop}%
\bibitem [{\citenamefont {{Stephen}}\ \emph {et~al.}(2019)\citenamefont {{Stephen}}, \citenamefont {{Dreyer}}, \citenamefont {{Iqbal}},\ and\ \citenamefont {{Schuch}}}]{Stephen2019}%
  \BibitemOpen
  \bibfield  {author} {\bibinfo {author} {\bibfnamefont {D.~T.}\ \bibnamefont {{Stephen}}}, \bibinfo {author} {\bibfnamefont {H.}~\bibnamefont {{Dreyer}}}, \bibinfo {author} {\bibfnamefont {M.}~\bibnamefont {{Iqbal}}},\ and\ \bibinfo {author} {\bibfnamefont {N.}~\bibnamefont {{Schuch}}},\ }\bibfield  {title} {\bibinfo {title} {{Detecting subsystem symmetry protected topological order via entanglement entropy}},\ }\href {https://doi.org/10.1103/PhysRevB.100.115112} {\bibfield  {journal} {\bibinfo  {journal} {Phys. Rev. B}\ }\textbf {\bibinfo {volume} {100}},\ \bibinfo {pages} {115112} (\bibinfo {year} {2019})}\BibitemShut {NoStop}%
\bibitem [{\citenamefont {{Kato}}\ and\ \citenamefont {{Brand{\~a}o}}(2020)}]{Kato2020}%
  \BibitemOpen
  \bibfield  {author} {\bibinfo {author} {\bibfnamefont {K.}~\bibnamefont {{Kato}}}\ and\ \bibinfo {author} {\bibfnamefont {F.~G.~S.~L.}\ \bibnamefont {{Brand{\~a}o}}},\ }\bibfield  {title} {\bibinfo {title} {{Toy model of boundary states with spurious topological entanglement entropy}},\ }\href {https://doi.org/10.1103/PhysRevResearch.2.032005} {\bibfield  {journal} {\bibinfo  {journal} {Phys. Rev. Research}\ }\textbf {\bibinfo {volume} {2}},\ \bibinfo {pages} {032005} (\bibinfo {year} {2020})}\BibitemShut {NoStop}%
\bibitem [{\citenamefont {{Berthiere}}\ and\ \citenamefont {{Witczak-Krempa}}(2022)}]{Berthiere2022}%
  \BibitemOpen
  \bibfield  {author} {\bibinfo {author} {\bibfnamefont {C.}~\bibnamefont {{Berthiere}}}\ and\ \bibinfo {author} {\bibfnamefont {W.}~\bibnamefont {{Witczak-Krempa}}},\ }\bibfield  {title} {\bibinfo {title} {{Entanglement of Skeletal Regions}},\ }\href {https://doi.org/10.1103/PhysRevLett.128.240502} {\bibfield  {journal} {\bibinfo  {journal} {Phys. Rev. Lett.}\ }\textbf {\bibinfo {volume} {128}},\ \bibinfo {pages} {240502} (\bibinfo {year} {2022})}\BibitemShut {NoStop}%
\bibitem [{\citenamefont {{Lyu}}\ \emph {et~al.}(2026)\citenamefont {{Lyu}}, \citenamefont {{Chandorkar}}, \citenamefont {{Kapoor}}, \citenamefont {{Takei}}, \citenamefont {{S{\o}rensen}},\ and\ \citenamefont {{Witczak-Krempa}}}]{Lyu2025}%
  \BibitemOpen
  \bibfield  {author} {\bibinfo {author} {\bibfnamefont {L.}~\bibnamefont {{Lyu}}}, \bibinfo {author} {\bibfnamefont {D.}~\bibnamefont {{Chandorkar}}}, \bibinfo {author} {\bibfnamefont {S.}~\bibnamefont {{Kapoor}}}, \bibinfo {author} {\bibfnamefont {S.}~\bibnamefont {{Takei}}}, \bibinfo {author} {\bibfnamefont {E.~S.}\ \bibnamefont {{S{\o}rensen}}},\ and\ \bibinfo {author} {\bibfnamefont {W.}~\bibnamefont {{Witczak-Krempa}}},\ }\bibfield  {title} {\bibinfo {title} {{Multiparty entanglement loops in quantum spin liquids}},\ }\href {https://doi.org/10.1038/s41467-026-71113-8} {\bibfield  {journal} {\bibinfo  {journal} {Nat. Commun.}\ }\textbf {\bibinfo {volume} {17}},\ \bibinfo {pages} {5650} (\bibinfo {year} {2026})}\BibitemShut {NoStop}%
\bibitem [{\citenamefont {{Qi}}\ \emph {et~al.}(2008)\citenamefont {{Qi}}, \citenamefont {{Hughes}},\ and\ \citenamefont {{Zhang}}}]{Qi2008a}%
  \BibitemOpen
  \bibfield  {author} {\bibinfo {author} {\bibfnamefont {X.-L.}\ \bibnamefont {{Qi}}}, \bibinfo {author} {\bibfnamefont {T.~L.}\ \bibnamefont {{Hughes}}},\ and\ \bibinfo {author} {\bibfnamefont {S.-C.}\ \bibnamefont {{Zhang}}},\ }\bibfield  {title} {\bibinfo {title} {{Topological field theory of time-reversal invariant insulators}},\ }\href {https://doi.org/10.1103/PhysRevB.78.195424} {\bibfield  {journal} {\bibinfo  {journal} {Phys. Rev. B}\ }\textbf {\bibinfo {volume} {78}},\ \bibinfo {pages} {195424} (\bibinfo {year} {2008})}\BibitemShut {NoStop}%
\bibitem [{\citenamefont {{Ryu}}\ \emph {et~al.}(2012)\citenamefont {{Ryu}}, \citenamefont {{Moore}},\ and\ \citenamefont {{Ludwig}}}]{Ryu2012b}%
  \BibitemOpen
  \bibfield  {author} {\bibinfo {author} {\bibfnamefont {S.}~\bibnamefont {{Ryu}}}, \bibinfo {author} {\bibfnamefont {J.~E.}\ \bibnamefont {{Moore}}},\ and\ \bibinfo {author} {\bibfnamefont {A.~W.~W.}\ \bibnamefont {{Ludwig}}},\ }\bibfield  {title} {\bibinfo {title} {{Electromagnetic and gravitational responses and anomalies in topological insulators and superconductors}},\ }\href {https://doi.org/10.1103/PhysRevB.85.045104} {\bibfield  {journal} {\bibinfo  {journal} {Phys. Rev. B}\ }\textbf {\bibinfo {volume} {85}},\ \bibinfo {pages} {045104} (\bibinfo {year} {2012})}\BibitemShut {NoStop}%
\bibitem [{\citenamefont {{Stone}}(2012)}]{Stone2012a}%
  \BibitemOpen
  \bibfield  {author} {\bibinfo {author} {\bibfnamefont {M.}~\bibnamefont {{Stone}}},\ }\bibfield  {title} {\bibinfo {title} {{Gravitational anomalies and thermal Hall effect in topological insulators}},\ }\href {https://doi.org/10.1103/PhysRevB.85.184503} {\bibfield  {journal} {\bibinfo  {journal} {Phys. Rev. B}\ }\textbf {\bibinfo {volume} {85}},\ \bibinfo {pages} {184503} (\bibinfo {year} {2012})}\BibitemShut {NoStop}%
\bibitem [{\citenamefont {{Ye}}\ and\ \citenamefont {{Wang}}(2013)}]{Ye2013c}%
  \BibitemOpen
  \bibfield  {author} {\bibinfo {author} {\bibfnamefont {P.}~\bibnamefont {{Ye}}}\ and\ \bibinfo {author} {\bibfnamefont {J.}~\bibnamefont {{Wang}}},\ }\bibfield  {title} {\bibinfo {title} {{Symmetry-protected topological phases with charge and spin symmetries: Response theory and dynamical gauge theory in two and three dimensions}},\ }\href {https://doi.org/10.1103/PhysRevB.88.235109} {\bibfield  {journal} {\bibinfo  {journal} {Phys. Rev. B}\ }\textbf {\bibinfo {volume} {88}},\ \bibinfo {pages} {235109} (\bibinfo {year} {2013})}\BibitemShut {NoStop}%
\bibitem [{\citenamefont {{Cheng}}\ and\ \citenamefont {{Gu}}(2014)}]{Cheng2014a}%
  \BibitemOpen
  \bibfield  {author} {\bibinfo {author} {\bibfnamefont {M.}~\bibnamefont {{Cheng}}}\ and\ \bibinfo {author} {\bibfnamefont {Z.-C.}\ \bibnamefont {{Gu}}},\ }\bibfield  {title} {\bibinfo {title} {{Topological Response Theory of Abelian Symmetry-Protected Topological Phases in Two Dimensions}},\ }\href {https://doi.org/10.1103/PhysRevLett.112.141602} {\bibfield  {journal} {\bibinfo  {journal} {Phys. Rev. Lett.}\ }\textbf {\bibinfo {volume} {112}},\ \bibinfo {pages} {141602} (\bibinfo {year} {2014})}\BibitemShut {NoStop}%
\bibitem [{\citenamefont {{Lapa}}\ \emph {et~al.}(2017)\citenamefont {{Lapa}}, \citenamefont {{Jian}}, \citenamefont {{Ye}},\ and\ \citenamefont {{Hughes}}}]{Lapa2017a}%
  \BibitemOpen
  \bibfield  {author} {\bibinfo {author} {\bibfnamefont {M.~F.}\ \bibnamefont {{Lapa}}}, \bibinfo {author} {\bibfnamefont {C.-M.}\ \bibnamefont {{Jian}}}, \bibinfo {author} {\bibfnamefont {P.}~\bibnamefont {{Ye}}},\ and\ \bibinfo {author} {\bibfnamefont {T.~L.}\ \bibnamefont {{Hughes}}},\ }\bibfield  {title} {\bibinfo {title} {{Topological electromagnetic responses of bosonic quantum Hall, topological insulator, and chiral semimetal phases in all dimensions}},\ }\href {https://doi.org/10.1103/PhysRevB.95.035149} {\bibfield  {journal} {\bibinfo  {journal} {Phys. Rev. B}\ }\textbf {\bibinfo {volume} {95}},\ \bibinfo {pages} {035149} (\bibinfo {year} {2017})}\BibitemShut {NoStop}%
\bibitem [{\citenamefont {{Nissinen}}\ and\ \citenamefont {{Volovik}}(2019)}]{Nissinen2019}%
  \BibitemOpen
  \bibfield  {author} {\bibinfo {author} {\bibfnamefont {J.}~\bibnamefont {{Nissinen}}}\ and\ \bibinfo {author} {\bibfnamefont {G.~E.}\ \bibnamefont {{Volovik}}},\ }\bibfield  {title} {\bibinfo {title} {{Elasticity tetrads, mixed axial-gravitational anomalies, and $(3+1)$-d quantum Hall effect}},\ }\href {https://doi.org/10.1103/PhysRevResearch.1.023007} {\bibfield  {journal} {\bibinfo  {journal} {Phys. Rev. Research}\ }\textbf {\bibinfo {volume} {1}},\ \bibinfo {pages} {023007} (\bibinfo {year} {2019})}\BibitemShut {NoStop}%
\bibitem [{\citenamefont {Han}\ \emph {et~al.}(2019)\citenamefont {Han}, \citenamefont {Wang},\ and\ \citenamefont {Ye}}]{Han2019}%
  \BibitemOpen
  \bibfield  {author} {\bibinfo {author} {\bibfnamefont {B.}~\bibnamefont {Han}}, \bibinfo {author} {\bibfnamefont {H.}~\bibnamefont {Wang}},\ and\ \bibinfo {author} {\bibfnamefont {P.}~\bibnamefont {Ye}},\ }\bibfield  {title} {\bibinfo {title} {{Generalized Wen-Zee terms}},\ }\href {https://doi.org/10.1103/PhysRevB.99.205120} {\bibfield  {journal} {\bibinfo  {journal} {Phys. Rev. B}\ }\textbf {\bibinfo {volume} {99}},\ \bibinfo {pages} {205120} (\bibinfo {year} {2019})}\BibitemShut {NoStop}%
\bibitem [{\citenamefont {{Bu{\v{c}}a}}\ and\ \citenamefont {{Prosen}}(2012)}]{Buca2012}%
  \BibitemOpen
  \bibfield  {author} {\bibinfo {author} {\bibfnamefont {B.}~\bibnamefont {{Bu{\v{c}}a}}}\ and\ \bibinfo {author} {\bibfnamefont {T.}~\bibnamefont {{Prosen}}},\ }\bibfield  {title} {\bibinfo {title} {{A note on symmetry reductions of the Lindblad equation: transport in constrained open spin chains}},\ }\href {https://doi.org/10.1088/1367-2630/14/7/073007} {\bibfield  {journal} {\bibinfo  {journal} {New J. Phys.}\ }\textbf {\bibinfo {volume} {14}},\ \bibinfo {pages} {073007} (\bibinfo {year} {2012})}\BibitemShut {NoStop}%
\bibitem [{\citenamefont {{de Groot}}\ \emph {et~al.}(2022)\citenamefont {{de Groot}}, \citenamefont {{Turzillo}},\ and\ \citenamefont {{Schuch}}}]{deGroot2022}%
  \BibitemOpen
  \bibfield  {author} {\bibinfo {author} {\bibfnamefont {C.}~\bibnamefont {{de Groot}}}, \bibinfo {author} {\bibfnamefont {A.}~\bibnamefont {{Turzillo}}},\ and\ \bibinfo {author} {\bibfnamefont {N.}~\bibnamefont {{Schuch}}},\ }\bibfield  {title} {\bibinfo {title} {{Symmetry Protected Topological Order in Open Quantum Systems}},\ }\href {https://doi.org/10.22331/q-2022-11-10-856} {\bibfield  {journal} {\bibinfo  {journal} {Quantum}\ }\textbf {\bibinfo {volume} {6}},\ \bibinfo {pages} {856} (\bibinfo {year} {2022})}\BibitemShut {NoStop}%
\bibitem [{\citenamefont {{Ma}}\ and\ \citenamefont {{Wang}}(2023)}]{Ma2023a}%
  \BibitemOpen
  \bibfield  {author} {\bibinfo {author} {\bibfnamefont {R.}~\bibnamefont {{Ma}}}\ and\ \bibinfo {author} {\bibfnamefont {C.}~\bibnamefont {{Wang}}},\ }\bibfield  {title} {\bibinfo {title} {{Average Symmetry-Protected Topological Phases}},\ }\href {https://doi.org/10.1103/PhysRevX.13.031016} {\bibfield  {journal} {\bibinfo  {journal} {Phys. Rev. X}\ }\textbf {\bibinfo {volume} {13}},\ \bibinfo {pages} {031016} (\bibinfo {year} {2023})}\BibitemShut {NoStop}%
\bibitem [{\citenamefont {{Ma}}\ \emph {et~al.}(2025)\citenamefont {{Ma}}, \citenamefont {{Zhang}}, \citenamefont {{Bi}}, \citenamefont {{Cheng}},\ and\ \citenamefont {{Wang}}}]{Ma2023}%
  \BibitemOpen
  \bibfield  {author} {\bibinfo {author} {\bibfnamefont {R.}~\bibnamefont {{Ma}}}, \bibinfo {author} {\bibfnamefont {J.-H.}\ \bibnamefont {{Zhang}}}, \bibinfo {author} {\bibfnamefont {Z.}~\bibnamefont {{Bi}}}, \bibinfo {author} {\bibfnamefont {M.}~\bibnamefont {{Cheng}}},\ and\ \bibinfo {author} {\bibfnamefont {C.}~\bibnamefont {{Wang}}},\ }\bibfield  {title} {\bibinfo {title} {{Topological Phases with Average Symmetries: The Decohered, the Disordered, and the Intrinsic}},\ }\href {https://doi.org/10.1103/PhysRevX.15.021062} {\bibfield  {journal} {\bibinfo  {journal} {Phys. Rev. X}\ }\textbf {\bibinfo {volume} {15}},\ \bibinfo {pages} {021062} (\bibinfo {year} {2025})}\BibitemShut {NoStop}%
\bibitem [{\citenamefont {{Lieu}}\ \emph {et~al.}(2020)\citenamefont {{Lieu}}, \citenamefont {{Belyansky}}, \citenamefont {{Young}}, \citenamefont {{Lundgren}}, \citenamefont {{Albert}},\ and\ \citenamefont {{Gorshkov}}}]{Lieu2020}%
  \BibitemOpen
  \bibfield  {author} {\bibinfo {author} {\bibfnamefont {S.}~\bibnamefont {{Lieu}}}, \bibinfo {author} {\bibfnamefont {R.}~\bibnamefont {{Belyansky}}}, \bibinfo {author} {\bibfnamefont {J.~T.}\ \bibnamefont {{Young}}}, \bibinfo {author} {\bibfnamefont {R.}~\bibnamefont {{Lundgren}}}, \bibinfo {author} {\bibfnamefont {V.~V.}\ \bibnamefont {{Albert}}},\ and\ \bibinfo {author} {\bibfnamefont {A.~V.}\ \bibnamefont {{Gorshkov}}},\ }\bibfield  {title} {\bibinfo {title} {{Symmetry Breaking and Error Correction in Open Quantum Systems}},\ }\href {https://doi.org/10.1103/PhysRevLett.125.240405} {\bibfield  {journal} {\bibinfo  {journal} {Phys. Rev. Lett.}\ }\textbf {\bibinfo {volume} {125}},\ \bibinfo {pages} {240405} (\bibinfo {year} {2020})}\BibitemShut {NoStop}%
\bibitem [{\citenamefont {{Albert}}\ and\ \citenamefont {{Jiang}}(2014)}]{Albert2014}%
  \BibitemOpen
  \bibfield  {author} {\bibinfo {author} {\bibfnamefont {V.~V.}\ \bibnamefont {{Albert}}}\ and\ \bibinfo {author} {\bibfnamefont {L.}~\bibnamefont {{Jiang}}},\ }\bibfield  {title} {\bibinfo {title} {{Symmetries and conserved quantities in Lindblad master equations}},\ }\href {https://doi.org/10.1103/PhysRevA.89.022118} {\bibfield  {journal} {\bibinfo  {journal} {Phys. Rev. A}\ }\textbf {\bibinfo {volume} {89}},\ \bibinfo {pages} {022118} (\bibinfo {year} {2014})}\BibitemShut {NoStop}%
\bibitem [{\citenamefont {{Ellison}}\ and\ \citenamefont {{Cheng}}(2025)}]{Ellison2025}%
  \BibitemOpen
  \bibfield  {author} {\bibinfo {author} {\bibfnamefont {T.~D.}\ \bibnamefont {{Ellison}}}\ and\ \bibinfo {author} {\bibfnamefont {M.}~\bibnamefont {{Cheng}}},\ }\bibfield  {title} {\bibinfo {title} {{Toward a Classification of Mixed-State Topological Orders in Two Dimensions}},\ }\href {https://doi.org/10.1103/PRXQuantum.6.010315} {\bibfield  {journal} {\bibinfo  {journal} {PRX Quantum}\ }\textbf {\bibinfo {volume} {6}},\ \bibinfo {pages} {010315} (\bibinfo {year} {2025})}\BibitemShut {NoStop}%
\bibitem [{\citenamefont {{Wang}}\ \emph {et~al.}(2025)\citenamefont {{Wang}}, \citenamefont {{Wu}},\ and\ \citenamefont {{Wang}}}]{Wang2025b}%
  \BibitemOpen
  \bibfield  {author} {\bibinfo {author} {\bibfnamefont {Z.}~\bibnamefont {{Wang}}}, \bibinfo {author} {\bibfnamefont {Z.}~\bibnamefont {{Wu}}},\ and\ \bibinfo {author} {\bibfnamefont {Z.}~\bibnamefont {{Wang}}},\ }\bibfield  {title} {\bibinfo {title} {{Intrinsic Mixed-State Topological Order}},\ }\href {https://doi.org/10.1103/PRXQuantum.6.010314} {\bibfield  {journal} {\bibinfo  {journal} {PRX Quantum}\ }\textbf {\bibinfo {volume} {6}},\ \bibinfo {pages} {010314} (\bibinfo {year} {2025})}\BibitemShut {NoStop}%
\bibitem [{\citenamefont {{Zhang}}\ \emph {et~al.}(2025)\citenamefont {{Zhang}}, \citenamefont {{Xu}}, \citenamefont {{Zhang}}, \citenamefont {{Xu}}, \citenamefont {{Bi}},\ and\ \citenamefont {{Luo}}}]{Zhang2025}%
  \BibitemOpen
  \bibfield  {author} {\bibinfo {author} {\bibfnamefont {C.}~\bibnamefont {{Zhang}}}, \bibinfo {author} {\bibfnamefont {Y.}~\bibnamefont {{Xu}}}, \bibinfo {author} {\bibfnamefont {J.-H.}\ \bibnamefont {{Zhang}}}, \bibinfo {author} {\bibfnamefont {C.}~\bibnamefont {{Xu}}}, \bibinfo {author} {\bibfnamefont {Z.}~\bibnamefont {{Bi}}},\ and\ \bibinfo {author} {\bibfnamefont {Z.-X.}\ \bibnamefont {{Luo}}},\ }\bibfield  {title} {\bibinfo {title} {{Strong-to-weak spontaneous breaking of 1-form symmetry and intrinsically mixed topological order}},\ }\href {https://doi.org/10.1103/PhysRevB.111.115137} {\bibfield  {journal} {\bibinfo  {journal} {Phys. Rev. B}\ }\textbf {\bibinfo {volume} {111}},\ \bibinfo {pages} {115137} (\bibinfo {year} {2025})}\BibitemShut {NoStop}%
\bibitem [{\citenamefont {{Lessa}}\ \emph {et~al.}(2025)\citenamefont {{Lessa}}, \citenamefont {{Ma}}, \citenamefont {{Zhang}}, \citenamefont {{Bi}}, \citenamefont {{Cheng}},\ and\ \citenamefont {{Wang}}}]{Lessa2025}%
  \BibitemOpen
  \bibfield  {author} {\bibinfo {author} {\bibfnamefont {L.~A.}\ \bibnamefont {{Lessa}}}, \bibinfo {author} {\bibfnamefont {R.}~\bibnamefont {{Ma}}}, \bibinfo {author} {\bibfnamefont {J.-H.}\ \bibnamefont {{Zhang}}}, \bibinfo {author} {\bibfnamefont {Z.}~\bibnamefont {{Bi}}}, \bibinfo {author} {\bibfnamefont {M.}~\bibnamefont {{Cheng}}},\ and\ \bibinfo {author} {\bibfnamefont {C.}~\bibnamefont {{Wang}}},\ }\bibfield  {title} {\bibinfo {title} {{Strong-to-Weak Spontaneous Symmetry Breaking in Mixed Quantum States}},\ }\href {https://doi.org/10.1103/PRXQuantum.6.010344} {\bibfield  {journal} {\bibinfo  {journal} {PRX Quantum}\ }\textbf {\bibinfo {volume} {6}},\ \bibinfo {pages} {010344} (\bibinfo {year} {2025})}\BibitemShut {NoStop}%
\bibitem [{\citenamefont {{Lee}}\ \emph {et~al.}(2025)\citenamefont {{Lee}}, \citenamefont {{You}},\ and\ \citenamefont {{Xu}}}]{Lee2022}%
  \BibitemOpen
  \bibfield  {author} {\bibinfo {author} {\bibfnamefont {J.~Y.}\ \bibnamefont {{Lee}}}, \bibinfo {author} {\bibfnamefont {Y.-Z.}\ \bibnamefont {{You}}},\ and\ \bibinfo {author} {\bibfnamefont {C.}~\bibnamefont {{Xu}}},\ }\bibfield  {title} {\bibinfo {title} {{Symmetry protected topological phases under decoherence}},\ }\href {https://doi.org/10.22331/q-2025-01-23-1607} {\bibfield  {journal} {\bibinfo  {journal} {Quantum}\ }\textbf {\bibinfo {volume} {9}},\ \bibinfo {pages} {1607} (\bibinfo {year} {2025})}\BibitemShut {NoStop}%
\bibitem [{\citenamefont {{Sala}}\ \emph {et~al.}(2024)\citenamefont {{Sala}}, \citenamefont {{Gopalakrishnan}}, \citenamefont {{Oshikawa}},\ and\ \citenamefont {{You}}}]{Sala2024}%
  \BibitemOpen
  \bibfield  {author} {\bibinfo {author} {\bibfnamefont {P.}~\bibnamefont {{Sala}}}, \bibinfo {author} {\bibfnamefont {S.}~\bibnamefont {{Gopalakrishnan}}}, \bibinfo {author} {\bibfnamefont {M.}~\bibnamefont {{Oshikawa}}},\ and\ \bibinfo {author} {\bibfnamefont {Y.}~\bibnamefont {{You}}},\ }\bibfield  {title} {\bibinfo {title} {{Spontaneous strong symmetry breaking in open systems: Purification perspective}},\ }\href {https://doi.org/10.1103/PhysRevB.110.155150} {\bibfield  {journal} {\bibinfo  {journal} {Phys. Rev. B}\ }\textbf {\bibinfo {volume} {110}},\ \bibinfo {pages} {155150} (\bibinfo {year} {2024})}\BibitemShut {NoStop}%
\bibitem [{\citenamefont {{Lu}}\ \emph {et~al.}(2026)\citenamefont {{Lu}}, \citenamefont {{Liu}}, \citenamefont {{Gopalakrishnan}},\ and\ \citenamefont {{You}}}]{Lu2025}%
  \BibitemOpen
  \bibfield  {author} {\bibinfo {author} {\bibfnamefont {T.-C.}\ \bibnamefont {{Lu}}}, \bibinfo {author} {\bibfnamefont {Y.-J.}\ \bibnamefont {{Liu}}}, \bibinfo {author} {\bibfnamefont {S.}~\bibnamefont {{Gopalakrishnan}}},\ and\ \bibinfo {author} {\bibfnamefont {Y.}~\bibnamefont {{You}}},\ }\bibfield  {title} {\bibinfo {title} {{Holographic Duality Between Bulk Topological Order and Boundary Mixed-State Order}},\ }\href {https://doi.org/10.1103/lrqp-pnl1} {\bibfield  {journal} {\bibinfo  {journal} {PRX Quantum}\ }\textbf {\bibinfo {volume} {7}},\ \bibinfo {pages} {033041} (\bibinfo {year} {2026})}\BibitemShut {NoStop}%
\bibitem [{\citenamefont {{Luo}}\ \emph {et~al.}(2025)\citenamefont {{Luo}}, \citenamefont {{Wang}},\ and\ \citenamefont {{Bi}}}]{Luo2025}%
  \BibitemOpen
  \bibfield  {author} {\bibinfo {author} {\bibfnamefont {R.}~\bibnamefont {{Luo}}}, \bibinfo {author} {\bibfnamefont {Y.-N.}\ \bibnamefont {{Wang}}},\ and\ \bibinfo {author} {\bibfnamefont {Z.}~\bibnamefont {{Bi}}},\ }\bibfield  {title} {\bibinfo {title} {{Topological Holography for Mixed-State Phases and Phase Transitions}},\ }\href {https://doi.org/10.1103/9kmh-gjf8} {\bibfield  {journal} {\bibinfo  {journal} {PRX Quantum}\ }\textbf {\bibinfo {volume} {6}},\ \bibinfo {pages} {040358} (\bibinfo {year} {2025})}\BibitemShut {NoStop}%
\bibitem [{\citenamefont {{Chatterjee}}\ and\ \citenamefont {{Wen}}(2023)}]{Chatterjee2023}%
  \BibitemOpen
  \bibfield  {author} {\bibinfo {author} {\bibfnamefont {A.}~\bibnamefont {{Chatterjee}}}\ and\ \bibinfo {author} {\bibfnamefont {X.-G.}\ \bibnamefont {{Wen}}},\ }\bibfield  {title} {\bibinfo {title} {{Symmetry as a shadow of topological order and a derivation of topological holographic principle}},\ }\href {https://doi.org/10.1103/PhysRevB.107.155136} {\bibfield  {journal} {\bibinfo  {journal} {Phys. Rev. B}\ }\textbf {\bibinfo {volume} {107}},\ \bibinfo {pages} {155136} (\bibinfo {year} {2023})}\BibitemShut {NoStop}%
\bibitem [{\citenamefont {Ji}\ and\ \citenamefont {Wen}(2020)}]{Ji2019}%
  \BibitemOpen
  \bibfield  {author} {\bibinfo {author} {\bibfnamefont {W.}~\bibnamefont {Ji}}\ and\ \bibinfo {author} {\bibfnamefont {X.-G.}\ \bibnamefont {Wen}},\ }\bibfield  {title} {\bibinfo {title} {{Categorical symmetry and noninvertible anomaly in symmetry-breaking and topological phase transitions}},\ }\href {https://doi.org/10.1103/PhysRevResearch.2.033417} {\bibfield  {journal} {\bibinfo  {journal} {Phys. Rev. Research}\ }\textbf {\bibinfo {volume} {2}},\ \bibinfo {pages} {033417} (\bibinfo {year} {2020})}\BibitemShut {NoStop}%
\bibitem [{\citenamefont {{Kong}}\ \emph {et~al.}(2020)\citenamefont {{Kong}}, \citenamefont {{Lan}}, \citenamefont {{Wen}}, \citenamefont {{Zhang}},\ and\ \citenamefont {{Zheng}}}]{Kong2020a}%
  \BibitemOpen
  \bibfield  {author} {\bibinfo {author} {\bibfnamefont {L.}~\bibnamefont {{Kong}}}, \bibinfo {author} {\bibfnamefont {T.}~\bibnamefont {{Lan}}}, \bibinfo {author} {\bibfnamefont {X.-G.}\ \bibnamefont {{Wen}}}, \bibinfo {author} {\bibfnamefont {Z.-H.}\ \bibnamefont {{Zhang}}},\ and\ \bibinfo {author} {\bibfnamefont {H.}~\bibnamefont {{Zheng}}},\ }\bibfield  {title} {\bibinfo {title} {{Algebraic higher symmetry and categorical symmetry: A holographic and entanglement view of symmetry}},\ }\href {https://doi.org/10.1103/PhysRevResearch.2.043086} {\bibfield  {journal} {\bibinfo  {journal} {Phys. Rev. Research}\ }\textbf {\bibinfo {volume} {2}},\ \bibinfo {pages} {043086} (\bibinfo {year} {2020})}\BibitemShut {NoStop}%
\bibitem [{\citenamefont {{Inamura}}\ and\ \citenamefont {{Wen}}(2023)}]{Inamura2023}%
  \BibitemOpen
  \bibfield  {author} {\bibinfo {author} {\bibfnamefont {K.}~\bibnamefont {{Inamura}}}\ and\ \bibinfo {author} {\bibfnamefont {X.-G.}\ \bibnamefont {{Wen}}},\ }\href@noop {} {\bibinfo {title} {{2+1D symmetry-topological-order from local symmetric operators in 1+1D}}} (\bibinfo {year} {2023}),\ \Eprint {https://arxiv.org/abs/2310.05790} {arXiv:2310.05790 [cond-mat.str-el]} \BibitemShut {NoStop}%
\bibitem [{\citenamefont {{O'Hern}}\ \emph {et~al.}(1999)\citenamefont {{O'Hern}}, \citenamefont {{Lubensky}},\ and\ \citenamefont {{Toner}}}]{OHern1999}%
  \BibitemOpen
  \bibfield  {author} {\bibinfo {author} {\bibfnamefont {C.~S.}\ \bibnamefont {{O'Hern}}}, \bibinfo {author} {\bibfnamefont {T.~C.}\ \bibnamefont {{Lubensky}}},\ and\ \bibinfo {author} {\bibfnamefont {J.}~\bibnamefont {{Toner}}},\ }\bibfield  {title} {\bibinfo {title} {{Sliding Phases in XY Models, Crystals, and Cationic Lipid-DNA Complexes}},\ }\href {https://doi.org/10.1103/PhysRevLett.83.2745} {\bibfield  {journal} {\bibinfo  {journal} {Phys. Rev. Lett.}\ }\textbf {\bibinfo {volume} {83}},\ \bibinfo {pages} {2745} (\bibinfo {year} {1999})}\BibitemShut {NoStop}%
\bibitem [{\citenamefont {{Mishra}}\ \emph {et~al.}(2004)\citenamefont {{Mishra}}, \citenamefont {{Ma}}, \citenamefont {{Zhang}}, \citenamefont {{Guertler}}, \citenamefont {{Tang}},\ and\ \citenamefont {{Wan}}}]{Mishra2004}%
  \BibitemOpen
  \bibfield  {author} {\bibinfo {author} {\bibfnamefont {A.}~\bibnamefont {{Mishra}}}, \bibinfo {author} {\bibfnamefont {M.}~\bibnamefont {{Ma}}}, \bibinfo {author} {\bibfnamefont {F.-C.}\ \bibnamefont {{Zhang}}}, \bibinfo {author} {\bibfnamefont {S.}~\bibnamefont {{Guertler}}}, \bibinfo {author} {\bibfnamefont {L.-H.}\ \bibnamefont {{Tang}}},\ and\ \bibinfo {author} {\bibfnamefont {S.}~\bibnamefont {{Wan}}},\ }\bibfield  {title} {\bibinfo {title} {{Directional Ordering of Fluctuations in a Two-Dimensional Compass Model}},\ }\href {https://doi.org/10.1103/PhysRevLett.93.207201} {\bibfield  {journal} {\bibinfo  {journal} {Phys. Rev. Lett.}\ }\textbf {\bibinfo {volume} {93}},\ \bibinfo {pages} {207201} (\bibinfo {year} {2004})}\BibitemShut {NoStop}%
\bibitem [{\citenamefont {Batista}\ and\ \citenamefont {Nussinov}(2005)}]{Batista2005}%
  \BibitemOpen
  \bibfield  {author} {\bibinfo {author} {\bibfnamefont {C.~D.}\ \bibnamefont {Batista}}\ and\ \bibinfo {author} {\bibfnamefont {Z.}~\bibnamefont {Nussinov}},\ }\bibfield  {title} {\bibinfo {title} {{Generalized Elitzur's theorem and dimensional reductions}},\ }\href {https://doi.org/10.1103/PhysRevB.72.045137} {\bibfield  {journal} {\bibinfo  {journal} {Phys. Rev. B}\ }\textbf {\bibinfo {volume} {72}},\ \bibinfo {pages} {045137} (\bibinfo {year} {2005})}\BibitemShut {NoStop}%
\bibitem [{\citenamefont {Nussinov}\ and\ \citenamefont {Fradkin}(2005)}]{Nussinov2005}%
  \BibitemOpen
  \bibfield  {author} {\bibinfo {author} {\bibfnamefont {Z.}~\bibnamefont {Nussinov}}\ and\ \bibinfo {author} {\bibfnamefont {E.}~\bibnamefont {Fradkin}},\ }\bibfield  {title} {\bibinfo {title} {{Discrete sliding symmetries, dualities, and self-dualities of quantum orbital compass models and $p+ip$ superconducting arrays}},\ }\href {https://doi.org/10.1103/PhysRevB.71.195120} {\bibfield  {journal} {\bibinfo  {journal} {Phys. Rev. B}\ }\textbf {\bibinfo {volume} {71}},\ \bibinfo {pages} {195120} (\bibinfo {year} {2005})}\BibitemShut {NoStop}%
\bibitem [{\citenamefont {{Seiberg}}\ and\ \citenamefont {{Shao}}(2021)}]{Seiberg2021a}%
  \BibitemOpen
  \bibfield  {author} {\bibinfo {author} {\bibfnamefont {N.}~\bibnamefont {{Seiberg}}}\ and\ \bibinfo {author} {\bibfnamefont {S.-H.}\ \bibnamefont {{Shao}}},\ }\bibfield  {title} {\bibinfo {title} {{Exotic symmetries, duality, and fractons in 2+1-dimensional quantum field theory}},\ }\href {https://doi.org/10.21468/SciPostPhys.10.2.027} {\bibfield  {journal} {\bibinfo  {journal} {SciPost Phys.}\ }\textbf {\bibinfo {volume} {10}},\ \bibinfo {pages} {027} (\bibinfo {year} {2021})}\BibitemShut {NoStop}%
\bibitem [{\citenamefont {{Distler}}\ \emph {et~al.}(2022)\citenamefont {{Distler}}, \citenamefont {{Karch}},\ and\ \citenamefont {{Raz}}}]{Distler2022}%
  \BibitemOpen
  \bibfield  {author} {\bibinfo {author} {\bibfnamefont {J.}~\bibnamefont {{Distler}}}, \bibinfo {author} {\bibfnamefont {A.}~\bibnamefont {{Karch}}},\ and\ \bibinfo {author} {\bibfnamefont {A.}~\bibnamefont {{Raz}}},\ }\bibfield  {title} {\bibinfo {title} {{Spontaneously broken subsystem symmetries}},\ }\href {https://doi.org/10.1007/JHEP03(2022)016} {\bibfield  {journal} {\bibinfo  {journal} {J. High Energy Phys.}\ }\textbf {\bibinfo {volume} {2022}}\bibinfo  {number} { (03)},\ \bibinfo {pages} {016}}\BibitemShut {NoStop}%
\bibitem [{\citenamefont {{Devakul}}\ \emph {et~al.}(2019)\citenamefont {{Devakul}}, \citenamefont {{You}}, \citenamefont {{Burnell}},\ and\ \citenamefont {{Sondhi}}}]{Devakul2018b}%
  \BibitemOpen
\bibfield  {number} {  }\bibfield  {author} {\bibinfo {author} {\bibfnamefont {T.}~\bibnamefont {{Devakul}}}, \bibinfo {author} {\bibfnamefont {Y.}~\bibnamefont {{You}}}, \bibinfo {author} {\bibfnamefont {F.~J.}\ \bibnamefont {{Burnell}}},\ and\ \bibinfo {author} {\bibfnamefont {S.~L.}\ \bibnamefont {{Sondhi}}},\ }\bibfield  {title} {\bibinfo {title} {{Fractal Symmetric Phases of Matter}},\ }\href {https://doi.org/10.21468/scipostphys.6.1.007} {\bibfield  {journal} {\bibinfo  {journal} {SciPost Phys.}\ }\textbf {\bibinfo {volume} {6}},\ \bibinfo {pages} {007} (\bibinfo {year} {2019})}\BibitemShut {NoStop}%
\bibitem [{\citenamefont {{Dennis}}\ \emph {et~al.}(2002)\citenamefont {{Dennis}}, \citenamefont {{Kitaev}}, \citenamefont {{Landahl}},\ and\ \citenamefont {{Preskill}}}]{Dennis2002}%
  \BibitemOpen
  \bibfield  {author} {\bibinfo {author} {\bibfnamefont {E.}~\bibnamefont {{Dennis}}}, \bibinfo {author} {\bibfnamefont {A.}~\bibnamefont {{Kitaev}}}, \bibinfo {author} {\bibfnamefont {A.}~\bibnamefont {{Landahl}}},\ and\ \bibinfo {author} {\bibfnamefont {J.}~\bibnamefont {{Preskill}}},\ }\bibfield  {title} {\bibinfo {title} {{Topological quantum memory}},\ }\href {https://doi.org/10.1063/1.1499754} {\bibfield  {journal} {\bibinfo  {journal} {J. Math. Phys.}\ }\textbf {\bibinfo {volume} {43}},\ \bibinfo {pages} {4452} (\bibinfo {year} {2002})}\BibitemShut {NoStop}%
\bibitem [{\citenamefont {Kitaev}(2003)}]{Kitaev2003}%
  \BibitemOpen
  \bibfield  {author} {\bibinfo {author} {\bibfnamefont {A.~Y.}\ \bibnamefont {Kitaev}},\ }\bibfield  {title} {\bibinfo {title} {{Fault-tolerant quantum computation by anyons}},\ }\href {https://doi.org/10.1016/s0003-4916(02)00018-0} {\bibfield  {journal} {\bibinfo  {journal} {Ann. Phys.}\ }\textbf {\bibinfo {volume} {303}},\ \bibinfo {pages} {2} (\bibinfo {year} {2003})}\BibitemShut {NoStop}%
\bibitem [{\citenamefont {{Bullock}}\ and\ \citenamefont {{Brennen}}(2007)}]{Bullock2007}%
  \BibitemOpen
  \bibfield  {author} {\bibinfo {author} {\bibfnamefont {S.~S.}\ \bibnamefont {{Bullock}}}\ and\ \bibinfo {author} {\bibfnamefont {G.~K.}\ \bibnamefont {{Brennen}}},\ }\bibfield  {title} {\bibinfo {title} {{Qudit surface codes and gauge theory with finite cyclic groups}},\ }\href {https://doi.org/10.1088/1751-8113/40/13/013} {\bibfield  {journal} {\bibinfo  {journal} {J. Phys. A: Math. Theor.}\ }\textbf {\bibinfo {volume} {40}},\ \bibinfo {pages} {3481} (\bibinfo {year} {2007})}\BibitemShut {NoStop}%
\bibitem [{\citenamefont {Watanabe}\ \emph {et~al.}(2023)\citenamefont {Watanabe}, \citenamefont {Cheng},\ and\ \citenamefont {Fuji}}]{Watanabe2023}%
  \BibitemOpen
  \bibfield  {author} {\bibinfo {author} {\bibfnamefont {H.}~\bibnamefont {Watanabe}}, \bibinfo {author} {\bibfnamefont {M.}~\bibnamefont {Cheng}},\ and\ \bibinfo {author} {\bibfnamefont {Y.}~\bibnamefont {Fuji}},\ }\bibfield  {title} {\bibinfo {title} {{Ground state degeneracy on torus in a family of $Z_N$ toric code}},\ }\href {https://doi.org/10.1063/5.0134010} {\bibfield  {journal} {\bibinfo  {journal} {J. Math. Phys.}\ }\textbf {\bibinfo {volume} {64}},\ \bibinfo {pages} {051901} (\bibinfo {year} {2023})}\BibitemShut {NoStop}%
\bibitem [{\citenamefont {Hamma}\ \emph {et~al.}(2005)\citenamefont {Hamma}, \citenamefont {Zanardi},\ and\ \citenamefont {Wen}}]{Hamma2005}%
  \BibitemOpen
  \bibfield  {author} {\bibinfo {author} {\bibfnamefont {A.}~\bibnamefont {Hamma}}, \bibinfo {author} {\bibfnamefont {P.}~\bibnamefont {Zanardi}},\ and\ \bibinfo {author} {\bibfnamefont {X.-G.}\ \bibnamefont {Wen}},\ }\bibfield  {title} {\bibinfo {title} {{String and membrane condensation on three-dimensional lattices}},\ }\href {https://doi.org/10.1103/PhysRevB.72.035307} {\bibfield  {journal} {\bibinfo  {journal} {Phys. Rev. B}\ }\textbf {\bibinfo {volume} {72}},\ \bibinfo {pages} {035307} (\bibinfo {year} {2005})}\BibitemShut {NoStop}%
\bibitem [{\citenamefont {{Sohal}}\ and\ \citenamefont {{Prem}}(2025)}]{Sohal2025}%
  \BibitemOpen
  \bibfield  {author} {\bibinfo {author} {\bibfnamefont {R.}~\bibnamefont {{Sohal}}}\ and\ \bibinfo {author} {\bibfnamefont {A.}~\bibnamefont {{Prem}}},\ }\bibfield  {title} {\bibinfo {title} {{Noisy Approach to Intrinsically Mixed-State Topological Order}},\ }\href {https://doi.org/10.1103/PRXQuantum.6.010313} {\bibfield  {journal} {\bibinfo  {journal} {PRX Quantum}\ }\textbf {\bibinfo {volume} {6}},\ \bibinfo {pages} {010313} (\bibinfo {year} {2025})}\BibitemShut {NoStop}%
\bibitem [{\citenamefont {Hu}\ \emph {et~al.}(2026)\citenamefont {Hu}, \citenamefont {Zhang},\ and\ \citenamefont {Ye}}]{HuZhangYe2026}%
  \BibitemOpen
  \bibfield  {author} {\bibinfo {author} {\bibfnamefont {Y.-T.}\ \bibnamefont {Hu}}, \bibinfo {author} {\bibfnamefont {J.-Y.}\ \bibnamefont {Zhang}},\ and\ \bibinfo {author} {\bibfnamefont {P.}~\bibnamefont {Ye}},\ }\href@noop {} {\bibinfo {title} {{Fracton Topological Holography}}} (\bibinfo {year} {2026}),\ \Eprint {https://arxiv.org/abs/2606.03582} {arXiv:2606.03582 [quant-ph]} \BibitemShut {NoStop}%
\bibitem [{\citenamefont {{Fattal}}\ \emph {et~al.}(2004)\citenamefont {{Fattal}}, \citenamefont {{Cubitt}}, \citenamefont {{Yamamoto}}, \citenamefont {{Bravyi}},\ and\ \citenamefont {{Chuang}}}]{Fattal2004}%
  \BibitemOpen
  \bibfield  {author} {\bibinfo {author} {\bibfnamefont {D.}~\bibnamefont {{Fattal}}}, \bibinfo {author} {\bibfnamefont {T.~S.}\ \bibnamefont {{Cubitt}}}, \bibinfo {author} {\bibfnamefont {Y.}~\bibnamefont {{Yamamoto}}}, \bibinfo {author} {\bibfnamefont {S.}~\bibnamefont {{Bravyi}}},\ and\ \bibinfo {author} {\bibfnamefont {I.~L.}\ \bibnamefont {{Chuang}}},\ }\href@noop {} {\bibinfo {title} {{Entanglement in the stabilizer formalism}}} (\bibinfo {year} {2004}),\ \Eprint {https://arxiv.org/abs/quant-ph/0406168} {arXiv:quant-ph/0406168 [quant-ph]} \BibitemShut {NoStop}%
\bibitem [{\citenamefont {{Hamma}}\ \emph {et~al.}(2005)\citenamefont {{Hamma}}, \citenamefont {{Ionicioiu}},\ and\ \citenamefont {{Zanardi}}}]{Hamma2005a}%
  \BibitemOpen
  \bibfield  {author} {\bibinfo {author} {\bibfnamefont {A.}~\bibnamefont {{Hamma}}}, \bibinfo {author} {\bibfnamefont {R.}~\bibnamefont {{Ionicioiu}}},\ and\ \bibinfo {author} {\bibfnamefont {P.}~\bibnamefont {{Zanardi}}},\ }\bibfield  {title} {\bibinfo {title} {{Bipartite entanglement and entropic boundary law in lattice spin systems}},\ }\href {https://doi.org/10.1103/PhysRevA.71.022315} {\bibfield  {journal} {\bibinfo  {journal} {Phys. Rev. A}\ }\textbf {\bibinfo {volume} {71}},\ \bibinfo {pages} {022315} (\bibinfo {year} {2005})}\BibitemShut {NoStop}%
\bibitem [{\citenamefont {Xu}\ and\ \citenamefont {Moore}(2004)}]{Xu2004}%
  \BibitemOpen
  \bibfield  {author} {\bibinfo {author} {\bibfnamefont {C.}~\bibnamefont {Xu}}\ and\ \bibinfo {author} {\bibfnamefont {J.~E.}\ \bibnamefont {Moore}},\ }\bibfield  {title} {\bibinfo {title} {{Strong-Weak Coupling Self-Duality in the Two-Dimensional Quantum Phase Transition of $p+ip$ Superconducting Arrays}},\ }\href {https://doi.org/10.1103/PhysRevLett.93.047003} {\bibfield  {journal} {\bibinfo  {journal} {Phys. Rev. Lett.}\ }\textbf {\bibinfo {volume} {93}},\ \bibinfo {pages} {047003} (\bibinfo {year} {2004})}\BibitemShut {NoStop}%
\bibitem [{\citenamefont {Raussendorf}\ and\ \citenamefont {Briegel}(2001)}]{Raussendorf2001}%
  \BibitemOpen
  \bibfield  {author} {\bibinfo {author} {\bibfnamefont {R.}~\bibnamefont {Raussendorf}}\ and\ \bibinfo {author} {\bibfnamefont {H.~J.}\ \bibnamefont {Briegel}},\ }\bibfield  {title} {\bibinfo {title} {{A One-Way Quantum Computer}},\ }\href {https://doi.org/10.1103/PhysRevLett.86.5188} {\bibfield  {journal} {\bibinfo  {journal} {Phys. Rev. Lett.}\ }\textbf {\bibinfo {volume} {86}},\ \bibinfo {pages} {5188} (\bibinfo {year} {2001})}\BibitemShut {NoStop}%
\bibitem [{\citenamefont {{Devakul}}\ \emph {et~al.}(2018)\citenamefont {{Devakul}}, \citenamefont {{Williamson}},\ and\ \citenamefont {{You}}}]{Devakul2018a}%
  \BibitemOpen
  \bibfield  {author} {\bibinfo {author} {\bibfnamefont {T.}~\bibnamefont {{Devakul}}}, \bibinfo {author} {\bibfnamefont {D.~J.}\ \bibnamefont {{Williamson}}},\ and\ \bibinfo {author} {\bibfnamefont {Y.}~\bibnamefont {{You}}},\ }\bibfield  {title} {\bibinfo {title} {{Classification of subsystem symmetry-protected topological phases}},\ }\href {https://doi.org/10.1103/PhysRevB.98.235121} {\bibfield  {journal} {\bibinfo  {journal} {Phys. Rev. B}\ }\textbf {\bibinfo {volume} {98}},\ \bibinfo {pages} {235121} (\bibinfo {year} {2018})}\BibitemShut {NoStop}%
\bibitem [{\citenamefont {Burnell}\ \emph {et~al.}(2022)\citenamefont {Burnell}, \citenamefont {Devakul}, \citenamefont {Gorantla}, \citenamefont {Lam},\ and\ \citenamefont {Shao}}]{Burnell2022}%
  \BibitemOpen
  \bibfield  {author} {\bibinfo {author} {\bibfnamefont {F.~J.}\ \bibnamefont {Burnell}}, \bibinfo {author} {\bibfnamefont {T.}~\bibnamefont {Devakul}}, \bibinfo {author} {\bibfnamefont {P.}~\bibnamefont {Gorantla}}, \bibinfo {author} {\bibfnamefont {H.~T.}\ \bibnamefont {Lam}},\ and\ \bibinfo {author} {\bibfnamefont {S.-H.}\ \bibnamefont {Shao}},\ }\bibfield  {title} {\bibinfo {title} {{Anomaly inflow for subsystem symmetries}},\ }\href {https://doi.org/10.1103/PhysRevB.106.085113} {\bibfield  {journal} {\bibinfo  {journal} {Phys. Rev. B}\ }\textbf {\bibinfo {volume} {106}},\ \bibinfo {pages} {085113} (\bibinfo {year} {2022})}\BibitemShut {NoStop}%
\bibitem [{\citenamefont {Zhou}\ \emph {et~al.}(2022)\citenamefont {Zhou}, \citenamefont {Li}, \citenamefont {Yan}, \citenamefont {Ye},\ and\ \citenamefont {Meng}}]{Zhou2022a}%
  \BibitemOpen
  \bibfield  {author} {\bibinfo {author} {\bibfnamefont {C.}~\bibnamefont {Zhou}}, \bibinfo {author} {\bibfnamefont {M.-Y.}\ \bibnamefont {Li}}, \bibinfo {author} {\bibfnamefont {Z.}~\bibnamefont {Yan}}, \bibinfo {author} {\bibfnamefont {P.}~\bibnamefont {Ye}},\ and\ \bibinfo {author} {\bibfnamefont {Z.~Y.}\ \bibnamefont {Meng}},\ }\bibfield  {title} {\bibinfo {title} {{Detecting subsystem symmetry protected topological order through strange correlators}},\ }\href {https://doi.org/10.1103/PhysRevB.106.214428} {\bibfield  {journal} {\bibinfo  {journal} {Phys. Rev. B}\ }\textbf {\bibinfo {volume} {106}},\ \bibinfo {pages} {214428} (\bibinfo {year} {2022})}\BibitemShut {NoStop}%
\bibitem [{\citenamefont {{Bravyi}}\ \emph {et~al.}(2011)\citenamefont {{Bravyi}}, \citenamefont {{Leemhuis}},\ and\ \citenamefont {{Terhal}}}]{Bravyi2011}%
  \BibitemOpen
  \bibfield  {author} {\bibinfo {author} {\bibfnamefont {S.}~\bibnamefont {{Bravyi}}}, \bibinfo {author} {\bibfnamefont {B.}~\bibnamefont {{Leemhuis}}},\ and\ \bibinfo {author} {\bibfnamefont {B.~M.}\ \bibnamefont {{Terhal}}},\ }\bibfield  {title} {\bibinfo {title} {{Topological order in an exactly solvable 3D spin model}},\ }\href {https://doi.org/10.1016/j.aop.2010.11.002} {\bibfield  {journal} {\bibinfo  {journal} {Ann. Phys.}\ }\textbf {\bibinfo {volume} {326}},\ \bibinfo {pages} {839} (\bibinfo {year} {2011})}\BibitemShut {NoStop}%
\bibitem [{\citenamefont {Walker}\ and\ \citenamefont {Wang}(2012)}]{Walker2012}%
  \BibitemOpen
  \bibfield  {author} {\bibinfo {author} {\bibfnamefont {K.}~\bibnamefont {Walker}}\ and\ \bibinfo {author} {\bibfnamefont {Z.}~\bibnamefont {Wang}},\ }\bibfield  {title} {\bibinfo {title} {{(3+1)-TQFTs and topological insulators}},\ }\href {https://doi.org/10.1007/s11467-011-0194-z} {\bibfield  {journal} {\bibinfo  {journal} {Front. Phys.}\ }\textbf {\bibinfo {volume} {7}},\ \bibinfo {pages} {150} (\bibinfo {year} {2012})}\BibitemShut {NoStop}%
\bibitem [{\citenamefont {Lan}\ \emph {et~al.}(2018)\citenamefont {Lan}, \citenamefont {Kong},\ and\ \citenamefont {Wen}}]{Lan2018}%
  \BibitemOpen
  \bibfield  {author} {\bibinfo {author} {\bibfnamefont {T.}~\bibnamefont {Lan}}, \bibinfo {author} {\bibfnamefont {L.}~\bibnamefont {Kong}},\ and\ \bibinfo {author} {\bibfnamefont {X.-G.}\ \bibnamefont {Wen}},\ }\bibfield  {title} {\bibinfo {title} {{Classification of $\mathbf{(}3+1\mathbf{)}\mathrm{D}$ Bosonic Topological Orders: The Case When Pointlike Excitations Are All Bosons}},\ }\href {https://doi.org/10.1103/PhysRevX.8.021074} {\bibfield  {journal} {\bibinfo  {journal} {Phys. Rev. X}\ }\textbf {\bibinfo {volume} {8}},\ \bibinfo {pages} {021074} (\bibinfo {year} {2018})}\BibitemShut {NoStop}%
\bibitem [{\citenamefont {Wen}\ \emph {et~al.}(2018)\citenamefont {Wen}, \citenamefont {He}, \citenamefont {Tiwari}, \citenamefont {Zheng},\ and\ \citenamefont {Ye}}]{Wen2018a}%
  \BibitemOpen
  \bibfield  {author} {\bibinfo {author} {\bibfnamefont {X.}~\bibnamefont {Wen}}, \bibinfo {author} {\bibfnamefont {H.}~\bibnamefont {He}}, \bibinfo {author} {\bibfnamefont {A.}~\bibnamefont {Tiwari}}, \bibinfo {author} {\bibfnamefont {Y.}~\bibnamefont {Zheng}},\ and\ \bibinfo {author} {\bibfnamefont {P.}~\bibnamefont {Ye}},\ }\bibfield  {title} {\bibinfo {title} {{Entanglement entropy for (3+1)-dimensional topological order with excitations}},\ }\href {https://doi.org/10.1103/PhysRevB.97.085147} {\bibfield  {journal} {\bibinfo  {journal} {Phys. Rev. B}\ }\textbf {\bibinfo {volume} {97}},\ \bibinfo {pages} {085147} (\bibinfo {year} {2018})}\BibitemShut {NoStop}%
\bibitem [{\citenamefont {Reiss}\ and\ \citenamefont {Schmidt}(2019)}]{Reiss2019}%
  \BibitemOpen
  \bibfield  {author} {\bibinfo {author} {\bibfnamefont {D.~A.}\ \bibnamefont {Reiss}}\ and\ \bibinfo {author} {\bibfnamefont {K.~P.}\ \bibnamefont {Schmidt}},\ }\bibfield  {title} {\bibinfo {title} {{Quantum robustness and phase transitions of the 3D toric code in a field}},\ }\href {https://doi.org/10.21468/scipostphys.6.6.078} {\bibfield  {journal} {\bibinfo  {journal} {SciPost Phys.}\ }\textbf {\bibinfo {volume} {6}},\ \bibinfo {pages} {078} (\bibinfo {year} {2019})}\BibitemShut {NoStop}%
\bibitem [{\citenamefont {Lan}\ and\ \citenamefont {Wen}(2019)}]{Lan2019}%
  \BibitemOpen
  \bibfield  {author} {\bibinfo {author} {\bibfnamefont {T.}~\bibnamefont {Lan}}\ and\ \bibinfo {author} {\bibfnamefont {X.-G.}\ \bibnamefont {Wen}},\ }\bibfield  {title} {\bibinfo {title} {{Classification of $3+1\mathrm{D}$ Bosonic Topological Orders (II): The Case When Some Pointlike Excitations Are Fermions}},\ }\href {https://doi.org/10.1103/PhysRevX.9.021005} {\bibfield  {journal} {\bibinfo  {journal} {Phys. Rev. X}\ }\textbf {\bibinfo {volume} {9}},\ \bibinfo {pages} {021005} (\bibinfo {year} {2019})}\BibitemShut {NoStop}%
\bibitem [{\citenamefont {Zhang}\ and\ \citenamefont {Ye}(2021)}]{Zhang2021}%
  \BibitemOpen
  \bibfield  {author} {\bibinfo {author} {\bibfnamefont {Z.-F.}\ \bibnamefont {Zhang}}\ and\ \bibinfo {author} {\bibfnamefont {P.}~\bibnamefont {Ye}},\ }\bibfield  {title} {\bibinfo {title} {{Compatible braidings with Hopf links, multiloop, and Borromean rings in $(3+1)$-dimensional spacetime}},\ }\href {https://doi.org/10.1103/PhysRevResearch.3.023132} {\bibfield  {journal} {\bibinfo  {journal} {Phys. Rev. Research}\ }\textbf {\bibinfo {volume} {3}},\ \bibinfo {pages} {023132} (\bibinfo {year} {2021})}\BibitemShut {NoStop}%
\bibitem [{\citenamefont {Chen}\ and\ \citenamefont {Hsin}(2023)}]{chenyuan4d_2023}%
  \BibitemOpen
  \bibfield  {author} {\bibinfo {author} {\bibfnamefont {Y.-A.}\ \bibnamefont {Chen}}\ and\ \bibinfo {author} {\bibfnamefont {P.-S.}\ \bibnamefont {Hsin}},\ }\bibfield  {title} {\bibinfo {title} {{Exactly solvable lattice Hamiltonians and gravitational anomalies}},\ }\href {https://doi.org/10.21468/SciPostPhys.14.5.089} {\bibfield  {journal} {\bibinfo  {journal} {SciPost Phys.}\ }\textbf {\bibinfo {volume} {14}},\ \bibinfo {pages} {089} (\bibinfo {year} {2023})}\BibitemShut {NoStop}%
\bibitem [{\citenamefont {{Huang}}\ \emph {et~al.}(2025)\citenamefont {{Huang}}, \citenamefont {{Zhang}},\ and\ \citenamefont {{Ye}}}]{Huang2025jhep}%
  \BibitemOpen
  \bibfield  {author} {\bibinfo {author} {\bibfnamefont {Y.}~\bibnamefont {{Huang}}}, \bibinfo {author} {\bibfnamefont {Z.-F.}\ \bibnamefont {{Zhang}}},\ and\ \bibinfo {author} {\bibfnamefont {P.}~\bibnamefont {{Ye}}},\ }\bibfield  {title} {\bibinfo {title} {{Diagrammatics, pentagon equations, and hexagon equations of topological orders with loop- and membrane-like excitations}},\ }\href {https://doi.org/10.1007/JHEP06(2025)238} {\bibfield  {journal} {\bibinfo  {journal} {J. High Energy Phys.}\ }\textbf {\bibinfo {volume} {2025}}\bibinfo  {number} { (06)},\ \bibinfo {pages} {238}}\BibitemShut {NoStop}%
\bibitem [{\citenamefont {Chamon}(2005)}]{Chamon2005}%
  \BibitemOpen
\bibfield  {number} {  }\bibfield  {author} {\bibinfo {author} {\bibfnamefont {C.}~\bibnamefont {Chamon}},\ }\bibfield  {title} {\bibinfo {title} {{Quantum Glassiness in Strongly Correlated Clean Systems: An Example of Topological Overprotection}},\ }\href {https://doi.org/10.1103/PhysRevLett.94.040402} {\bibfield  {journal} {\bibinfo  {journal} {Phys. Rev. Lett.}\ }\textbf {\bibinfo {volume} {94}},\ \bibinfo {pages} {040402} (\bibinfo {year} {2005})}\BibitemShut {NoStop}%
\bibitem [{\citenamefont {Haah}(2013)}]{Haah2013}%
  \BibitemOpen
  \bibfield  {author} {\bibinfo {author} {\bibfnamefont {J.}~\bibnamefont {Haah}},\ }\bibfield  {title} {\bibinfo {title} {{Commuting Pauli Hamiltonians as Maps between Free Modules}},\ }\href {https://doi.org/10.1007/s00220-013-1810-2} {\bibfield  {journal} {\bibinfo  {journal} {Commun. Math. Phys.}\ }\textbf {\bibinfo {volume} {324}},\ \bibinfo {pages} {351} (\bibinfo {year} {2013})}\BibitemShut {NoStop}%
\bibitem [{\citenamefont {Pretko}(2018)}]{Pretko2018}%
  \BibitemOpen
  \bibfield  {author} {\bibinfo {author} {\bibfnamefont {M.}~\bibnamefont {Pretko}},\ }\bibfield  {title} {\bibinfo {title} {{The fracton gauge principle}},\ }\href {https://doi.org/10.1103/PhysRevB.98.115134} {\bibfield  {journal} {\bibinfo  {journal} {Phys. Rev. B}\ }\textbf {\bibinfo {volume} {98}},\ \bibinfo {pages} {115134} (\bibinfo {year} {2018})}\BibitemShut {NoStop}%
\bibitem [{\citenamefont {Shirley}\ \emph {et~al.}(2018)\citenamefont {Shirley}, \citenamefont {Slagle}, \citenamefont {Wang},\ and\ \citenamefont {Chen}}]{Shirley2018}%
  \BibitemOpen
  \bibfield  {author} {\bibinfo {author} {\bibfnamefont {W.}~\bibnamefont {Shirley}}, \bibinfo {author} {\bibfnamefont {K.}~\bibnamefont {Slagle}}, \bibinfo {author} {\bibfnamefont {Z.}~\bibnamefont {Wang}},\ and\ \bibinfo {author} {\bibfnamefont {X.}~\bibnamefont {Chen}},\ }\bibfield  {title} {\bibinfo {title} {{Fracton Models on General Three-Dimensional Manifolds}},\ }\href {https://doi.org/10.1103/PhysRevX.8.031051} {\bibfield  {journal} {\bibinfo  {journal} {Phys. Rev. X}\ }\textbf {\bibinfo {volume} {8}},\ \bibinfo {pages} {031051} (\bibinfo {year} {2018})}\BibitemShut {NoStop}%
\bibitem [{\citenamefont {{Slagle}}\ and\ \citenamefont {{Kim}}(2018)}]{Slagle2018}%
  \BibitemOpen
  \bibfield  {author} {\bibinfo {author} {\bibfnamefont {K.}~\bibnamefont {{Slagle}}}\ and\ \bibinfo {author} {\bibfnamefont {Y.~B.}\ \bibnamefont {{Kim}}},\ }\bibfield  {title} {\bibinfo {title} {{X-cube model on generic lattices: Fracton phases and geometric order}},\ }\href {https://doi.org/10.1103/PhysRevB.97.165106} {\bibfield  {journal} {\bibinfo  {journal} {Phys. Rev. B}\ }\textbf {\bibinfo {volume} {97}},\ \bibinfo {pages} {165106} (\bibinfo {year} {2018})}\BibitemShut {NoStop}%
\bibitem [{\citenamefont {Prem}\ \emph {et~al.}(2019)\citenamefont {Prem}, \citenamefont {Huang}, \citenamefont {Song},\ and\ \citenamefont {Hermele}}]{Prem2019a}%
  \BibitemOpen
  \bibfield  {author} {\bibinfo {author} {\bibfnamefont {A.}~\bibnamefont {Prem}}, \bibinfo {author} {\bibfnamefont {S.-J.}\ \bibnamefont {Huang}}, \bibinfo {author} {\bibfnamefont {H.}~\bibnamefont {Song}},\ and\ \bibinfo {author} {\bibfnamefont {M.}~\bibnamefont {Hermele}},\ }\bibfield  {title} {\bibinfo {title} {{Cage-Net Fracton Models}},\ }\href {https://doi.org/10.1103/PhysRevX.9.021010} {\bibfield  {journal} {\bibinfo  {journal} {Phys. Rev. X}\ }\textbf {\bibinfo {volume} {9}},\ \bibinfo {pages} {021010} (\bibinfo {year} {2019})}\BibitemShut {NoStop}%
\bibitem [{\citenamefont {Seiberg}\ and\ \citenamefont {Shao}(2020)}]{Seiberg2020c}%
  \BibitemOpen
  \bibfield  {author} {\bibinfo {author} {\bibfnamefont {N.}~\bibnamefont {Seiberg}}\ and\ \bibinfo {author} {\bibfnamefont {S.-H.}\ \bibnamefont {Shao}},\ }\bibfield  {title} {\bibinfo {title} {{Exotic $U(1)$ Symmetries, Duality, and Fractons in 3+1-Dimensional Quantum Field Theory}},\ }\href {https://doi.org/10.21468/SciPostPhys.9.4.046} {\bibfield  {journal} {\bibinfo  {journal} {SciPost Phys.}\ }\textbf {\bibinfo {volume} {9}},\ \bibinfo {pages} {046} (\bibinfo {year} {2020})}\BibitemShut {NoStop}%
\bibitem [{\citenamefont {Yuan}\ \emph {et~al.}(2020)\citenamefont {Yuan}, \citenamefont {Chen},\ and\ \citenamefont {Ye}}]{Yuan2020}%
  \BibitemOpen
  \bibfield  {author} {\bibinfo {author} {\bibfnamefont {J.-K.}\ \bibnamefont {Yuan}}, \bibinfo {author} {\bibfnamefont {S.~A.}\ \bibnamefont {Chen}},\ and\ \bibinfo {author} {\bibfnamefont {P.}~\bibnamefont {Ye}},\ }\bibfield  {title} {\bibinfo {title} {{Fractonic superfluids}},\ }\href {https://doi.org/10.1103/PhysRevResearch.2.023267} {\bibfield  {journal} {\bibinfo  {journal} {Phys. Rev. Research}\ }\textbf {\bibinfo {volume} {2}},\ \bibinfo {pages} {023267} (\bibinfo {year} {2020})}\BibitemShut {NoStop}%
\bibitem [{\citenamefont {Li}\ and\ \citenamefont {Ye}(2020)}]{Li2020}%
  \BibitemOpen
  \bibfield  {author} {\bibinfo {author} {\bibfnamefont {M.-Y.}\ \bibnamefont {Li}}\ and\ \bibinfo {author} {\bibfnamefont {P.}~\bibnamefont {Ye}},\ }\bibfield  {title} {\bibinfo {title} {{Fracton physics of spatially extended excitations}},\ }\href {https://doi.org/10.1103/PhysRevB.101.245134} {\bibfield  {journal} {\bibinfo  {journal} {Phys. Rev. B}\ }\textbf {\bibinfo {volume} {101}},\ \bibinfo {pages} {245134} (\bibinfo {year} {2020})}\BibitemShut {NoStop}%
\bibitem [{\citenamefont {Li}\ and\ \citenamefont {Ye}(2021)}]{Li2021}%
  \BibitemOpen
  \bibfield  {author} {\bibinfo {author} {\bibfnamefont {M.-Y.}\ \bibnamefont {Li}}\ and\ \bibinfo {author} {\bibfnamefont {P.}~\bibnamefont {Ye}},\ }\bibfield  {title} {\bibinfo {title} {{Fracton physics of spatially extended excitations. II. Polynomial ground state degeneracy of exactly solvable models}},\ }\href {https://doi.org/10.1103/PhysRevB.104.235127} {\bibfield  {journal} {\bibinfo  {journal} {Phys. Rev. B}\ }\textbf {\bibinfo {volume} {104}},\ \bibinfo {pages} {235127} (\bibinfo {year} {2021})}\BibitemShut {NoStop}%
\bibitem [{\citenamefont {{Shen}}\ \emph {et~al.}(2022)\citenamefont {{Shen}}, \citenamefont {{Wu}}, \citenamefont {{Li}}, \citenamefont {{Qin}},\ and\ \citenamefont {{Yao}}}]{Shen2022}%
  \BibitemOpen
  \bibfield  {author} {\bibinfo {author} {\bibfnamefont {X.}~\bibnamefont {{Shen}}}, \bibinfo {author} {\bibfnamefont {Z.}~\bibnamefont {{Wu}}}, \bibinfo {author} {\bibfnamefont {L.}~\bibnamefont {{Li}}}, \bibinfo {author} {\bibfnamefont {Z.}~\bibnamefont {{Qin}}},\ and\ \bibinfo {author} {\bibfnamefont {H.}~\bibnamefont {{Yao}}},\ }\bibfield  {title} {\bibinfo {title} {{Fracton topological order at finite temperature}},\ }\href {https://doi.org/10.1103/PhysRevResearch.4.L032008} {\bibfield  {journal} {\bibinfo  {journal} {Phys. Rev. Research}\ }\textbf {\bibinfo {volume} {4}},\ \bibinfo {pages} {L032008} (\bibinfo {year} {2022})}\BibitemShut {NoStop}%
\bibitem [{\citenamefont {{Li}}\ and\ \citenamefont {{Ye}}(2023)}]{Li2023}%
  \BibitemOpen
  \bibfield  {author} {\bibinfo {author} {\bibfnamefont {M.-Y.}\ \bibnamefont {{Li}}}\ and\ \bibinfo {author} {\bibfnamefont {P.}~\bibnamefont {{Ye}}},\ }\bibfield  {title} {\bibinfo {title} {{Hierarchy of entanglement renormalization and long-range entangled states}},\ }\href {https://doi.org/10.1103/PhysRevB.107.115169} {\bibfield  {journal} {\bibinfo  {journal} {Phys. Rev. B}\ }\textbf {\bibinfo {volume} {107}},\ \bibinfo {pages} {115169} (\bibinfo {year} {2023})}\BibitemShut {NoStop}%
\bibitem [{\citenamefont {{Hu}}\ \emph {et~al.}(2025)\citenamefont {{Hu}}, \citenamefont {{Li}},\ and\ \citenamefont {{Ye}}}]{Hu2025}%
  \BibitemOpen
  \bibfield  {author} {\bibinfo {author} {\bibfnamefont {Y.-T.}\ \bibnamefont {{Hu}}}, \bibinfo {author} {\bibfnamefont {M.-Y.}\ \bibnamefont {{Li}}},\ and\ \bibinfo {author} {\bibfnamefont {P.}~\bibnamefont {{Ye}}},\ }\bibfield  {title} {\bibinfo {title} {{Preparing code states via seed-entangler-enriched sequential quantum circuits: Application to tetradigit topological error-correcting codes}},\ }\href {https://doi.org/10.1103/d8gs-fnwt} {\bibfield  {journal} {\bibinfo  {journal} {Phys. Rev. B}\ }\textbf {\bibinfo {volume} {112}},\ \bibinfo {pages} {165139} (\bibinfo {year} {2025})}\BibitemShut {NoStop}%
\bibitem [{\citenamefont {{Li}}\ and\ \citenamefont {{Wu}}(2026)}]{Li2026}%
  \BibitemOpen
  \bibfield  {author} {\bibinfo {author} {\bibfnamefont {M.-Y.}\ \bibnamefont {{Li}}}\ and\ \bibinfo {author} {\bibfnamefont {Y.}~\bibnamefont {{Wu}}},\ }\href@noop {} {\bibinfo {title} {{Fragmented Topological Excitations in Generalized Hypergraph Product Codes}}} (\bibinfo {year} {2026}),\ \Eprint {https://arxiv.org/abs/2601.09850} {arXiv:2601.09850 [quant-ph]} \BibitemShut {NoStop}%
\bibitem [{\citenamefont {{Castelnovo}}\ and\ \citenamefont {{Chamon}}(2007)}]{Castelnovo2007}%
  \BibitemOpen
  \bibfield  {author} {\bibinfo {author} {\bibfnamefont {C.}~\bibnamefont {{Castelnovo}}}\ and\ \bibinfo {author} {\bibfnamefont {C.}~\bibnamefont {{Chamon}}},\ }\bibfield  {title} {\bibinfo {title} {{Topological order and topological entropy in classical systems}},\ }\href {https://doi.org/10.1103/PhysRevB.76.174416} {\bibfield  {journal} {\bibinfo  {journal} {Phys. Rev. B}\ }\textbf {\bibinfo {volume} {76}},\ \bibinfo {pages} {174416} (\bibinfo {year} {2007})}\BibitemShut {NoStop}%
\bibitem [{\citenamefont {{Helmes}}\ \emph {et~al.}(2015)\citenamefont {{Helmes}}, \citenamefont {{St{\'e}phan}},\ and\ \citenamefont {{Trebst}}}]{Helmes2015}%
  \BibitemOpen
  \bibfield  {author} {\bibinfo {author} {\bibfnamefont {J.}~\bibnamefont {{Helmes}}}, \bibinfo {author} {\bibfnamefont {J.-M.}\ \bibnamefont {{St{\'e}phan}}},\ and\ \bibinfo {author} {\bibfnamefont {S.}~\bibnamefont {{Trebst}}},\ }\bibfield  {title} {\bibinfo {title} {{R{\'e}nyi entropy perspective on topological order in classical toric code models}},\ }\href {https://doi.org/10.1103/PhysRevB.92.125144} {\bibfield  {journal} {\bibinfo  {journal} {Phys. Rev. B}\ }\textbf {\bibinfo {volume} {92}},\ \bibinfo {pages} {125144} (\bibinfo {year} {2015})}\BibitemShut {NoStop}%
\bibitem [{\citenamefont {{Liu}}\ \emph {et~al.}(2025)\citenamefont {{Liu}}, \citenamefont {{Chen}}, \citenamefont {{Zhang}}, \citenamefont {{Zhou}},\ and\ \citenamefont {{Zhang}}}]{Liu2025}%
  \BibitemOpen
  \bibfield  {author} {\bibinfo {author} {\bibfnamefont {Z.}~\bibnamefont {{Liu}}}, \bibinfo {author} {\bibfnamefont {L.}~\bibnamefont {{Chen}}}, \bibinfo {author} {\bibfnamefont {Y.}~\bibnamefont {{Zhang}}}, \bibinfo {author} {\bibfnamefont {S.}~\bibnamefont {{Zhou}}},\ and\ \bibinfo {author} {\bibfnamefont {P.}~\bibnamefont {{Zhang}}},\ }\bibfield  {title} {\bibinfo {title} {{Diagnosing strong-to-weak symmetry breaking via Wightman correlators}},\ }\href {https://doi.org/10.1038/s42005-025-02199-7} {\bibfield  {journal} {\bibinfo  {journal} {Commun. Phys.}\ }\textbf {\bibinfo {volume} {8}},\ \bibinfo {pages} {274} (\bibinfo {year} {2025})}\BibitemShut {NoStop}%
\bibitem [{\citenamefont {Zhu}\ \emph {et~al.}(2023)\citenamefont {Zhu}, \citenamefont {Chen}, \citenamefont {Ye},\ and\ \citenamefont {Trebst}}]{Zhu2022}%
  \BibitemOpen
  \bibfield  {author} {\bibinfo {author} {\bibfnamefont {G.-Y.}\ \bibnamefont {Zhu}}, \bibinfo {author} {\bibfnamefont {J.-Y.}\ \bibnamefont {Chen}}, \bibinfo {author} {\bibfnamefont {P.}~\bibnamefont {Ye}},\ and\ \bibinfo {author} {\bibfnamefont {S.}~\bibnamefont {Trebst}},\ }\bibfield  {title} {\bibinfo {title} {{Topological Fracton Quantum Phase Transitions by Tuning Exact Tensor Network States}},\ }\href {https://doi.org/10.1103/PhysRevLett.130.216704} {\bibfield  {journal} {\bibinfo  {journal} {Phys. Rev. Lett.}\ }\textbf {\bibinfo {volume} {130}},\ \bibinfo {pages} {216704} (\bibinfo {year} {2023})}\BibitemShut {NoStop}%
\bibitem [{\citenamefont {{Zhou}}\ \emph {et~al.}(2022)\citenamefont {{Zhou}}, \citenamefont {{Li}}, \citenamefont {{Yan}}, \citenamefont {{Ye}},\ and\ \citenamefont {{Meng}}}]{Zhou2022}%
  \BibitemOpen
  \bibfield  {author} {\bibinfo {author} {\bibfnamefont {C.}~\bibnamefont {{Zhou}}}, \bibinfo {author} {\bibfnamefont {M.-Y.}\ \bibnamefont {{Li}}}, \bibinfo {author} {\bibfnamefont {Z.}~\bibnamefont {{Yan}}}, \bibinfo {author} {\bibfnamefont {P.}~\bibnamefont {{Ye}}},\ and\ \bibinfo {author} {\bibfnamefont {Z.~Y.}\ \bibnamefont {{Meng}}},\ }\bibfield  {title} {\bibinfo {title} {{Evolution of dynamical signature in the X-cube fracton topological order}},\ }\href {https://doi.org/10.1103/PhysRevResearch.4.033111} {\bibfield  {journal} {\bibinfo  {journal} {Phys. Rev. Research}\ }\textbf {\bibinfo {volume} {4}},\ \bibinfo {pages} {033111} (\bibinfo {year} {2022})}\BibitemShut {NoStop}%
\bibitem [{\citenamefont {Ding}\ \emph {et~al.}(2025)\citenamefont {Ding}, \citenamefont {Tang}, \citenamefont {Wang}, \citenamefont {Wang}, \citenamefont {Mao},\ and\ \citenamefont {Yan}}]{Ding2025}%
  \BibitemOpen
  \bibfield  {author} {\bibinfo {author} {\bibfnamefont {Y.-M.}\ \bibnamefont {Ding}}, \bibinfo {author} {\bibfnamefont {Y.}~\bibnamefont {Tang}}, \bibinfo {author} {\bibfnamefont {Z.}~\bibnamefont {Wang}}, \bibinfo {author} {\bibfnamefont {Z.}~\bibnamefont {Wang}}, \bibinfo {author} {\bibfnamefont {B.-B.}\ \bibnamefont {Mao}},\ and\ \bibinfo {author} {\bibfnamefont {Z.}~\bibnamefont {Yan}},\ }\bibfield  {title} {\bibinfo {title} {{Tracking the variation of entanglement R\'enyi negativity: A quantum Monte Carlo study}},\ }\href {https://doi.org/10.1103/PhysRevB.111.L241108} {\bibfield  {journal} {\bibinfo  {journal} {Phys. Rev. B}\ }\textbf {\bibinfo {volume} {111}},\ \bibinfo {pages} {L241108} (\bibinfo {year} {2025})}\BibitemShut {NoStop}%
\bibitem [{\citenamefont {Mao}\ \emph {et~al.}(2025)\citenamefont {Mao}, \citenamefont {Ding}, \citenamefont {Wang}, \citenamefont {Hu},\ and\ \citenamefont {Yan}}]{Mao2025}%
  \BibitemOpen
  \bibfield  {author} {\bibinfo {author} {\bibfnamefont {B.-B.}\ \bibnamefont {Mao}}, \bibinfo {author} {\bibfnamefont {Y.-M.}\ \bibnamefont {Ding}}, \bibinfo {author} {\bibfnamefont {Z.}~\bibnamefont {Wang}}, \bibinfo {author} {\bibfnamefont {S.}~\bibnamefont {Hu}},\ and\ \bibinfo {author} {\bibfnamefont {Z.}~\bibnamefont {Yan}},\ }\bibfield  {title} {\bibinfo {title} {{Sampling reduced density matrix to extract fine levels of entanglement spectrum and restore entanglement Hamiltonian}},\ }\href {https://doi.org/10.1038/s41467-025-58058-0} {\bibfield  {journal} {\bibinfo  {journal} {Nat. Commun.}\ }\textbf {\bibinfo {volume} {16}},\ \bibinfo {pages} {2880} (\bibinfo {year} {2025})}\BibitemShut {NoStop}%
\bibitem [{\citenamefont {{Calabrese}}\ and\ \citenamefont {{Cardy}}(2004)}]{Calabrese2004}%
  \BibitemOpen
  \bibfield  {author} {\bibinfo {author} {\bibfnamefont {P.}~\bibnamefont {{Calabrese}}}\ and\ \bibinfo {author} {\bibfnamefont {J.}~\bibnamefont {{Cardy}}},\ }\bibfield  {title} {\bibinfo {title} {{Entanglement entropy and quantum field theory}},\ }\href {https://doi.org/10.1088/1742-5468/2004/06/P06002} {\bibfield  {journal} {\bibinfo  {journal} {J. Stat. Mech.: Theory Exp.}\ }\textbf {\bibinfo {volume} {2004}}\bibinfo  {number} { (06)},\ \bibinfo {pages} {P06002}}\BibitemShut {NoStop}%
\bibitem [{\citenamefont {Castelnovo}\ and\ \citenamefont {Chamon}(2007)}]{Castelnovo2007a}%
  \BibitemOpen
\bibfield  {number} {  }\bibfield  {author} {\bibinfo {author} {\bibfnamefont {C.}~\bibnamefont {Castelnovo}}\ and\ \bibinfo {author} {\bibfnamefont {C.}~\bibnamefont {Chamon}},\ }\bibfield  {title} {\bibinfo {title} {{Entanglement and topological entropy of the toric code at finite temperature}},\ }\href {https://doi.org/10.1103/PhysRevB.76.184442} {\bibfield  {journal} {\bibinfo  {journal} {Phys. Rev. B}\ }\textbf {\bibinfo {volume} {76}},\ \bibinfo {pages} {184442} (\bibinfo {year} {2007})}\BibitemShut {NoStop}%
\bibitem [{\citenamefont {Castelnovo}\ and\ \citenamefont {Chamon}(2008)}]{Castelnovo2008}%
  \BibitemOpen
  \bibfield  {author} {\bibinfo {author} {\bibfnamefont {C.}~\bibnamefont {Castelnovo}}\ and\ \bibinfo {author} {\bibfnamefont {C.}~\bibnamefont {Chamon}},\ }\bibfield  {title} {\bibinfo {title} {{Topological order in a three-dimensional toric code at finite temperature}},\ }\href {https://doi.org/10.1103/PhysRevB.78.155120} {\bibfield  {journal} {\bibinfo  {journal} {Phys. Rev. B}\ }\textbf {\bibinfo {volume} {78}},\ \bibinfo {pages} {155120} (\bibinfo {year} {2008})}\BibitemShut {NoStop}%
\bibitem [{\citenamefont {{Eisert}}\ \emph {et~al.}(2010)\citenamefont {{Eisert}}, \citenamefont {{Cramer}},\ and\ \citenamefont {{Plenio}}}]{Eisert2010}%
  \BibitemOpen
  \bibfield  {author} {\bibinfo {author} {\bibfnamefont {J.}~\bibnamefont {{Eisert}}}, \bibinfo {author} {\bibfnamefont {M.}~\bibnamefont {{Cramer}}},\ and\ \bibinfo {author} {\bibfnamefont {M.~B.}\ \bibnamefont {{Plenio}}},\ }\bibfield  {title} {\bibinfo {title} {{Colloquium: Area laws for the entanglement entropy}},\ }\href {https://doi.org/10.1103/RevModPhys.82.277} {\bibfield  {journal} {\bibinfo  {journal} {Rev. Mod. Phys.}\ }\textbf {\bibinfo {volume} {82}},\ \bibinfo {pages} {277} (\bibinfo {year} {2010})}\BibitemShut {NoStop}%
\bibitem [{\citenamefont {Kalis}\ \emph {et~al.}(2012)\citenamefont {Kalis}, \citenamefont {Klagges}, \citenamefont {Or\'us},\ and\ \citenamefont {Schmidt}}]{Kalis2012}%
  \BibitemOpen
  \bibfield  {author} {\bibinfo {author} {\bibfnamefont {H.}~\bibnamefont {Kalis}}, \bibinfo {author} {\bibfnamefont {D.}~\bibnamefont {Klagges}}, \bibinfo {author} {\bibfnamefont {R.}~\bibnamefont {Or\'us}},\ and\ \bibinfo {author} {\bibfnamefont {K.~P.}\ \bibnamefont {Schmidt}},\ }\bibfield  {title} {\bibinfo {title} {{Fate of the cluster state on the square lattice in a magnetic field}},\ }\href {https://doi.org/10.1103/PhysRevA.86.022317} {\bibfield  {journal} {\bibinfo  {journal} {Phys. Rev. A}\ }\textbf {\bibinfo {volume} {86}},\ \bibinfo {pages} {022317} (\bibinfo {year} {2012})}\BibitemShut {NoStop}%
\bibitem [{\citenamefont {{Li}}\ and\ \citenamefont {{Mong}}(2019)}]{Li2019}%
  \BibitemOpen
  \bibfield  {author} {\bibinfo {author} {\bibfnamefont {Z.}~\bibnamefont {{Li}}}\ and\ \bibinfo {author} {\bibfnamefont {R.~S.~K.}\ \bibnamefont {{Mong}}},\ }\href@noop {} {\bibinfo {title} {{Finite-temperature topological entanglement entropy for CSS codes}}} (\bibinfo {year} {2019}),\ \Eprint {https://arxiv.org/abs/1910.07545} {arXiv:1910.07545 [cond-mat.str-el]} \BibitemShut {NoStop}%
\bibitem [{\citenamefont {M\"uhlhauser}\ \emph {et~al.}(2020)\citenamefont {M\"uhlhauser}, \citenamefont {Walther}, \citenamefont {Reiss},\ and\ \citenamefont {Schmidt}}]{Muehlhauser2020}%
  \BibitemOpen
  \bibfield  {author} {\bibinfo {author} {\bibfnamefont {M.}~\bibnamefont {M\"uhlhauser}}, \bibinfo {author} {\bibfnamefont {M.~R.}\ \bibnamefont {Walther}}, \bibinfo {author} {\bibfnamefont {D.~A.}\ \bibnamefont {Reiss}},\ and\ \bibinfo {author} {\bibfnamefont {K.~P.}\ \bibnamefont {Schmidt}},\ }\bibfield  {title} {\bibinfo {title} {{Quantum robustness of fracton phases}},\ }\href {https://doi.org/10.1103/PhysRevB.101.054426} {\bibfield  {journal} {\bibinfo  {journal} {Phys. Rev. B}\ }\textbf {\bibinfo {volume} {101}},\ \bibinfo {pages} {054426} (\bibinfo {year} {2020})}\BibitemShut {NoStop}%
\bibitem [{\citenamefont {Zhou}\ and\ \citenamefont {Ye}(2024)}]{ZhouYe2024}%
  \BibitemOpen
  \bibfield  {author} {\bibinfo {author} {\bibfnamefont {Y.}~\bibnamefont {Zhou}}\ and\ \bibinfo {author} {\bibfnamefont {P.}~\bibnamefont {Ye}},\ }\bibfield  {title} {\bibinfo {title} {{Entanglement fractalization}},\ }\href {https://doi.org/10.1103/physrevresearch.6.043145} {\bibfield  {journal} {\bibinfo  {journal} {Phys. Rev. Research}\ }\textbf {\bibinfo {volume} {6}},\ \bibinfo {pages} {043145} (\bibinfo {year} {2024})}\BibitemShut {NoStop}%
\bibitem [{\citenamefont {Huang}\ \emph {et~al.}(2025)\citenamefont {Huang}, \citenamefont {Zhou},\ and\ \citenamefont {Ye}}]{HuangZhouYe2025}%
  \BibitemOpen
  \bibfield  {author} {\bibinfo {author} {\bibfnamefont {X.-Y.}\ \bibnamefont {Huang}}, \bibinfo {author} {\bibfnamefont {Y.}~\bibnamefont {Zhou}},\ and\ \bibinfo {author} {\bibfnamefont {P.}~\bibnamefont {Ye}},\ }\bibfield  {title} {\bibinfo {title} {{Entanglement scaling behaviors of free fermions on hyperbolic lattices}},\ }\href {https://doi.org/10.1103/physrevresearch.7.023098} {\bibfield  {journal} {\bibinfo  {journal} {Phys. Rev. Research}\ }\textbf {\bibinfo {volume} {7}},\ \bibinfo {pages} {023098} (\bibinfo {year} {2025})}\BibitemShut {NoStop}%
\bibitem [{\citenamefont {Li}\ and\ \citenamefont {Haldane}(2008)}]{Li2008}%
  \BibitemOpen
  \bibfield  {author} {\bibinfo {author} {\bibfnamefont {H.}~\bibnamefont {Li}}\ and\ \bibinfo {author} {\bibfnamefont {F.~D.~M.}\ \bibnamefont {Haldane}},\ }\bibfield  {title} {\bibinfo {title} {{Entanglement Spectrum as a Generalization of Entanglement Entropy: Identification of Topological Order in Non-Abelian Fractional Quantum Hall Effect States}},\ }\href {https://doi.org/10.1103/PhysRevLett.101.010504} {\bibfield  {journal} {\bibinfo  {journal} {Phys. Rev. Lett.}\ }\textbf {\bibinfo {volume} {101}},\ \bibinfo {pages} {010504} (\bibinfo {year} {2008})}\BibitemShut {NoStop}%
\bibitem [{\citenamefont {{Popescu}}\ \emph {et~al.}(2006)\citenamefont {{Popescu}}, \citenamefont {{Short}},\ and\ \citenamefont {{Winter}}}]{Popescu2006}%
  \BibitemOpen
  \bibfield  {author} {\bibinfo {author} {\bibfnamefont {S.}~\bibnamefont {{Popescu}}}, \bibinfo {author} {\bibfnamefont {A.~J.}\ \bibnamefont {{Short}}},\ and\ \bibinfo {author} {\bibfnamefont {A.}~\bibnamefont {{Winter}}},\ }\bibfield  {title} {\bibinfo {title} {{Entanglement and the foundations of statistical mechanics}},\ }\href {https://doi.org/10.1038/nphys444} {\bibfield  {journal} {\bibinfo  {journal} {Nat. Phys.}\ }\textbf {\bibinfo {volume} {2}},\ \bibinfo {pages} {754} (\bibinfo {year} {2006})}\BibitemShut {NoStop}%
\end{thebibliography}

\appendix

\end{document}